\documentclass[10pt,fleqn]{article}
\usepackage{times,latexsym,amssymb,amsmath,theorem,epsfig}
\usepackage{mathptm}  
\usepackage[colorlinks=true]{hyperref}

\newtheorem{theorem}{Theorem}[section]
\newtheorem{algorithm}[theorem]{Algorithm}

\newtheorem{corollary}[theorem]{Corollary}
\newtheorem{definition}[theorem]{Definition}
{\theorembodyfont{\rmfamily} \newtheorem{example}[theorem]{Example}}
\newtheorem{lemma}[theorem]{Lemma}
\newtheorem{problem}[theorem]{Problem}
\newtheorem{procedure}[theorem]{Procedure}

\newtheorem{proposition}[theorem]{Proposition}

{\theorembodyfont{\rmfamily} \newtheorem{remark}[theorem]{Remark}}

\begin{document}

\newcommand{\CDC}[1]{\textcolor{blue}{{#1}}}
\newcommand{\CDCC}[1]{\textcolor{green}{{#1}}}

\newcommand{\JVS}[1]{\textcolor{red}{{#1}}}

\newcommand{\sign}{\mbox{sign}}

\newcommand{\mathbbd}{\mbox{${\mathbb{D}}$}} 
\newcommand{\DS}{\mbox{${\mathbb{DS}}$}} 
\newcommand{\DSC}{\mbox{${\mathbb{DSC}}$}} 
\newcommand{\M}{\mbox{${\mathbb{M}}$}} 
\newcommand{\N}{\mbox{${\mathbb{N}}$}} 
\newcommand{\PP}{\mbox{${\mathbb{P}}$}} 
\newcommand{\Q}{\mbox{${\mathbb{Q}}$}} 
\newcommand{\W}{\mbox{${\mathbb{W}}$}} 
\newcommand{\C}{\mbox{$C$}} 
\newcommand{\R}{\mbox{$R$}} 
\newcommand{\T}{\mbox{$T$}} 
\newcommand{\Z}{\mbox{$Z$}}

\newcommand{\boundary}{\mbox{$\rm bndry$}}
\newcommand{\card}{\mbox{$\rm card$}}
\newcommand{\cisetsmin}{\mbox{$\rm CISets_{\min}$}}
\newcommand{\cisets}{\mbox{$\rm CISets$}}
\newcommand{\cix}{\mbox{$\rm CIX$}}
\newcommand{\colpwr}{\mbox{Colpwr}}
\newcommand{\closure}{\mbox{closure}}
\newcommand{\da}{\mbox{$\rm DA$}}
\newcommand{\domainofattraction}{\mbox{$DA$}}
\newcommand{\ds}{\mbox{$\rm DS$}}
\newcommand{\graph}{\mbox{Graph}}
\newcommand{\init}{\mbox{Init}}
\newcommand{\interior}{\mbox{int}}
\newcommand{\Mp}{\mbox{\bf M}}
\newcommand{\powerset}{\mbox{Pwrset}}
\newcommand{\pwr}{\mbox{Pwrset}}

\newcommand{\orderlexg}{\mbox{$>_{{\rm lex}}$}}
\newcommand{\orderlexl}{\mbox{$<_{{\rm lex}}$}}
\newcommand{\nm}{\mbox{$\mathbb{N}_m$}}
\newcommand{\relationconnect}{\mbox{${\rm R_{connect}}$}}
\newcommand{\zk}{\mbox{$\mathbb{Z}_k$}}
\newcommand{\zn}{\mbox{$\mathbb{Z}_n$}}
\newcommand{\zm}{\mbox{$\mathbb{Z}_m$}}
\newcommand{\zo}{\mbox{${\rm O}$}} 
\newcommand{\zr}{\mbox{$\mathbb{Z}_r$}}
\newcommand{\zp}{\mbox{$\mathbb{Z}_p$}}
\newcommand{\zpos}{\mbox{$\mathbb{Z}_{+}$}}
\newcommand{\zposn}{\mbox{$\mathbb{Z}_{+}^n$}}

\newcommand{\almosteverywhere}{\mbox{$\forall E$}}
\newcommand{\realnumbersextended}{\mbox{$\overline{\mathbb{R}}$}}
\newcommand{\real}{\mbox{$\mathbb{R}$}}
\newcommand{\reals}{\mbox{$\mathbb{R}$}}

\newcommand{\cdiscclosed}{\mbox{$\mathbb{D}_c$}}
\newcommand{\cdiscopen}{\mbox{$\mathbb{D}_o$}}
\newcommand{\cdiscoutopen}{\mbox{$(\mathbb{D}^c)_o$}}
\newcommand{\cminus}{\mbox{$\mathbb{C}^-$}}
\newcommand{\cplus}{\mbox{$\mathbb{C}^+$}}
\newcommand{\im}{\mbox{\rm Im}}
\newcommand{\re}{\mbox{\rm Re}}

\newcommand{\cilspaces}{\mbox{${\rm CILSpaces}$}} 
\newcommand{\cilssp}{\mbox{${\rm CILSP}$}} 
\newcommand{\cilsp}{\mbox{${\rm CILSp}$}} 
\newcommand{\cli}{\mbox{${\rm CLI}$}} 
\newcommand{\climin}{\mbox{${\rm CLI_{min}}$}} 
\newcommand{\cn}{\mbox{$\mathbb{C}^n$}} 
\newcommand{\cmcm}{\mbox{$\mathbb{C}^m$}} 
\newcommand{\linearsubspaces}{\mbox{${\rm LinSubspaces}$}} 
\newcommand{\lat}{\mbox{${\rm Lat}$}} 
\newcommand{\quotientspaces}{\mbox{${\rm Quotientspaces}$}} 
\newcommand{\rinfty}{\mbox{$\mathbb{R}^{\infty}$}}
\newcommand{\rk}{\mbox{$\mathbb{R}^k$}}
\newcommand{\rmrm}{\mbox{$\mathbb{R}^m$}}
\newcommand{\rn}{\mbox{$\mathbb{R}^n$}} 
\newcommand{\rp}{\mbox{$\mathbb{R}^p$}}
\newcommand{\rpos}{\mbox{$\mathbb{R}_{+}$}}
\newcommand{\rposk}{\mbox{$\mathbb{R}_{+}^k$}}
\newcommand{\rposm}{\mbox{$\mathbb{R}_{+}^m$}}
\newcommand{\rposn}{\mbox{$\mathbb{R}_{+}^n$}}
\newcommand{\rposp}{\mbox{$\mathbb{R}_{+}^p$}}
\newcommand{\rr}{\mbox{$\mathbb{R}^r$}}
\newcommand{\rspos}{\mbox{$\mathbb{R}_{s+}$}}
\newcommand{\rsposn}{\mbox{$\mathbb{R}_{s+}^n$}}
\newcommand{\simplexn}{\mbox{$\mathbb{S}_{+}^n$}}
\newcommand{\sn}{\mbox{$\mathbb{S}_{+}^n$}}

\newcommand{\cimod}{\mbox{${\rm CIModules}$}} 

\newcommand{\switch}{\mbox{${\rm Switch}$}} 
\newcommand{\threshold}{\mbox{${\rm Threshold}$}}

\newcommand{\ara}{\mbox{${\rm ARA}$}} 
\newcommand{\dom}{\mbox{\rm Dom}}
\newcommand{\Fc}{\mbox{${\bf F}$}} 
\newcommand{\hara}{\mbox{${\rm HARA}$}} 
\newcommand{\range}{\mbox{\rm Range}}
\newcommand{\rra}{\mbox{${\rm RRA}$}} 
\newcommand{\U}{\mbox{${\rm U}$}} 
\newcommand{\A}{\mbox{${\rm A}$}} 

\newcommand{\fanal}{\mathcal{C}^{\omega}}
\newcommand{\fsmooth}{\mathcal{C}^{\infty}}
\newcommand{\fcont}{\mathcal{C}}
\newcommand{\fpcont}{\mathcal{PC}}
\newcommand{\fpconst}{\mathcal{PC}onst}

\newcommand{\arrow}{\mbox{${\rm arrow}$}}
\newcommand{\blockdiagonal}{\mbox{${\rm Bdiag}$}}
\newcommand{\cnm}{\mbox{$\mathbb{C}^{n \times m}$}}
\newcommand{\cnn}{\mbox{$\mathbb{C}^{n \times n}$}}
\newcommand{\cnp}{\mbox{$\mathbb{C}^{n \times p}$}}
\newcommand{\cpn}{\mbox{$\mathbb{C}^{p \times n}$}}
\newcommand{\cpp}{\mbox{$\mathbb{C}^{p \times p}$}}
\newcommand{\cum}{\mbox{${\rm cum}$}}
\newcommand{\dnn}{\mbox{$\mathbb{R}_{\rm diag}^{n \times n}$}}
\newcommand{\dnnpos}{\mbox{$\mathbb{R}_{\rm +, diag}^{n \times n}$}}
\newcommand{\dnnspos}{\mbox{$\mathbb{R}_{\rm diag, +}^{n \times n}$}}
\newcommand{\imp}{\mbox{${\rm Imprim}$}}
\newcommand{\invertible}{\mbox{${\rm inv}$}}
\newcommand{\rkk}{\mbox{$\mathbb{R}^{k \times k}$}}
\newcommand{\rkm}{\mbox{$\mathbb{R}^{k \times m}$}}
\newcommand{\rkn}{\mbox{$\mathbb{R}^{k \times n}$}}
\newcommand{\rmk}{\mbox{$\mathbb{R}^{m \times k}$}}
\newcommand{\rmm}{\mbox{$\mathbb{R}^{m \times m}$}}
\newcommand{\rmn}{\mbox{$\mathbb{R}^{m \times n}$}}
\newcommand{\rmnarrow}{\mbox{$\mathbb{R}_{\arrow}^{m \times n}$}}
\newcommand{\rmnblockdiagonal}{\mbox{$\mathbb{R}_{{\rm Bdiag}}^{m \times n}$}}
\newcommand{\rmncoordinated}{\mbox{$\mathbb{R}_{{\rm c}}^{m \times n}$}}
\newcommand{\rmr}{\mbox{$\mathbb{R}^{m \times r}$}}
\newcommand{\rnk}{\mbox{$\mathbb{R}^{n \times k}$}}
\newcommand{\rnm}{\mbox{$\mathbb{R}^{n \times m}$}}
\newcommand{\rnn}{\mbox{$\mathbb{R}^{n \times n}$}}
\newcommand{\rnnarrow}{\mbox{$\mathbb{R}_{\arrow}^{n \times n}$}}
\newcommand{\rnncoordinated}{\mbox{$\mathbb{R}_{{\rm c}}^{n \times n}$}}
\newcommand{\rnnspd}{\mbox{$\mathbb{R}_{spd}^{n \times n}$}}
\newcommand{\rnnsspd}{\mbox{$\mathbb{R}_{sspd}^{n \times n}$}}
\newcommand{\rnp}{\mbox{$\mathbb{R}^{n \times p}$}}
\newcommand{\rnr}{\mbox{$\mathbb{R}^{n \times r}$}}
\newcommand{\rpk}{\mbox{$\mathbb{R}^{p \times k}$}}
\newcommand{\rpm}{\mbox{$\mathbb{R}^{p \times m}$}}
\newcommand{\rpn}{\mbox{$\mathbb{R}^{p \times n}$}}
\newcommand{\rpp}{\mbox{$\mathbb{R}^{p \times p}$}}
\newcommand{\rpr}{\mbox{$\mathbb{R}^{p \times r}$}}
\newcommand{\rrn}{\mbox{$\mathbb{R}^{r \times n}$}}
\newcommand{\rrr}{\mbox{$\mathbb{R}^{r \times r}$}}
\newcommand{\sposdefnn}{\mbox{$\mathbb{RSPD}^{n \times n}$}}
\newcommand{\spd}{\mbox{$\mathbb{R}_{\rm spd}^{n \times n}$}}

\newcommand{\dposnn}{\mbox{$\mathbb{R}_{+,{\rm diag}}^{n \times n}$}}
\newcommand{\dpossnn}{\mbox{$\mathbb{R}_{s+,{\rm diag}}^{n \times n}$}}
\newcommand{\dsposnn}{\mbox{$\mathbb{R}_{s+,{\rm diag}}^{n \times n}$}}
\newcommand{\dscnn}{\mbox{$\mathbb{DSC}_{+}^{n \times n}$}}
\newcommand{\dsnn}{\mbox{$\mathbb{DS}_{+}^{n \times n}$}}
\newcommand{\indeximprim}{\mbox{\rm imprim}}
\newcommand{\indexprim}{\mbox{\rm indexprim}}
\newcommand{\imprim}{\mbox{\rm imprim}}
\newcommand{\matrixrowtrunc}{\mbox{$\rm matrowtrunc$}}
\newcommand{\matrixdiagtrunc}{\mbox{$\rm mdtrunc$}}
\newcommand{\mnn}{\mbox{$\mathbb{M}_{+}^{n \times n}$}}
\newcommand{\mposkk}{\mbox{$M_{+}^{k \times k}$}}
\newcommand{\mposmm}{\mbox{$M_{+}^{m \times m}$}}
\newcommand{\mposnn}{\mbox{$M_{+}^{n \times n}$}}
\newcommand{\nn}{\mbox{$\mathbb{N}_n$}}
\newcommand{\permnn}{\mbox{$\mathbb{P}^{n \times n}$}}
\newcommand{\permpp}{\mbox{$\mathbb{P}^{p \times p}$}}
\newcommand{\posR}[2]{\mbox{$R_{+}^{#1\times#2}$}}
\newcommand{\rposik}{\mbox{$\mathbb{R}_{+}^{\infty\times k}$}}
\newcommand{\rposim}{\mbox{$\mathbb{R}_{+}^{\infty\times m}$}}
\newcommand{\rposin}{\mbox{$\mathbb{R}_{+}^{\infty\times n}$}}
\newcommand{\rposiq}{\mbox{$\mathbb{R}_{+}^{\infty\times q}$}}
\newcommand{\rposkk}{\mbox{$\mathbb{R}_{+}^{k \times k}$}}
\newcommand{\rposkm}{\mbox{$\mathbb{R}_{+}^{k \times m}$}}
\newcommand{\rposkn}{\mbox{$\mathbb{R}_{+}^{k \times n}$}}
\newcommand{\rposkp}{\mbox{$\mathbb{R}_{+}^{k \times p}$}}
\newcommand{\rposmm}{\mbox{$\mathbb{R}_{+}^{m \times m}$}}
\newcommand{\rposmk}{\mbox{$\mathbb{R}_{+}^{m \times k}$}}
\newcommand{\rposmn}{\mbox{$\mathbb{R}_{+}^{m \times n}$}}
\newcommand{\rposnk}{\mbox{$\mathbb{R}_{+}^{n \times k}$}}
\newcommand{\rposnm}{\mbox{$\mathbb{R}_{+}^{n \times m}$}}
\newcommand{\rposnn}{\mbox{$\mathbb{R}_{+}^{n \times n}$}}
\newcommand{\rposnp}{\mbox{$\mathbb{R}_{+}^{n \times p}$}}
\newcommand{\rpospm}{\mbox{$\mathbb{R}_{+}^{p \times m}$}}
\newcommand{\rpospn}{\mbox{$\mathbb{R}_{+}^{p \times n}$}}
\newcommand{\rposqm}{\mbox{$\mathbb{R}_{+}^{q \times m}$}}
\newcommand{\rposqn}{\mbox{$\mathbb{R}_{+}^{q \times n}$}}
\newcommand{\rposqq}{\mbox{$\mathbb{R}_{+}^{q \times q}$}}
\newcommand{\rsposnn}{\mbox{$\mathbb{R}_{s+}^{n \times n}$}}
\newcommand{\Table}{\mbox{$\rm table$}}
\newcommand{\wnn}{\mbox{$\W^{n \times n}$}}

\newcommand{\adjoint}{\mbox{$\rm Adj$}}
\newcommand{\affine}{\mbox{$\rm Affine$}}
\newcommand{\affineb}{\mbox{$\rm AffB$}}
\newcommand{\blockdiag}{\mbox{{\rm Block-diag}}}
\newcommand{\circulant}{\mbox{$\rm Circulant$}}
\newcommand{\cols}{\mbox{$\rm col$}}
\newcommand{\diag}{\mbox{$\rm Diag$}}
\newcommand{\dsc}{doubly stochastic circulant }
\newcommand{\dscs}{doubly stochastic circulants }
\newcommand{\eig}{\mbox{$\rm eig$}}
\newcommand{\lspan}{\mbox{\rm span}}
\newcommand{\posr}[1]{#1\mbox{\rm -pos-rank}}
\newcommand{\posrank}{\mbox{$\rm pos-rank$}}
\newcommand{\projection}{\mbox{$\rm proj$}}
\newcommand{\rank}{\mbox{$\rm rank$}}
\newcommand{\rowrankz}{\mbox{$\rm row-rank_Z$}}
\newcommand{\specrad}{\mbox{$\rm specrad$}}
\newcommand{\spec}{\mbox{$\rm spec$}}
\newcommand{\spectrum}{\mbox{$\rm spec$}}
\newcommand{\tr}{\mbox{$\rm tr$}}
\newcommand{\trace}{\mbox{\rm trace}}

\newcommand{\affinehull}{\mbox{$\rm affh$}}
\newcommand{\convexhull}{\mbox{$\rm convh$}}
\newcommand{\ri}{\mbox{\rm ri}}

\newcommand{\cekk}{\mbox{$CE_{k,k}$}}
\newcommand{\cekm}{\mbox{$CE_{k,m}$}}
\newcommand{\cekn}{\mbox{$CE_{k,n}$}}
\newcommand{\cenn}{\mbox{$CE_{n,n}$}}
\newcommand{\ckk}{\mbox{$\mathbb{C}_{k,k}$}}
\newcommand{\ckm}{\mbox{$\mathbb{C}_{k,m}$}}
\newcommand{\cone}{\mbox{\rm cone}} 
\newcommand{\faces}{\mbox{\rm Faces}}
\newcommand{\hp}{\mbox{\rm HyperPlane}}
\newcommand{\normalvector}{\mbox{$V_{normal}$}}
\newcommand{\phs}{\mbox{$PHS$}}
\newcommand{\plsets}{\mbox{$PLSets$}}
\newcommand{\polyhedralcone}{\mbox{\rm Polyhcone}}
\newcommand{\shp}{\mbox{\rm SupportHyperPlane}}
\newcommand{\subpolytope}{\mbox{\rm SubPolytope}}
\newcommand{\subrectangle}{\mbox{\rm SubRectangle}}
\newcommand{\vertices}{\mbox{$V_{vertices}$}}

\newcommand{\qy}{\mbox{$Q_y$}}
\newcommand{\qlsdp}{\mbox{${\bf \partial Q_{lsdp}}$}}
\newcommand{\qp}{\mbox{${\bf Q_{lsp}}$}}
\newcommand{\qpd}{\mbox{${\bf Q_{lsdp}}$}}
\newcommand{\qpr}{\mbox{${\bf Q_{lsp,r}}$}}
\newcommand{\qprs}{\mbox{${\bf Q_{lsp,s}}$}}
\newcommand{\qpdr}{\mbox{${\bf Q_{lsdp,r}}$}}
\newcommand{\dqps}{\mbox{${\bf \partial Q_{lsp,r}}$}}
\newcommand{\dqpss}{\mbox{${\bf \partial Q_{lsp,s}}$}}
\newcommand{\dqpdr}{\mbox{${\bf \partial Q_{lsdp,r}}$}}
\newcommand{\dqprs}{\mbox{${\bf \partial Q_{lsp,r,s}}$}}
\newcommand{\dqpdrs}{\mbox{${\bf \partial Q_{lsdp,r,s}}$}}
\newcommand{\dqp}{\mbox{${\bf \partial Q_{lsp}}$}}
\newcommand{\dqpd}{\mbox{${\bf \partial Q_{lsdp}}$}}

\newcommand{\con}{\mbox{${\rm con}$}}
\newcommand{\conmat}{\mbox{${\rm conmat}$}}
\newcommand{\conset}{\mbox{${\rm Conset}$}}
\newcommand{\coconset}{\mbox{${\rm co-Conset}$}}
\newcommand{\controllablepair}{\mbox{${\rm conpair}$}}
\newcommand{\controllabilitymatrix}{\mbox{${\rm conmat}$}}
\newcommand{\controllableset}{\mbox{${\rm conset}$}}
\newcommand{\cocontrollableset}{\mbox{${\rm co-conset}$}}
\newcommand{\controlset}{\mbox{$\rm controlset$}}
\newcommand{\ls}{\mbox{${\rm LS}$}}
\newcommand{\lsp}{\mbox{${\rm LSP}$}}
\newcommand{\lspmin}{\mbox{${\rm LSP_{min}}$}}
\newcommand{\reachm}{\mbox{${\rm reachm}$}}
\newcommand{\obsm}{\mbox{${\rm obsm}$}}
\newcommand{\obsmat}{\mbox{${\rm obsmat}$}}
\newcommand{\obsmap}{\mbox{${\rm obsmap}$}}
\newcommand{\realization}{\mbox{${\rm realiz}$}}
\newcommand{\reconmap}{\mbox{${\rm reconmap}$}}

\newcommand{\argmin}{\mbox{${\rm argmin}$}}
\newcommand{\argmax}{\mbox{${\rm argmax}$}}
\newcommand{\rai}{\mbox{${\rm RAI}$}}

\newcommand{\obs}{\mbox{${\rm obs}$}}
\newcommand{\aobs}{\mbox{${\rm A_{obs}}$}}
\newcommand{\qobs}{\mbox{${\rm Q_{obs}}$}}
\newcommand{\svdtruncation}{\mbox{${\rm SVDtrunc}$}}
\newcommand{\zclosure}{\mbox{${\rm \mathcal{Z}-cl}$}}

\newcommand{\act}{\mbox{$\rm act$}}
\newcommand{\aux}{\mbox{$\rm Aux$}}
\newcommand{\cat}{\mbox{$\rm cat$}}
\newcommand{\cig}{\mbox{$\rm CIG$}}
\newcommand{\cigmin}{\mbox{$\rm CIG_{\min}$}}
\newcommand{\cil}{\mbox{$\rm CIL$}}
\newcommand{\child}{\mbox{$\rm Chi$}}
\newcommand{\co}{\mbox{$\rm CO$}}
\newcommand{\codes}{\mbox{$\rm CODES$}}
\newcommand{\controllabletriple}{\mbox{$\rm ConTriple$}}
\newcommand{\coreach}{\mbox{$\rm coreach$}}
\newcommand{\coreachcomponent}{\mbox{$\rm coreachco$}}
\newcommand{\coreachgen}{\mbox{$\rm coreachgen$}}
\newcommand{\coreachset}{\mbox{$\rm coreachset$}}
\newcommand{\csublanguage}{\mbox{$\rm C$}}
\newcommand{\cpcsublanguage}{\mbox{$\rm C_{pc}$}}
\newcommand{\csuplanguage}{\mbox{$\rm CSupL$}}
\newcommand{\cpcsuplanguage}{\mbox{$\rm CSupL_{pc}$}}
\newcommand{\decdes}{\mbox{$\rm decDES$}}
\newcommand{\infcsuplanguage}{\mbox{$\inf {\rm CSupL}$}}
\newcommand{\infcsuppclanguage}{\mbox{$\inf {\rm CSupL_{pc}}$}}
\newcommand{\last}{\mbox{$\rm last$}}
\newcommand{\length}{\mbox{$\rm length$}}
\newcommand{\markact}{\mbox{$\rm markact$}}
\newcommand{\moddes}{\mbox{$\rm modDES$}}
\newcommand{\nextact}{\mbox{$\rm nextact$}}
\newcommand{\normaltuple}{\mbox{$\rm NormalTuple$}}
\newcommand{\parent}{\mbox{$\rm Par$}}
\newcommand{\prefix}{\mbox{$\rm prefix$}}
\newcommand{\ps}{\mbox{$\rm PS$}}
\newcommand{\qeq}{\mbox{$\rm QEQ$}}
\newcommand{\reachcomponent}{\mbox{$\rm reachco$}}
\newcommand{\reachgen}{\mbox{$\rm reachgen$}}
\newcommand{\reachset}{\mbox{$\rm reachset$}}
\newcommand{\runs}{\mbox{$\rm Runs$}}
\newcommand{\selfloop}{\mbox{$\rm selfloop$}}
\newcommand{\setldfg}{\mbox{$\rm SetL_{DFG}$}}
\newcommand{\setllc}{\mbox{$\rm SetL_{lc}$}}
\newcommand{\setllcn}{\mbox{$\rm SetL_{lcn}$}}
\newcommand{\setln}{\mbox{$\rm SetL_N$}}
\newcommand{\setlnfg}{\mbox{$\rm SetL_{NFG}$}}
\newcommand{\setlreg}{\mbox{$\rm SetL_{reg}$}}
\newcommand{\suffix}{\mbox{$\rm suffix$}}
\newcommand{\shuffle}{\mbox{$\rm shuffle$}}
\newcommand{\size}{\mbox{$\rm Size$}}
\newcommand{\starrr}{\mbox{$\rm Star$}}
\newcommand{\supap}{\mbox{$\sup {\rm AP}$}}
\newcommand{\supc}{\mbox{$\sup {\rm C}$}}
\newcommand{\supcc}{\mbox{$\sup {\rm cC}$}}
\newcommand{\supccn}{\mbox{$\sup {\rm cCN}$}}
\newcommand{\supcsublanguage}{\mbox{$\sup {\rm C}$}}
\newcommand{\supcpcsublanguage}{\mbox{$\sup {\rm C_{pc}}$}}
\newcommand{\supcnsublanguage}{\mbox{$\sup {\rm CN}$}}
\newcommand{\supcn}{\mbox{$\sup {\rm CN}$}}
\newcommand{\supmccn}{\mbox{$\sup {\rm mcCN}$}}
\newcommand{\suppc}{\mbox{$\sup {\rm PC}$}}
\newcommand{\suppn}{\mbox{$\sup {\rm PN}$}}
\newcommand{\supn}{\mbox{$\sup {\rm N}$}}
\newcommand{\TIME}{\mbox{$\rm TIME$}}
\newcommand{\transrel}{\mbox{$\rm Tr$}}
\newcommand{\trim}{\mbox{$\rm trim$}}
\newcommand{\trimgen}{\mbox{$\rm trimgen$}}
\newcommand{\triple}{\mbox{$\rm tri$}}
\newcommand{\uc}{\mbox{$\rm uc$}}

\newcommand{\cogstocsp}{\mbox{${\rm COGStocSP}$}}
\newcommand{\cvf}{\mbox{${\rm cvf}$}}
\newcommand{\gstocs}{\mbox{${\rm GStocS}$}}
\newcommand{\gstocsp}{\mbox{${\rm GStocSP}$}}
\newcommand{\gstoccsp}{\mbox{${\rm GStocCSP}$}}

\newcommand{\fstocs}{\mbox{${\rm FStocS}$}}
\newcommand{\fss}{\mbox{${\rm FSS}$}}
\newcommand{\fstocsp}{\mbox{${\rm FStocSP}$}}

\newcommand{\stoccs}{\mbox{${\rm StocCS}$}}
\newcommand{\stocs}{\mbox{${\rm StocS}$}}

\newcommand{\is}{\mbox{${\rm IS}$}}
\newcommand{\isdsrs}{\mbox{${\rm ISDSrs}$}}
\newcommand{\isdsones}{\mbox{${\rm ISDS1s}$}}
\newcommand{\isdsrscommon}{\mbox{${\rm ISDSrsCommon}$}}
\newcommand{\isdsonescommon}{\mbox{${\rm ISDS1sCommon}$}}
\newcommand{\isdsrsprivate}{\mbox{${\rm ISDSrsPrivate}$}}
\newcommand{\isnn}{\mbox{${\rm ISNN}$}}
\newcommand{\inn}{\mbox{${\rm I_{nn}}$}}

\newcommand{\bits}{\mbox{$\rm bits$}}

\newcommand{\dtime}{\mbox{$\rm DTIME$}}
\newcommand{\exptime}{\mbox{$\rm EXPTIME$}}
\newcommand{\np}{\mbox{$\rm NP$}}
\newcommand{\ntime}{\mbox{$\rm NTIME$}}
\newcommand{\polylog}{\mbox{$\rm polylog$}}
\newcommand{\timecomplexity}{\mbox{$\rm TIME$}}

\newcommand{\glnr}{\mbox{$Gl_{n}(\mathbb{R})$}}

\newcommand{\aut}{\mbox{${\bf Aut}$}}
\newcommand{\catset}{\mbox{${\bf Set}$}}
\newcommand{\comp}{\mbox{$comp$}}
\newcommand{\Grp}{\mbox{${\bf Grp}$}}
\newcommand{\kaut}{\mbox{${\bf K-Aut}$}}
\newcommand{\kmedv}{\mbox{${\bf K-Medv}$}}
\newcommand{\kmedvio}{\mbox{$({\bf K-Medv} \downarrow <I^+,O>)$}}
\newcommand{\moduler}{\mbox{${\bf Module_R}$}}
\newcommand{\ob}{\mbox{$ob$}}
\newcommand{\Set}{\mbox{${\bf Set}$}}
\newcommand{\cattopo}{\mbox{${\bf Topo}$}}

\newcommand{\cont}{\mbox{${\rm cont}$}}
\newcommand{\Deg}{\mbox{${\rm Deg}$}}
\newcommand{\degree}{\mbox{${\rm deg}$}}
\newcommand{\diff}{\mbox{${\rm diff}$}}
\newcommand{\gmon}{\mbox{${\rm G_{mon}}$}}
\newcommand{\diffrpos}{\mbox{${\rm Diff} \mathbb{R}_+$}}
\newcommand{\lcm}{\mbox{${\rm lcm}$}}
\newcommand{\mon}{\mbox{${\rm mon}$}}
\newcommand{\mnm}{\mbox{${\rm mnm}$}}
\newcommand{\order}{\mbox{${\rm order}$}}
\newcommand{\rpoly}[2]{\mbox{$R_+[#1]/(#1^{#2}-1)$}}
\newcommand{\slsp}{\mbox{${\rm SL}\Sigma{\rm P}$}}
\newcommand{\spoly}[2]{\mbox{$S_+[#1]/(#1^{#2}-1)$}}
\newcommand{\sqfree}{\mbox{${\rm sqfree}$}}
\newcommand{\support}{\mbox{${\rm support}$}}
\newcommand{\trdeg}{\mbox{${\rm trdeg}$}}

\newcommand{\as}{\mbox{{\rm $a.s.$}}} 
\newcommand{\aslim}{\mbox{{\rm $a.s.-\lim$}}} 
\newcommand{\ci}{\mbox{{\rm $CI$}}} 
\newcommand{\ciffg}{\mbox{$(F_1,F_2 | G ) \in \ci}} 
\newcommand{\dlim}{\mbox{{\rm $D-\lim$}}} 
\newcommand{\essinf}{\mbox{{\rm $essinf$}}} 
\newcommand{\esssup}{\mbox{{\rm $esssup$}}} 
\newcommand{\foralmostall}{\mbox{{\rm $\mbox{almost all}$}}} 
\newcommand{\ift}{\mbox{{\rm $ift$}}} 
\newcommand{\ltwolim}{\mbox{{\rm $L_2-\lim$}}} 
\newcommand{\pdf}{\mbox{{\rm $pdf$}}} 
\newcommand{\pessinf}{\mbox{{\rm $P-essinf$}}} 
\newcommand{\pesssup}{\mbox{{\rm $P-esssup$}}} 
\newcommand{\plim}{\mbox{{\rm $P-\lim$}}} 

\newcommand{\aloc}{\mbox{{\rm ${\bf A_{loc}}$}}}
\newcommand{\alocplus}{\mbox{{\rm ${\bf A_{loc}^+}$}}}
\newcommand{\aone}{\mbox{{\rm ${\bf A_1}$}}}
\newcommand{\aplus}{\mbox{{\rm ${\bf A^+}$}}}
\newcommand{\bvar}{\mbox{{\rm $BV$}}}
\newcommand{\bvarc}{\mbox{{\rm $BV^c$}}}
\newcommand{\cadlag}{\mbox{{\rm c\`{a}dl\`{a}g}}}
\newcommand{\DL}{\mbox{{\rm $DL$}}}
\newcommand{\mone}{\mbox{{\rm $M_1$}}}
\newcommand{\monec}{\mbox{{\rm $M_1^c$}}}
\newcommand{\monepos}{\mbox{{\rm $M_{+,1}$}}}
\newcommand{\moneu}{\mbox{{\rm $M_{1u}$}}}
\newcommand{\moneuc}{\mbox{{\rm $M_{1u}^c$}}}
\newcommand{\moneuloc}{\mbox{{\rm $M_{1uloc}$}}}
\newcommand{\moneulocc}{\mbox{{\rm $M_{1uloc}^c$}}}
\newcommand{\mtwo}{\mbox{{\rm $M_2$}}}
\newcommand{\mtwos}{\mbox{{\rm $M_{2s}$}}}
\newcommand{\mtwosc}{\mbox{{\rm $M_{2s}^c$}}}
\newcommand{\mtwosd}{\mbox{{\rm $M_{2s}^d$}}}
\newcommand{\mtwosloc}{\mbox{{\rm $M_{2sloc}$}}}
\newcommand{\mtwoslocc}{\mbox{{\rm $M_{2sloc}^c$}}}
\newcommand{\mtwoslocd}{\mbox{{\rm $M_{2sloc}^d$}}}
\newcommand{\mtwoc}{\mbox{{\rm $M_2^c$}}}
\newcommand{\var}{\mbox{{\rm $Var$}}}
\newcommand{\semim}{\mbox{{\rm $SemM$}}}
\newcommand{\semimc}{\mbox{{\rm $SemM^c$}}}
\newcommand{\semimone}{\mbox{{\rm $SemM_1$}}}
\newcommand{\semimonec}{\mbox{{\rm $SemM_1^c$}}}
\newcommand{\semimspecial}{\mbox{{\rm $SemM_s$}}}
\newcommand{\semimtwo}{\mbox{{\rm $SemM_2$}}}
\newcommand{\semimtwoc}{\mbox{{\rm $SemM_2^c$}}}
\newcommand{\semimloc}{\mbox{{\rm $SemM_{loc}$}}}
\newcommand{\semimlocc}{\mbox{{\rm $SemM_{loc}^c$}}}
\newcommand{\stoppingtimes}{\mbox{{\rm $T_{st}$}}}
\newcommand{\stoppingtimesinfty}{\mbox{{\rm $T_{st \uparrow \infty}$}}}
\newcommand{\submone}{\mbox{{\rm $SubM_1$}}}
\newcommand{\submonec}{\mbox{{\rm $SubM_1^c$}}}
\newcommand{\submonepos}{\mbox{{\rm $SubM_{+,1}$}}}
\newcommand{\submpos}{\mbox{{\rm $SubM_+$}}}
\newcommand{\submposc}{\mbox{{\rm $SubM_+^c$}}}
\newcommand{\supmone}{\mbox{{\rm $SupM_1$}}}

\newcommand{\deficiency}{\mbox{{\rm dfc}}}
\newcommand{\reactionnet}{\mbox{{\rm rnet}}}
\newcommand{\rnet}{\mbox{{\rm $rnet$}}}
\newcommand{\netc}{\mbox{{\rm $net_c$}}}
\newcommand{\nets}{\mbox{{\rm $net_s$}}}

\newcommand{\lane}{\mbox{$\rm lane$}}
\newcommand{\od}{\mbox{$\rm OD$}}
\newcommand{\pldim}{\mbox{$\rm m$} \times \mbox{$\rm km/h$} \times \mbox{$\rm veh}}
\newcommand{\roadnet}{\mbox{$\rm RoadNet$}}
\newcommand{\roadsection}{\mbox{$\rm Secs$}}
\newcommand{\subnet}{\mbox{$\rm SubNet$}}
\newcommand{\subnetin}{\mbox{$\rm SubNet_{in}$}}
\newcommand{\subnetout}{\mbox{$\rm SubNet_{out}$}}
\newcommand{\subnetlink}{\mbox{$\rm R_{link}$}}

\newcommand{\dl}{D_{\lambda}}
\newcommand{\occ}{\mbox{\rm Occ}}

\newcommand{\ta}{\widetilde{\alpha}}
\newcommand{\tb}{\widetilde{\beta}}
\newcommand{\tg}{\widetilde{\gamma}}
\newcommand{\td}{\widetilde{\delta}}
\newcommand{\te}{\widetilde{\varepsilon}}
\newcommand{\tx}{\widetilde{\xi}}
\newcommand{\tp}{\widetilde{p}}
\newcommand{\tm}{\widetilde{M}}
\newcommand{\wt}[1]{\widetilde{#1}}
\newcommand{\wh}[1]{\widehat{#1}}

\newcommand{\TT}{T\!\!\!\! I}
\newcommand{\DD}{D\!\!\!\! I}
\newcommand{\RR}{R\!\!\!\! I}
\newcommand{\CC}{C\!\!\!\! I}
\newcommand{\NN}{N\!\!\!\!\!\! I\;}      
 


\title{\LARGE\sf A New Approach to Lossy Network Compression of a Tuple of Correlated  Multivariate 
Gaussian RVs
}


\author
{
{\large\sf Charalambos D. Charalambous}\\ 
{\small\sf Department of Electrical and Computer Engineering,
             University of Cyprus}\\
{\small\sf P.O.Box 20537, CY-1678 Nicosia, Cyprus}\\
{\small\sf chadcha@ucy.ac.cy}\\ [0.3cm]
{\large\sf Jan H. van Schuppen}\\ 
{\small\sf Van Schuppen Control Research}\\
{\small\sf Gouden Leeuw 143, 1103 KB Amsterdam, The Netherlands}\\
{\small\sf jan.h.van.schuppen@xs4all.nl}\\ [0.3cm]
{\small\sf \today}
}
\date{}
\maketitle
\begin{quotation}
\begin{small}
\noindent
The classical Gray and Wyner source coding for a simple network for sources that generate  a tuple of multivariate, correlated  Gaussian random variables $Y_1 : \Omega \rightarrow {\mathbb R}^{p_1}$ and $Y_2 : \Omega \rightarrow {\mathbb R}^{p_2}$,  is re-examined using  the geometric approach of Gaussian random variables, and  the weak stochastic realization of correlated Gaussian random variables. New results are:

(1) The formulation, methods and algorithms to  parametrize all random variables $W : \Omega \rightarrow {\mathbb R}^n $  which make the two components of the tuple $(Y_1,Y_2)$ conditionally independent, according to the weak stochastic realization of $(Y_1, Y_2)$. Use is made of the transformation of random variables $(Y_1,Y_2)$    via non-singular transformations $(S_1,S_2)$,  into   their canonical variable form as 
$S_1Y_1=(V_1, Y_1^\prime)=((Y_{11},Y_{12}), Y_{13})$,  $S_2Y_2=(V_2,Y_2^\prime)=((Y_{21},Y_{22}),Y_{23}),$ where $Y_{11}=Y_{21}-$a.s., $Y_{13}$ and $Y_{23}$ are independent  and each of these has independent components, $Y_{12}$ and $Y_{22}$ are correlated and each of these has independent components, ${\bf E}[Y_{12} Y_{22}^T]=  D$, 
  for some diagonal matrix $D$ with diagonal entries $1 > d_1 \geq d_2 \geq \ldots \geq d_n>0$  in $(0,1)$ called {\em the canonical correlation coefficients}. 

(2) A formula for Wyner's lossy common information  for joint decoding with mean-square error distortions ${\mathbb E}||Y_1-\hat{Y}_1||_{{\mathbb R}^{p_1}}^2 \leq \Delta_1 \in [0,\infty]$ and  ${\mathbb E}||Y_2-\hat{Y}_2||_{{\mathbb R}^{p_2}}^2 \leq \Delta_2 \in [0,\infty]$, where $(\hat{Y}_1, \hat{Y}_2)$ are the reproductions of $(Y_1,Y_2)$, given by 
$C_{W}(Y_1,Y_2)
   =  \frac{1}{2} \sum_{j=1}^n 
        \ln
        \left(
        \frac{1+d_j}{1-d_j}
        \right),$ 
where  the distortion region is defined by $ 0\leq \Delta_1 \leq n(1-d_1)$, $0\leq \Delta_2 \leq n(1-d_1)$.  

(3) Parametrization of the lossy rate region of the Gray and Wyner source coding problem, and the calculation of the smallest common message rate $R_0$ on the Gray and Wyner source problem, when the sum rate $R_0+R_1+R_2$ is arbitrary close to the joint rate distortion function $R_{Y_1, Y_2}(\Delta_1, \Delta_2)$ of joint decoding. 

The methods and algorithms may be  applicable to other problems of multi-user communication, such as, the multiple access channel, etc. The discussion is largely self-contained and proceeds from first principles.

\par\vspace{1\baselineskip}\par\noindent

\par\vspace{1\baselineskip}\par\noindent

{\em Preliminary results} are accepted for publication in  \cite{charalambous:vanschuppe:2020}. This paper provides an extensive analysis and  proofs.\\
This paper is submitted to IEEE Transactions on Information Theory, 30 July 2020.

\end{small}
\end{quotation}

\ \


\section{Introduction}\label{sec:intro}
\label{sect:intro}

\begin{figure}
\setlength{\unitlength}{0.24cm}
\begin{center}
\begin{picture}(48,18)(0,0)
\put(0,0){\framebox(48,18)}
\put(2,8){\framebox(4,4)}
\put(12,5){\framebox(4,10)}
\put(24,5){\framebox(6,2)}
\put(24,9){\framebox(6,2)}
\put(24,13){\framebox(6,2)}
\put(36,5){\framebox(4,2)}
\put(36,13){\framebox(4,2)}
\put(10,4){\dashbox{.5}(8,12)}
\put(22,4){\dashbox{.5}(10,12)}
\put(34,4){\dashbox{.5}(8,12)}
\put(6,10){\vector(1,0){6}}
	\put(16,6){\vector(1,0){8}}
	\put(16,10){\vector(1,0){8}}
        \put(16,14){\vector(1,0){8}}
	\put(30,6){\vector(1,0){6}}
	\put(30,10){\vector(1,0){8}}
        \put(30,14){\vector(1,0){6}}
	\put(38,10){\vector(0,-1){3}}
	\put(38,10){\vector(0,1){3}}
        \put(40,6){\vector(1,0){6}}
        \put(40,14){\vector(1,0){6}}
	\put(5,13){$(Y_1^N, Y_2^N)$}
	\put(43,7){$\hat{Y}_2^N$}
	\put(43,15){$\hat{Y}_1^N$}
	\put(24.2,5.5){{\small Channel 2}}
	\put(24.2,9.5){{\small Channel 0}}
	\put(24.2,13.5){{\small Channel 1}}
\put(3,2){Source}
\put(12,2){Encoder}
\put(24,2){Channels}
\put(36,2){Decoders}
\end{picture}
\end{center}
\caption{The Gray and Wyner source coding for a simple network \cite{gray-wyner:1974} $(Y_{1,i}, Y_{2,i})\sim {\bf P}_{Y_1,Y_2}, i=1, \ldots, N.$
}
\label{fig:gwn}
\end{figure}
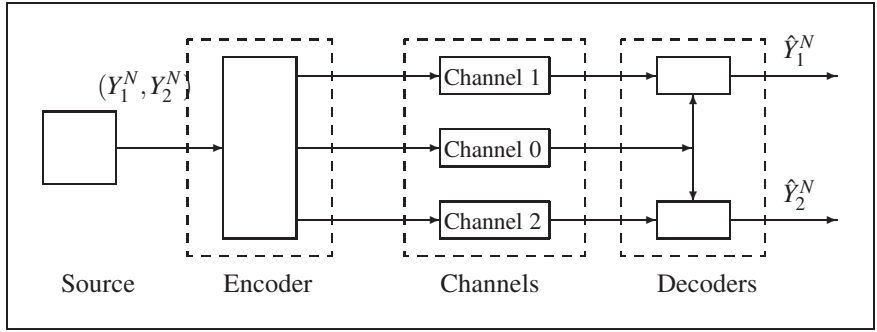
%

An important class of theoretical and practical problems in communication is of a  multi-user nature. Such problems are: lossless and lossy network source coding for compression of data generated by multiple sources, and  network channel coding  for  transmission of data generated by multiple sources over noisy channels. A sub-class of network source coding problems deals with two sources that generate at each time instant,  symbols that are  stationary memoryless,   multivariate, correlated,  and jointly Gaussian distributed. 
This paper is focused on a new approach, which is applied to such network source coding problems, although  the method apply  to network  channel coding problems.

Gray and Wyner  in their seminal paper,  {\it source coding for a simple network} \cite{gray-wyner:1974},   characterized the lossless rate region   for a tuple of finite-valued random variables, and the lossy rate region  for a tuple of arbitrary distributed random variables. 
Many extensions and generalizations followed Gray and Wyner's fundamental work. Wyner \cite{wyner:1975}, introduced an operational definition of the common information between a  tuple of sources that generate   symbols with   values in finite spaces. Wyner's operational definition of common information is defined  as the minimum achievable common message rate   on  the  Gray and Wyner lossless  rate  region.  Witsenhausen  \cite{witsenhausen:1976b} has investigated bounds for Wyner's common information, and sequences of pairs of random variables in this regard  \cite{witsenhausen:1975:pairsrv}.
  G\'{a}cs and K\"{o}rner \cite{gacs-korner:1973} introduced another definition of common randomness between a tuple of jointly
independent and identically distributed random variables.  Viswanatha, Akyol and Rose \cite{viswanatha:akyol:rose:2014}, and Xu, Liu, and Chen \cite{xu:liu:chen:2016:ieeetit}, explored the connection of Wyner's common information and the Gray and Wyner lossy rate region, to generalize Wyner's common information to its lossy counterpart, for random variables taking values in arbitrary spaces. They characterized Wyner's lossy common information, as   the minimum common message rate    on  the Gray and Wyner lossy rate region, when the sum rate is arbitrary close to the rate distortion function with joint decoding for the Gray and Wyner lossy network. Applications  to encryption and secret key generation are discussed by Viswanatha, Akyol and Rose in \cite{viswanatha:akyol:rose:2014} (and references therein).


Present methods and algorithms to compute such rates, for specific application examples  are subject to a number of limitations which often prevent their practical usefulness:

(1) Rates that lie in the Gray and Wyner rate region are only known for the
special case of a tuple of scalar-valued Gaussian random variables,  with square error distortion, and for the  special case of  the
 doubly symmetric binary source with Hamming distance distortion \cite{gray-wyner:1974} (Section 2.5). 
 
(2) Wyner's lossy common information is only computed in closed form,  for the special cases of a tuple of scalar-valued Gaussian random variables,  with square error distortion, and for the  special case of 
the doubly symmetric binary source  \cite{viswanatha:akyol:rose:2014} (Section III.C), and  \cite{xu:liu:chen:2016:ieeetit} (Section IV.C).


(3) Important generalizations, of the above  application examples to a tuple of sources that generate multivariate Gaussian symbols,  require new derivations often of considerable difficulty. The mathematics of the derivations of closed form expressions  for such generalizations are not transparent, and often require new mathematical tools.

This paper introduces a new approach at the assemblage of problems of multi-user information theory, in which achievable rate regions or rates,  are parametrized by Gaussian  {\em auxiliary random variables} $W: \Omega \rightarrow {\mathbb R}^n$,  that  make two multivariate correlated Gaussian random variables     $Y_{1} : \Omega \rightarrow {\mathbb R}^{p_1},  Y_{2} : \Omega \rightarrow {\mathbb R}^{p_2}$ conditionally independent
\begin{align}
{\bf P}_{Y_1,Y_2|W}= {\bf P}_{Y_1|W}{\bf P}_{Y_2|W}. \label{CI_G}
\end{align}
The focus  of the current paper  is on  multi-user problems of information theory, for which the operational definitions of achievable rate regions or rates are characterized by their information theoretic definitions, via  mutual information, conditional mutual information, etc.,  that depend on the joint distribution ${\bf P}_{Y_1,Y_2,W}$ of the random variables $(W,Y_1,Y_2)$.

One application of the methods and algorithms that emerge from the new approach,  is the Gray and Wyner source coding for a simple network \cite{gray-wyner:1974}.   For two sources that generate symbols, according to the model of   jointly independent and identically distributed multivariate correlated Gaussian random variables      $(Y_{1}, Y_{2})$,  the Gray and Wyner rate region can be  parametrized by an auxiliary random variable $W$ that satisfies conditional independence (\ref{CI_G}). 

The following are the highlights of the methods and algorithms of the new approach of the  paper:\\

\noindent {\it Method 1.}\\
This method utilizes \cite{putten:schuppen:1985} to\\
(i) parametrize the family of Gaussian probability distributions ${\bf P}_{Y_1,Y_2,W}(y_1,y_2,w)$ by the multidimensional
random variable $W$ such that $(Y_1,Y_2)$ are conditionally independent, conditioned on $W$, that is, (\ref{CI_G}) holds, and the marginal distribution ${\bf P}_{Y_1,Y_2,W}(y_1,y_2,\infty) = {\bf P}_{Y_1,Y_2} (y_1,y_2)$ coincides with the
distribution of $(Y_1,Y_2)$, and to\\
(ii) represent the random variables $(Y_1,Y_2)$ using the weak stochastic realization, expressed in terms of $W$ and two independent random variables $V_{1} : \Omega \rightarrow {\mathbb R}^{p_1},  V_{2} : \Omega \rightarrow {\mathbb R}^{p_2}$, which are independent of $W$.\\

%

\noindent {\it Method 2.}\\
This method 
utilizes the geometric approach to Gaussian random variables \cite{hotelling:1936,anderson:1958,gittens:1985}, where the underlying
geometric object of a Gaussian random variable $Y : \Omega \rightarrow {\mathbb R}^p$ is the $\sigma-$algebra $F^Y$ generated by $Y$.
 A {\it basis transformation}
of such a random variable is then the transformation defined by a non-singular matrix $S \in {\mathbb R}^{p\times p}$, and directly follows that  
$F^Y = F^{SY}$. For a tuple of jointly Gaussian multivariate random variables $(Y_1,Y_2)$, a basis transformation of
this tuple consists of a matrix
\begin{align}
 &S = \blockdiag ( S_1 , S_2 ),  \label{meth_1a} \\
&S_1 , S_2 \ \ \mbox{square and non-singular matrices, and hence  the spaces satisfy} \label{meth_1b}\\
&F^{ Y_1} = F^{S_1 Y_1}, \ \ F^{Y_2} = F^{ S_2 Y_2 }.\label{meth_1c}
\end{align}
 This transformation yields (the full specification is given in  Definition~\ref{def:grvcommoncorrelatedprivateinfo})
  \begin{align}
& S_1Y_1=(V_1, Y_1^\prime)=((Y_{11},Y_{12}), Y_{13}), \ \ S_2Y_2=(V_2,Y_2^\prime)=((Y_{21},Y_{22}),Y_{23}),  \label{cvf_1}   \\
 &Y_{11}=Y_{21}-a.s., \\
&  Y_{13} \ \ \mbox{and}\ \  Y_{23} \ \ \mbox{are independent  and each of these has independent components},\label{cvf_2}  \\
&  Y_{12} \ \ \mbox{and}\ \  Y_{22} \ \ \mbox{are correlated and each of these has independent components},\label{cvf_2a}  \\
&{\bf E}[Y_{12} Y_{22}^T]=  D\label{cvf_3}
 \end{align}
  for some diagonal matrix $D$ with diagonal entries in $(0,1)$ called  {\em the canonical correlation coefficients}. Component $Y_1^\prime=Y_{13}$
 is the private component of $S_1Y_1$, and component $V_1=(Y_{11}, Y_{12})$ is the
correlated component of $S_1Y_1$, with respect to $S_2Y_2$, and similarly for $V_2=(Y_{21}, Y_{22})$.  This method is equivalent to pre-processing
of the tuple of correlated random variables $(Y_{1,i},Y_{2,i}) \sim {\bf P}_{Y_1,Y_2},  i = 1, \ldots, N$, with the aid of a
linear pre-encoder transformation 
$S = \blockdiag ( S_1 , S_2 )$.
The power of methods 1 and 2 is more apparent in the actual calculations of rates and the development of  algorithms to compute such rates. 

The rest of the section serves mainly to review the Gray and Wyner characterization of lossy rate region and the characterization of  Wyner's lossy common information, for the purpose of linking these to Methods 1 and 2.   Then the power of  Methods 1 and 2 with respect to the computations of rates that lie on the Gray and Wyner characterization of lossy rate region is  further discussed. 

\subsection{Literature Review}
\label{sect:lite}
\noindent {\it (a) The Gray and Wyner source coding for a simple network \cite{gray-wyner:1974}. }\\
Consider the Gray and Wyner source coding for a simple network shown Fig.~\ref{fig:gwn}, for  a tuple of jointly independent and identically distributed multivariate Gaussian random
variables $(Y_1^N, Y_2^N)= \{(Y_{1,i}, Y_{2,i}): i=1,2, \ldots,N\}$,   
\begin{align}
Y_{1,i} : \Omega \rightarrow {\mathbb R}^{p_1}= {\mathbb Y}_1, \ \ \ \ Y_{2,i} : \Omega \rightarrow {\mathbb R}^{p_2}={\mathbb Y}_2, \ \ i = 1, \ldots, N
\end{align}
 with square error distortion
functions at the two decoders,
\begin{align}
 D_{Y_1} (y_1^N, \hat{y}_1^N)= \frac{1}{N} \sum_{i=1}^N ||y_{1,i}-\hat{y}_{1,i}||_{{\mathbb R}^{p_1}}^2,  \ \  \ \  D_{Y_2} (y_2^N, \hat{y}_2^N)= \frac{1}{N} \sum_{i=1}^N ||y_{2,i}-\hat{y}_{2,i}||_{{\mathbb R}^{p_2}}^2
\end{align} 
where $||\cdot||_{{\mathbb R}^{p_i}}^2$ are Euclidean distances on ${\mathbb R}^{p_i}, i=1,2$.  \\
The encoder  takes as its input the data sequences $(Y_1^N, Y_2^N)$ and produces at its output  three messages \\
$(S_0, S_1, S_2)$, with binary bit representations $(NR_0, NR_1, NR_2)$.  There are three channels, Channel $0$, Channel $1$, Channel $2$,  with capacities $(C_0, C_1, C_2)$ (in bits per second), respectively,  to  transmit the messages  to two decoders. Channel $0$ is a public channel and channel $1$ and channel $2$ are the private channels, which connect the encoder to each of the two decoders. Message  $S_0$ is a  {\em common} or {\em public} message that is transmitted through the public channel $0$  with capacity $C_0$ to  decoder $1$ and decoder $2$,  $S_1$ is a {\em private} message, which is transmitted through the {\em private}  channel $1$ with capacity $C_1$ to decoder $1$,  and  $S_2$ is a {\em private} message, which is transmitted through the {\em private}  channel $2$ with capacity $C_2$ to decoder $2$.\\
Decoder $1$ has as objective to reproduce $Y_1^N$ by $\hat{Y}_1^N$ subject to an average distortion   and decoder $2$  has as objective to  reproduce $Y_2^N$ by $\hat{Y}_2^N$,  subject to an average distortion, where  $(\hat{Y}_{1,i},\hat{Y}_{2,i})=(\hat{y}_{1,i}, \hat{y}_{2,i}) \in \hat{\mathbb Y}_1 \times \hat{\mathbb Y}_2 \subseteq  {\mathbb Y}_1 \times {\mathbb Y}_2, i=1, \ldots,N$, that is, 
\begin{align}
{\bf E}\Big\{D_{Y_1} (Y_1^N, \hat{Y}_1^N)\Big\}\leq \Delta_1, \ \ \ \ {\bf E}\Big\{D_{Y_2} (Y_2^N, \hat{Y}_2^N)\Big\}\leq \Delta_2, \ \ \ \ (\Delta_1, \Delta_2) \in [0,\infty] \times [0,\infty].
\end{align}

Gray and Wyner characterized the rate region, denoted by  ${\cal R}_{GW}(\Delta_1, \Delta_2)$,   by a coding scheme that uses the auxiliary random variable $W$, as described below. Define the family of probability distributions
\begin{align}
{\cal P} \triangleq \Big\{ {\bf P}_{Y_1, Y_2, W}(y_1,y_2,w), \ \ y_1 \in {\mathbb Y}_1, y_2 \in {\mathbb Y}_2, w \in {\mathbb W}: \ \  {\bf P}_{Y_1, Y_2, W}(y_1,y_2,\infty)={\bf P}_{Y_1, Y_2}(y_1, y_2)\Big\}
\end{align}
for some auxiliary random variable  $W$, i.e.,  such that the joint probability distribution ${\bf P}_{Y_1,Y_2,W}(y_1,y_2,w)$ on ${\mathbb Y}_1\times {\mathbb Y}_2\times {\mathbb W}$, has a $(Y_1, Y_2)-$marginal  probability distribution  ${\bf P}_{Y_1,Y_2}(y_1,y_2)$ on  ${\mathbb Y}_1\times {\mathbb Y}_2$ that coincides with the probability distribution  of $(Y_1, Y_2)$. \\
The characterization of ${\cal R}_{GW}(\Delta_1, \Delta_2)$ is described  in terms of an auxiliary random variable, as follows. 

\begin{theorem}
\label{theorem_8} (Theorem 8 in \cite{gray-wyner:1974})\\
Let ${\cal R}_{GW}(\Delta_1, \Delta_2)$ denote the Gray and Wyner rate regionm of the simple network shown in Fig.~\ref{fig:gwn}.\\
 Suppose there exists $\hat{y}_i \in \hat{\mathbb Y}_i$ such that ${\bf E}\{d_{Y_i}(Y_i, \hat{y}_i)\}< \infty$, for  $i=1,2$.\\
For each ${\bf P}_{Y_1, Y_2, W} \in {\cal P}$ and $\Delta_1 \geq 0, \Delta_2 \geq 0$, define the subset of Euclidean $3-$dimensional space 
 \begin{align}
  {\cal R}_{GW}^{{\bf P}_{Y_1,Y_2,W}}(\Delta_1, \Delta_2) = \Big\{\Big(R_0,R_1,R_2\Big): \ \       R_0 \geq I(Y_1, Y_2; W), \ \ 
 R_1 \geq R_{Y_1|W}(\Delta_1), \ \ R_2 \geq R_{Y_2|W}(\Delta_2) \Big\} \label{eq_32} 
\end{align} 
where $R_{Y_i|W}(\Delta_i)$ is the conditional rate distortion function of $Y_i^N$, conditioned on $W^N$, at decoder $i$, for $i=1,2$,  and $R_{Y_1,Y_2}(\Delta_1,\Delta_2)$ is the joint rate distortion function of joint decoding of $(Y_1^N, Y_2^N)$ (all single letter).
Let
\begin{align}
{\cal R}_{GW}^{*}(\Delta_1, \Delta_2)= \Big(\bigcup_{ {\bf P}_{Y_1,Y_2, W} \in {\cal P}} {\cal R}_{GW}^{{\bf P}_{Y_1,Y_2,W}}(\Delta_1, \Delta_2)\Big)^c
\end{align}
where $\big(\cdot\big)^c$ denotes the closure of the  indicated set. 
Then the achievable Gray-Wyner lossy rate region  is given by
\begin{align}
{\cal R}_{GW}(\Delta_1, \Delta_2)={\cal R}_{GW}^{*}(\Delta_1, \Delta_2).  \label{eq_31}
\end{align}
\end{theorem}
Gray and Wyner \cite{gray-wyner:1974} (Theorem 6) also showed that, if   $(R_0, R_1, R_2) \in {\cal R}_{GW}(\Delta_1, \Delta_2)$, then
\begin{align}
& R_0 + R_1 +R_2 \geq R_{Y_1, Y_2}(\Delta_1, \Delta_2),   \label{eq_32a}  \\
& R_0 +R_1  \geq R_{Y_1}(\Delta_1), \label{eq_32b} \\ 
& R_0 +R_2  \geq R_{Y_2}(\Delta_2) \label{eq_32c} 
\end{align}
where $R_{Y_i}(\Delta_i)$ is the rate distortion function of $Y_i^N$ at decoder $i$, for $i=1,2$.
The inequality in (\ref{eq_32a}) is called the {\em Pangloss Bound} of the Gray-Wyner lossy rate region ${\cal R}_{GW}(\Delta_1, \Delta_2)$. The set of triples $(R_0, R_1, R_2) \in {\cal R}_{GW}(\Delta_1, \Delta_2)$ that satisfy the equality $R_0+R_1+R_2= R_{Y_1, Y_2}(\Delta_1, \Delta_2)$ 
 is called the {\em Pangloss Plane} of the Gray-Wyner lossy rate region ${\cal R}_{GW}(\Delta_1, \Delta_2)$.

For the
special case of a tuple of scalar-valued (bivariate) Gaussian random variables, i.e., $p_1=p_2=1$,  with square error distortion and $\Delta_1=\Delta_2=\Delta$, Gray and Wyner \cite{gray-wyner:1974} (Section 2.5, (B)), showed that the choice of scalar-valued Gaussian random variable $W$ that satisfies conditional independence (\ref{CI_G}), ensures a corresponding   rate triple $(R_0,R_1,R_2) \in {\cal 
R}_{GW}(\Delta_1, \Delta_2)$  that lies on Pangloss Plane, and they  derived explicitly the formulaes of these rates.  Similarly for the
the doubly symmetric binary source with Hamming distance distortion function.\\

\noindent {\it (b) Wyner's common Information of finite-valued random variables.}\\
Wyner \cite{wyner:1975}, introduced an operational definition of the common information between a  tuple of   random variables $(Y_1^N, Y_2^N)$,  that   takes  values in finite spaces. \\
 The {\it first approach}  of Wyner's operational definition of common information between sequences $Y_1^N$ and $Y_2^N$ is defined  as the minimum achievable common message rate  $R_0$ on  the  Gray-Wyner Network  of Fig.~\ref{fig:gwn}.

Wyner's single letter information theoretic characterization of the infimum of all achievable message rates $R_0$, called Wyner's  common information,  is defined  by,  \\
\begin{align}
C(Y_1, Y_2)= \inf_{{\bf P}_{Y_1, Y_2, W}:\: {\bf P}_{Y_1, Y_2|W}={\bf P}_{Y_1|W} {\bf P}_{Y_2|W}} I(Y_1, Y_2; W). \label{eq_41}
\end{align}
Here ${\bf P}_{Y_1, Y_2, W}$ is any joint  probability distribution on ${\mathbb Y}_1 \times {\mathbb Y}_2 \times {\mathbb W}$ with $(Y_1, Y_2)-$marginal ${\bf P}_{Y_1, Y_2}$, such that $W$ makes $Y_1$ and $Y_2$ conditionally independent, that is ${\bf P}_{Y_1, Y_2, W}\in {\cal P}$. \\


\noindent{\it (c) Minimum common message rate  and  Wyner's lossy  common information for arbitrary random variables.}\\
Viswanatha, Akyol and Rose \cite{viswanatha:akyol:rose:2014}, and Xu, Liu, and Chen \cite{xu:liu:chen:2016:ieeetit}, explored the connection of Wyner's common information and the Gray-Wyner lossy rate region, to provide a new interpretation of Wyner's common information to its lossy counterpart. They   first defined and characterized the minimum common message rate $R_0$   on  the Gray-Wyner lossy rate region, when the sum rate is arbitrary close to the rate distortion function with joint decoding for the Gray-Wyner lossy network.

The following characterization  is derived by Xu, Liu, and Chen \cite{xu:liu:chen:2016:ieeetit} (an equivalent characterization is also derived by Viswanatha, Akyol and Rose \cite{viswanatha:akyol:rose:2014}).
 
\begin{theorem}(Theorem 4 in \cite{xu:liu:chen:2016:ieeetit})\\
\label{theorem_4}
Suppose  there exists $\hat{y}_i \in \hat{\mathbb Y}_i$ such that ${\bf E}\{d_{Y_i}(Y_i, \hat{y}_i)\}< \infty$, for $i=1,2$. \\Let $C_{GW}(Y_1, Y_2; \Delta_1, \Delta_2)$ denote the minimum common message rate $R_0$ on the Gray-Wyner lossy rate region ${\cal R}_{GW}(\Delta_1,\Delta_2)$, with sum rate not exceeding the joint rate distortion function $R_{Y_1,Y_2}(\Delta_1, \Delta_2)$, while satisfying the average distortions. \\
Then  $C_{GW}(Y_1, Y_2; \Delta_1, \Delta_2)$  is characterized by the optimization problem  
\begin{align}
C_{GW}(Y_1, Y_2; \Delta_1, \Delta_2)=\inf\: I(Y_1, Y_2; W)
\end{align}
such that the following identity holds
\begin{align}
R_{Y_1|W}(\Delta_1)+R_{Y_2|W}(\Delta_2)+ I(Y_1, Y_2; W)=R_{Y_1, Y_2}(\Delta_1, \Delta_2) \label{equality_1}
\end{align}
where the infimum is over all random
variables $W$ taking values in ${\mathbb W}$, which parametrize the source distribution via ${\bf P}_{Y_1,Y_2,W}$, having a ${\mathbb Y}_1\times {\mathbb Y}_2-$marginal  source
distribution ${\bf P}_{Y_1,Y_2}$, and induce joint distributions ${\bf P}_{W,Y_1,Y_2,\hat{Y}_1,\hat{Y}_2} $ which satisfy the constraint.\\
\end{theorem}

A   necessary condition for the equality constraint (\ref{equality_1}) to hold  (see Appendix B in \cite{xu:liu:chen:2016:ieeetit}) is 
$R_{Y_1, Y_2|W}(\Delta_1, \Delta_2)=R_{Y_1|W}(\Delta_1)+ R_{Y_2|W}(\Delta_2)$, and  sufficient condition for this equality  to hold is the conditional independence condition \cite{xu:liu:chen:2016:ieeetit}: ${\bf P}_{Y_1,Y_2|W}={\bf P}_{Y_1|W} {\bf P}_{Y_2|W}$.  Hence, a sufficient condition for  any rate $(R_0,R_1,R_2) \in  {\cal R}_{GW}(\Delta_1,\Delta_2)$ to lie on the Pangloss plane is the  conditional independence.

Further, it is shown in \cite{viswanatha:akyol:rose:2014,xu:liu:chen:2016:ieeetit}, that there exists a distortion region such that $C_{GW}(Y_1, Y_2; \Delta_1, \Delta_2) = C_W(Y_1, Y_2)$, i.e., it is independent of the distortions $(\Delta_1, \Delta_2)$, and $C_W(Y_1, Y_2)=C(Y_1,Y_2)$ i.e. it equals to the Wyner's  information theoretic characterization of  common information between $Y_1$ and $Y_2$, defined by (\ref{eq_41}). 

The next theorem is derived by Xu, Liu, and Chen \cite{xu:liu:chen:2016:ieeetit}.

\begin{theorem}(Theorem 5 in  \cite{xu:liu:chen:2016:ieeetit})\\
\label{them_5_xlc}
Let $(Y_1, Y_2)$ be a pair  of random variables with distribution  ${\bf P}_{Y_1, Y_2}$ on the alphabet space ${\mathbb Y}_1\times {\mathbb Y}_2$, where ${\mathbb Y}_1$ and ${\mathbb Y}_2$ are arbitrary measurable spaces that can be discrete or continuous.  \\ Let $W$ be any random variable achieving $C(Y_1, Y_2)$ defined by (\ref{eq_41}).\\ Let the reproduction alphabet $\hat{\mathbb Y}_1={\mathbb Y}_1$, $\hat{\mathbb Y}_2={\mathbb Y}_2$ and two per-letter distortion measures $d_{Y_1}(y_1,\hat{y}_1), d_{Y_2}(y_2,\hat{y}_2)$ satisfy
\begin{align}
d_{Y_i}(y_i, \hat{y}_i)> d_{Y_i}(y_i,y_i)=0, \ \  y_i \neq \hat{y}_i, \ \ i=1,2 \label{eq_200}
\end{align}
If the following conditions are satisfied:\\
1)  for any $y_1\in {\mathbb Y}_1, y_2\in {\mathbb Y}_2$  and  $w \in {\mathbb W}$,  ${\bf P}_{W|Y_1, Y_2}>0$; \\
2) there  exists an $\hat{y}_i \in \hat{\mathbb Y}_i$, such that
\begin{align}
{\bf E}\big\{d_{Y_i}(Y_i, \hat{y}_i)\big\}< \infty, \ \  \ \ i=1,2 \label{eq_201}
\end{align}
then there exists a strictly positive vector $\gamma = (\gamma_1, \gamma_2)\in (0,\infty)\times (0,\infty)$, such that, for $0\leq (\Delta_1, \Delta_2) \leq \gamma$,  
\begin{align}
C_{GW}(Y_1, Y_2; \Delta_1, \Delta_2) = C_W(Y_1, Y_2)=C(Y_1,Y_2). \label{eq_202}
\end{align} 
Moreover, $C_{GW}(Y_1, Y_2;\Delta_1,\Delta_2)$ is constant on ${\cal D}_W = \big\{(\Delta_1,\Delta_2) \in [0,\infty]\times [0,\infty]:  0\leq (\Delta_1, \Delta_2) \leq \gamma\big\}$.
\end{theorem}
   
It should be mentioned that the analog of the above theorem is also derived by  Viswanatha, Akyol and Rose in \cite{viswanatha:akyol:rose:2014} in Lemma 1. Hence, by the work of Viswanatha, Akyol and Rose \cite{viswanatha:akyol:rose:2014}, and Xu, Liu, and Chen \cite{xu:liu:chen:2016:ieeetit}, then  Wyner's lossy common information  is precisely the smallest message rate $C_{GW}(Y_1, Y_2; \Delta_1, \Delta_2)$ on the Gray-Wyner  lossy rate region, for certain distortion levels, when the total rate, $R_0+R_1+R_2$,  is arbitrary close to the rate distortion function with joint decoding, i.e., $R_{Y_1, Y_2}(\Delta_1, \Delta_2)$. 

For a bivariate Gaussian random variables with square-error distortion, the rate-triple \\$(R_0,R_1,R_2) \in  {\cal R}_{GW}(\Delta_1,\Delta_2)$ that lies on the Pangloss plane is computed by Gray and Wyner \cite{gray-wyner:1974} (see Section~2.5, (B) and \cite{viswanatha:akyol:rose:2014,xu:liu:chen:2016:ieeetit}). 
The formulae for $C(Y_1,Y_2)$ is given below for the purpose of comparing it to Wyner's lossy common information formulae of a tuple of multivariate Gaussian random variables derived in this paper.

\begin{theorem}\cite{gray-wyner:1974,viswanatha:akyol:rose:2014,xu:liu:chen:2016:ieeetit}) \\
\label{the:scalar-ci}
Consider two jointly Gaussian scalar-valued  random variables $(Y_1, Y_2)$ with zero mean ${\bf E}[Y_1]={\bf E}[Y_2]=0$, unit variance ${\bf E}[Y_1^2]={\bf E}[Y_2^2]=1$, and ${\bf E}[Y_1 Y_2]=\rho \in [-1,1]$, that is, the variance matrix\footnote{Gray and Wyner \cite{gray-wyner:1974} discussed the case $\rho \in [0,1]$.} is 
\begin{align}
Q_{(Y_1,Y_2)} = \left[ \begin{array}{cc}   \sigma_{Y_1}^2 & \rho \sigma_{Y_2} \sigma_{Y_2} \\
\rho \sigma_{Y_2} \sigma_{Y_2} & \sigma_{Y_2}^2 \end{array} \right]= \left[ \begin{array}{cc} 1 & \rho  \\
\rho  & 1 \end{array} \right] \label{cvf_s}
\end{align} 
with eigenvalues $\lambda_1=1-|\rho|, \lambda_2=1+|\rho|$, 
and square error distortion $d_{Y_i}(y_i,\hat{y}_i)=(y_i-\hat{y}_i)^2, i=1,2.$\\
For $\rho \in [0,1)$ then 
\begin{align}
C_{GW}(Y_1, Y_2; \Delta_1,\Delta_2)= C_W(Y_1, Y_2)=C(Y_1,Y_2)= \frac{1}{2}\log\frac{1+\rho}{1-\rho},\ \ \ \ 0 \leq \Delta_i\leq 1-\rho, \ \ i=1,2. 
\end{align}
\end{theorem}

It should be mentioned that  for the Gaussian example of Theorem~\ref{the:scalar-ci},   Gray and Wyner \cite{gray-wyner:1974} in Section~2.5, (B), computed the rate-triple $(R_0, R_1, R_2)\in {\cal R}_{GW}(Y_1,Y_2; \Delta, \Delta)$,  i.e., for the symmetric case, that lies on the Pangloss plane, and that  
 $C_{GW}(Y_1, Y_2; \Delta_1,\Delta_2)= C_W(Y_1, Y_2)=R_{Y_1,Y_2}(1-\rho, 1-\rho)$ for $0 \leq \Delta  \leq 1-\rho$. \\
Viswanatha, Akyol and Rose in \cite{viswanatha:akyol:rose:2014}, and Xu, Liu, and Chen \cite{xu:liu:chen:2016:ieeetit} computed $C_{GW}(Y_1, Y_2; \Delta_1, \Delta_2)$  for all distortion regions, by using    the expression of joint rate distortion function $R_{Y_1, Y_2}(\Delta_1,\Delta_2)$ derived in \cite{xiao-luo:2005}.


Several applications of Wyner's common information are
found in the characterizations of rate regions of multi-user information theory by Kramer in \cite{kramer:2008} (and references therein). Specific examples are given in the paper
\cite{lapidoth:wigger:2016}, where the authors refer to the concept of Wyner's common information and provide a bound on
the transmission over multi-access channels.  

But the formula to compute any rate triple $(R_0,R_1, R_2)$ that lies on the Gray-Wyner rate region ${\cal R}_{GW}(\Delta_1,\Delta_2)$,  $C_{GW}(Y_1,Y_2; \Delta_1, \Delta_2)$, and Wyner's
lossy common information $C_W(Y_1,Y_2)$, based on Theorems~\ref{theorem_8}-\ref{them_5_xlc}, for the general case of multivariate Gaussian random variables was
never derived  to the best of the authors' knowledge.\\

\subsection{Main Theorems and Discussion of Method 1 and Method 2}
\label{sect:mo}
Method 2 
is equivalent to a pre-processing
 at the encoder.
Method 2 is used in this paper prior to applying Method 1 to derive a parametrization of the family of jointly Gaussian
probability distributions ${\bf P}_{Y_1,Y_2,W}$, having the properties of Method 2. 

Method 1 and Method 2 are used (in this paper)  to compute the marginal rate distortion functions $R_{Y_i} (\Delta_i), i=1,2$, of multivariate  jointly Gaussian random variables, subject to square-error distortions, the conditional rate distortion functions $R_{Y_i|W}(\Delta_1)$ and $R_{Y_2|W}(\Delta_2), i=1,2$, and joint rate distortion function $R_{Y_1,Y_2} (\Delta_1,\Delta_2)$, all expressed in terms of the elements of the covariance matrix $Q_{\cvf}$.

Method 1 and Method 2 are used to construct the weak stochastic realizations of
the reproductions $(\hat{Y}_1, \hat{Y}_2)$ of $(Y_1, Y_2)$, which achieve the  rate distortion functions, $R_{Y_1, Y_2}(\Delta_1, \Delta_2), R_{Y_i} (\Delta_i)$ and $R_{Y_i|W}(\Delta_1), i=1,2$,  and the sum rate $R_{Y_1,Y_2} (\Delta_1,\Delta_2)=R_{Y_1|W}(\Delta_1)+ R_{Y_2|W}(\Delta_2)+I(Y_1,Y_2;W)$, in which the realizations are described by additive Gaussian noise channels.

Method 1 and  Method 2, are used to compute  a rate triple that lie on the Gray-Wyner rate region, $(R_0, R_1, R_2) \in {\cal R}_{GW}(\Delta_1, \Delta_2)$, to compute the
minimum common message rate on the Gray-Wyner lossy rate region $R_0 \in {\cal R}_{GW}(\Delta_1, \Delta_2)$, for the rate triple $(R_0, R_1, R_2) \in {\cal R}_{GW}(\Delta_1, \Delta_2)$, with sum rate $R_0 +R_1 +R_2$ that is equal to the the joint rate distortion function of joint decoding, that is, the sum rate lies on the  {\em Pangloss Plane} \cite{gray-wyner:1974} of ${\cal R}_{GW}(\Delta_1, \Delta_2)$.

Method 1 and method 2, are used to identify the range of values of $(\Delta_1, \Delta_2)$ such that \\ $C_{GW}(Y_1,Y_2;\Delta_1,\Delta_2) =C_W(Y_1,Y_2) = I(Y_1,Y_2;W^*)$ is independent of the distortion levels
$(\Delta_1, \Delta_2)$ and corresponds to Wyner's lossy common information. Further, these methods can be used to determine  the weak stochastic
realizations of the random variables $(W^*,Y_1,Y_2,\hat{Y}_1,\hat{Y}_2)$, which achieve $C_{GW}(Y_1,Y_2;\Delta_1,\Delta_2)$ and  $C_W(Y_1,Y_2)$.  These are given in the current paper (with the exception of $C_{GW}(Y_1,Y_2;\Delta_1,\Delta_2)$). 
%
%
%

For other applications, such as data base applications, secret key generation, and the lossy extension of
the G\'{a}cs and K\"{o}rner common information, the reader is advised to read Viswanatha, Akyol and Rose \cite{viswanatha:akyol:rose:2014},
specifically Section IV.


This paper, however is devoted to the development of the methods, and their   applications to the solution of the optimization problem $C(Y_1,Y_2)=C_W(Y_1,Y_2)$ for the region ${\cal D}_W$, for a tuple of multivariate random variables, 
$Y_{i} : \Omega \rightarrow {\mathbb R}^{p_i}, i = 1,2,$. 



 
What follows  are two of the main theorems derived in this paper.

 Assume that the tuple is already transformed
to the canonical variable representation,
see Def. \ref{def:grvcommoncorrelatedprivateinfo}.
Note that then  
\begin{align}
\mbox{the three tuples
of random variables} \ \
(Y_{11},  Y_{21}), \ \
(Y_{12},  Y_{22}), \ \
(Y_{13},  Y_{23}) \ \ \mbox{are independent}.
\end{align}

The first main theorem is Wyner's common information $C(Y_1, Y_2)$ defined by (\ref{eq_41}) and the realization of the triple $(Y_1, Y_2, W^*)=(Y_1, Y_2, W^*)$ that achieves  $C(Y_1, Y_2)$ (the notation is given in Section~\ref{ap:notationprob}). 
%
%
\begin{theorem}
\label{thm_cia}
Consider a tuple of Gaussian random variables
$(Y_1,Y_2) \in G(0,Q_{\cvf})$ as described and decomposed 
according to Algorithm \ref{alg:wcommoninfogrvs}.
\begin{itemize}
\item[(a)]
Then,
\begin{align}
      C(Y_1,Y_2)
  = & C(Y_{11},Y_{21}) 
        + C(Y_{12},Y_{22})
        + C(Y_{13},Y_{23}) \\
   = & \left\{
        \begin{array}{lll}
          0, & \mbox{if} &  p_{13} > 0, ~ p_{23} > 0, ~
                p_{11} = p_{12} = p_{21} = p_{22} = 0, \\
          \frac{1}{2} \sum_{i=1}^n \ln 
            \left(
              \frac{1+d_i}{1-d_i}
            \right), & \mbox{if} &  p_{12} = p_{22} >0, ~ p_{11} = p_{21} =0, ~
                        p_{13} \geq 0, ~ p_{23} \geq 0, \\
          +\infty, & \mbox{if} & p_{11} = p_{21} > 0. 
        \end{array}
        \right. \nonumber
\end{align}
In particular cases one computes the canonical variable
decomposition of the tuple $(Y_1,Y_2)$ and 
obtains the indices,
$(p_{11}, p_{12}, p_{13})$ and $(p_{21}, p_{22}, p_{23})$. 
Then,
\begin{align}
      C(Y_{11},Y_{21})
   = & +\infty, \ \ \mbox{if} \ \ p_{11} = p_{21} > 0; \\
      C(Y_{13},Y_{23})
   = & 0, \ \ \mbox{if} \ \ p_{13} > 0 \ \ \mbox{and} \ \ p_{23} > 0; \\
      C(Y_{12}, Y_{22})
   = & \frac{1}{2} \sum_{i=1}^n \ln 
            \left(
              \frac{1+d_i}{1-d_i}
            \right), \ \ 
            \mbox{if} \ \ p_{12} = p_{22} >0 .  \label{coinf_3}
\end{align}
%
Thus $C(Y_{12},Y_{22})$ is the most interesting value if defined.
\item[(b)]
The random variable $W^*$ defined below,
is such that $C(Y_1, Y_2)$ of part (a) is attained.
\begin{align} 
    & W^*: \Omega \rightarrow \mathbb{R}^n, \ \ n \in \zpos, \nonumber \\
     & n_1 = p_{11} = p_{21}, \ \
        n_2 = p_{12} = p_{22}, \ \     
        n_1 + n_2 = n, \nonumber \\
     & W^*
  =  \left(
        \begin{array}{l}
          W_1^* \\ W_2^*
        \end{array}
        \right), ~~~
        W_1^*: \Omega \rightarrow \mathbb{R}^{n_1}, \ \
        W_2^*: \Omega \rightarrow \mathbb{R}^{n_2}, \ \ \nonumber \\
    &  W_1^*
   =  Y_{11} = Y_{21}, \\
    &  W_2^*
   =  L_1 Y_{12} + L_2 Y_{22} + L_3 V, \ \
        \mbox{see Theorem \ref{th:commoninfogrvcorrelated_new}.(b) for the formulas of $L_1, L_2, L_3$;} \\
&\mbox{then}\nonumber \\        
    & (Y_1,  Y_2,  W^*) \in G(0,Q_s(I)), ~
        (F^{Y_{11},Y_{12},Y_{13}}, 
         F^{Y_{21},Y_{22},Y_{23}}| F^{W_1^*,W_2^*} ) \in \ci, \ \  \mbox{see (\ref{eq:gtriple_1}) for $Q_s(I)$,} \\
    & F^{W_1^*} \subseteq (F^{Y_{11}} \vee F^{Y_{21}}), 
        F^{W_2^*} \subseteq (F^{Y_{12}} \vee F^{Y_{22}});\\
& \mbox{then also,} \nonumber  \\       
   &   C(Y_1,Y_2)
  =  I(Y_1, Y_2;W^*).
\end{align}
\item[(c)] The following operations are made, using (a),
\begin{align}
      W^*
   = & \left(
        \begin{array}{l}
          W_1^* \\ W_2^*
        \end{array}
        \right),  \\
      W_1^*
   = & Y_{11} = Y_{21}, \\
      W_2^*
   = & L_1 Y_{12} + L_2 Y_{22} + L_3 V,  \ \
        \mbox{see (\ref{eq:l1l2}),(\ref{eq:l3}) 
           for the formulas of} ~ L_1, ~ L_2, ~ L_3; \\
      Z_{12}
   = & Y_{12} - {\bf E}[Y_{12} | F^{W_2^*} ] 
        = Y_{12} - Q_{Y_{12},W_2^*} Q_{W_2^*}^{-1} W_2^*, ~ \\
      Z_{22} 
   = & Y_{22} - {\bf E}[Y_{22} | F^{W_2^*} ] = Y_{22} - Q_{Y_{22},W_2^*} Q_{W_2^*}^{-1} W_2^*, \\
      Z_{13}
   = & Y_{13}, \ \ Z_{23} = Y_{23},  \ \
        \mbox{(the components $Z_{11}$ and $Z_{21}$ do not exist),} \\
      Z_1
   = & \left(
        \begin{array}{l}
          Z_{12} \\ Z_{13}
        \end{array}
        \right), ~
      Z_2
      = \left(
        \begin{array}{l}
          Z_{22} \\ Z_{23}
        \end{array}
        \right).
\end{align}
Hence, 
\begin{align}
     Y_{11}
   =&  W_{1}^* = Y_{21}, \\
     Y_{12}
   = & Q_{Y_{12},W_2^*} Q_{W_2^*}^{-1} W_2^* +Z_{12}, \ \ \ \
       Y_{22} =  Q_{Y_{22},W_2^*} Q_{W_2^*}^{-1} W_2^*+Z_{22}, \\
      Y_{13}
   = & Z_{13}, \ \ \ \ Y_{23} = Z_{23}.
\end{align}
equivalently
\begin{align*}
&\left( \begin{array}{ll} Y_{11} \\Y_{12} \\ Y_{13} \end{array} \right) = \left( \begin{array}{lll}  I_{n_1}  & 0   \\   0  &  Q_{Y_{12},W_2^*} Q_{W_2^*}^{-1} \\ 0 & 0 \end{array} \right) \left( \begin{array}{ll} W_1^* \\ W_2^*\end{array} \right) + \left( \begin{array}{lll}  0  & 0   \\     I_{n_2} & 0 \\ 0 & I_{n-n_1-n_2} \end{array} \right) \left( \begin{array}{ll} Z_{12} \\ Z_{13}\end{array} \right)  \\
&\left( \begin{array}{ll} Y_{21} \\Y_{22} \\ Y_{23} \end{array} \right) = \left( \begin{array}{lll}  I_{n_1}  & 0   \\   0  &  Q_{Y_{22},W_2^*} Q_{W_2^*}^{-1} \\ 0 & 0 \end{array} \right) \left( \begin{array}{ll} W_1^* \\ W_2^*\end{array} \right) + \left( \begin{array}{lll}  0  & 0   \\     I_{n_2} & 0 \\ 0 & I_{n-n_1-n_2} \end{array} \right) \left( \begin{array}{ll} Z_{22} \\ Z_{23}\end{array} \right) . 
\end{align*}

\end{itemize}
\end{theorem}

The derivation of Theorem~\ref{thm_cia} is  presented in Section~\ref{subsec:wciarbitrarygrvs},  after several of the tools are presented, such as, weak stochastic realizations and minimal realizations.

\begin{remark}
\label{rem_lit_1}
 Although, Corollary 1 in  \cite{satpathy:cuff:2012} gives an  expression analogous to the case (\ref{coinf_3}), this  is expressed in terms of the correlation coefficients, $\rho_i \in (-1,1)$ and not the canonical correlation coefficients $d_i \in (0,1)$. Moreover, no realization of the triple $(Y_1, Y_2, W^*)$ is given,  as in Theorem~\ref{thm_cia} (which is based on applying the paremetrization of Theorem~\ref{them_putten:schuppen:1983}). 
 Corollary 1 of  \cite{satpathy:cuff:2012} is  also repeated in \cite{veld-gastpar:2016}, under Lemma 1, with the correlation coefficients $\rho_i$ replaced by their absolute values $|\rho_i|$. Moreover, 
 the proof in \cite{satpathy:cuff:2012,veld-gastpar:2016}  is judged  incomplete, 
because there is no optimization over realizations of a triple of random variable $(Y_1, Y_2, W)$ , which induce distributions ${\bf P}_{Y_1, Y_2, W}$, such that the conditional independence ${\bf P}_{Y_1, Y_2|W}={\bf P}_{Y_1|W} {\bf P}_{ Y_2|W}$  holds, and that the minimum in (\ref{eq_41}) occurs at   $(Y_1, Y_2, W^*)$. A technical issue  of the derivation  in  \cite{satpathy:cuff:2012}, is the assumption that there exist nonsingular matrices $S_{Y_1}, S_{Y_2}, S_W$ of $Y_1, Y_2, W$ such the three cross-covariances $Q_{Y_1,Y_2}, Q_{Y_1, W}, Q_{Y_2, W}$ can be simultaneously diagonalized, 
which is not true. Further, in the {\it Relaxed Wyner's Common Information} in    \cite{sula:gastpar:2019:arxiv} (see eqn(6)), it is not recognized that, for jointly Gaussian random variables $(Y_1, Y_2)$,  there always  exist realizations of $(Y_1, Y_2)$,  parametrized  by a Gaussian auxiliary random variable $W$, such that  ${\bf P}_{Y_1, Y_2|W}={\bf P}_{Y_1|W} {\bf P}_{ Y_2|W}$ (see  Section~\ref{ap:ci}); this means the minimization in    \cite{sula:gastpar:2019:arxiv} (see eqn(6)) occurs at $I(Y_1; Y_2|W)=0$, hence the constraint is always satisfied.
  These points are  further discussed in Remark~\ref{rem_lit_2} and Remark~\ref{rem_lit_rel}.
\end{remark}

The second main theorem is the formula of  Wyner's lossy common information $C_{GW}(Y_1, Y_2; \Delta_1, \Delta_2)$.

\begin{theorem}\label{thm:r0}
Consider a tuple $(Y_1,Y_2)$ of Gaussian random variables
in the canonical variable form of  Def. \ref{def:grvcommoncorrelatedprivateinfo}.
Restrict attention to the correlated parts of these random variables, as defined in Theorem~\ref{them_putten:schuppen:1983},  by (\ref{ci_par_1})-(\ref{ci_par_3}). \\
Then  Wyner's lossy common information  is given by
\begin{align}
C_{GW}(Y_1,Y_2;\Delta_1,\Delta_2)=&\; C_{W}(Y_1,Y_2)\\
=& \;C(Y_1,Y_2)
   =  \frac{1}{2} \sum_{j=1}^n 
        \ln
        \left(
        \frac{1+d_j}{1-d_j}
        \right), \ \   (\Delta_1, \Delta_2)\in {\cal D}_W
\end{align}
where  the distortion region is defined by 
\begin{align}
{\cal D}_W =& \Big\{(\Delta_1, \Delta_2)\in [0,\infty]\times [0,\infty]\Big| \ \ 0\leq \Delta_1 \leq n(1-d_1), \ \ 0\leq \Delta_2 \leq n(1-d_1)\Big\}, \\
 &\forall j \in {\mathbb Z}_n, \ \ {d_j \in (0,1)}. \label{dist_reg_ci}
\end{align} 
\end{theorem}

Additional, main contributions  include:  Theorem~\ref{th:jrdf_g}, which gives the joint RDF $R_{Y_1, Y_2}(\Delta_1, \Delta_2)$ and the realization of the optimal test channel that achives it, Theorem~\ref{th:jrdf}, which gives the conditional  RDFs $R_{Y_i|W}(\Delta_i,), i=1, 2$ and the realizations of the optimal test channels that achieve them, and  Theorem~\ref{thm_par}, which 
parametrizes the Gray and Wyner rate region  ${\cal R}_{GW}(\Delta_1, \Delta_2)$.

The derivation of Theorem~\ref{thm:r0} is presented in  Section~\ref{sect:com_r0} and makes use of a degenerate version of the  realization of the triple $(Y_1,Y_2,W^*)$ given in Theorem~\ref{thm_cia},  and the joint RDF $R_{Y_1, Y_2}(\Delta_1, \Delta_2)$,   given in   Theorem~\ref{th:jrdf_g}.

\subsection{Structure of the Paper}
\label{sect:stru} 
The terminology of the paper is simplified. When a tuple of Gaussian random variables is mentioned it
refers to a tuple of multivariate jointly-Gaussian random variables. The long expressions have been abbreviated
to save on space of the paper.

Section~\ref{sec:problem} introduces the mathematical tools of the geometric approach to Gaussian random variables (Section~\ref{ap:grvs}), the weak stochastic realization of conditional independence (Section~\ref{ap:ci}). 

 Section~\ref{sect:wciinformation} contains the problem statement,
the solution procedure (Section~\ref{sect:sol-pro}), and the weak realization of a tuple of multivariable random
variables $(Y_1,Y_2)$ such that another multivariate Gaussian random variable $W$ makes $Y_1$ and $Y_2$ conditionally
independent (Section~\ref{sect:ci}).
%
%
Section~\ref{sec:commoninfogrv} derives the formulae for Wyner’s lossy common information
$C_W(Y_1,Y_2)=C(Y_1,Y_2)$. This section also 
 provides the calculations of rate distortion functions 
$ R_{Y_i} (\Delta_i), i=1,2$,
the weak stochastic realizations of the random variables $(Y_1,Y_2,\hat{Y}_1,\hat{Y}_2,W)$ which achieve these rate distortion
functions, for jointly multivariate Gaussian random variables, with square-error distortion
functions.

Section~\ref{sect:wcilossy} contains the derivation of {Wyner's lossy Common information. 


Section~\ref{sec:concludingremarks} includes remarks on possible extensions.

Appendix~\ref{ap:inequality} makes use of a matrix equality and a determinant inequality first obtained by Hua LooKeng in
1952, which are used to carry out the optimization problem of Wyner's lossy common
information $C_W(Y_1,Y_2)=C(Y_1,Y_2)$.
 
Finally it is noted that an extended version of the current paper is found in \cite{charalambous:schuppen:2019:arxiv}.


\section{Probabilistic Properties of Tuples of Random Variables}
\label{sec:problem}
The reader finds in this section the solution procedure   for  two fundamental concepts of probability theory. 

1) The transformation of tuple of Gaussian multivariate random variables $(Y_1, Y_2)$, via nonsingular transformations $(S_1, S_2)$ such that the transform random variables $Y_1 \mapsto S_1 Y_1, Y_2 \mapsto S_2 Y_2$ are represented in their canonical variable form.

2) The parametrization of all  jointly Gaussian distributions  ${\bf P}_{Y_1, Y_2, W}(y_1, y_2,w)$ by a zero mean Gaussian  random variables $W: \Omega \rightarrow \mathbb{R}^{k}\equiv {\mathbb W}$ such that \\
(a) $W$ makes the multivariate random variables  $(Y_1, Y_2)$ conditional independent, and\\ 
(b) the marginal distribution ${\bf P}_{Y_1, Y_2, W}(y_1, y_1, \infty)={\bf P}_{Y_1, Y_2}(y_1, y_2)$ coincides with the joint distribution of the multivariare random variables $(Y_1, Y_2)$.\\

From  the mathematical concepts 1) and 2) one can formulate   Method 1  and Method 2, discussed in Section~\ref{sect:mo}.\\
These two mathematical concepts are introduced and discussed in Section~\ref{ap:ci} and Section~\ref{ap:grvs}.

\subsection{Notation of Elements of Probability Theory}\label{ap:notationprob}
%
The notation used in the paper is briefly specified.
Denote by $\mathbb{Z}_+ = \{ 1,2, \ldots, \}$ 
the set of the integers
and by $\mathbb{N} = \{ 0,1,2, \ldots, \}$
the set of the natural integers.
For $n \in \zpos$ denote the following finite subsets 
of the above defined sets by
$\mathbb{Z}_n = \{ 1,2, \ldots, n \}$
and $\mathbb{N}_n = \{ 0,1,2, \ldots, n\}$.
\par
Denote the real numbers by
$\mathbb{R}$ and the set of
positive and of strictly positive real numbers, respectively, by
$\rpos = [0,\infty)$ and ${\mathbb R}_{++}=(0,\infty) \subset \mathbb{R}$.
The vector space of $n$-tuples of real numbers is denoted 
by $\rn$.
Denote the Borel $\sigma$-algebra on this vector space
by $B(\mathbb{R}^n)$
hence $(\mathbb{R}^n,B(\mathbb{R}^n))$
is a measurable space.
\par
The expression $\mathbb{R}^{n \times m}$
denotes the set of $n$ by $m$ matrices with elements
in the real numbers, for $n, ~ m \in \zpos$.
For the symmetric matrix $Q \in \rnn$ 
the inequality $Q \geq 0$ denotes
that for all vectors $u \in \rn$ the inequality
$u^T Q u \geq 0$ holds.
Similar, 
$Q > 0$ denotes that for all 
$u \in \rn \backslash \{0\}$,
$u^T Q u > 0$.
The notation $Q_1 \leq Q_2$  denotes that
$Q_2 - Q_1 \geq 0$.
\par
Consider a probability space denoted by
$(\Omega,F,{\mathbb P})$
consisting of a set $\Omega$, 
a $\sigma$-algebra $F$ of subsets of $\Omega$,
and a probability measure
${\mathbb P}: F \rightarrow [0,1]$.
\par
A real-valued random variable
is a function $X: \Omega \rightarrow \mathbb{R}$
such that 
the following set belongs to the indicated
$\sigma$-algebra,
$\{ \omega \in \Omega | X(\omega) \in (-\infty,u] \} \in F$
for all $u \in \mathbb{R}$.
A random variable taking values in an arbitrary  measurable space $({\mathbb X},B({\mathbb X}))$
is defined correspondingly
by $X: \Omega \rightarrow {\mathbb X}$
and $X^{-1}(A) = \{ \omega \in \Omega | X(\omega) \in A \} \in B({\mathbb X})$,
for all $A \in B({\mathbb X})$.  The measure (or distribution if ${\mathbb X}$ is a Euclidean space) induced by the random variable on $({\mathbb X},B({\mathbb X}))$ is denoted by ${\bf P}_X$ or ${\bf P}(dx)$. The $\sigma$-algebra generated by a random variable
$X: \Omega \rightarrow {\mathbb X}$ is defined as the smallest
$\sigma$-algebra containing the 
subsets $X^{-1}(A) \in F$ for all $A \in B({\mathbb X})$.
It is denoted by $F^X$.
The real-valued random variable $X$ is called $G$-measurable
for a $\sigma$-algebra $G \subseteq F$
if the subset
$\{ \omega \in \Omega | X(\omega) \in (-\infty,u] \} \in G$
for all $u \in \mathbb{R}$.
Denote the set of positive random variables
which are measurable on a sub-$\sigma$-algebra
$G \subseteq F$ by,
\[
  L_+(G) =
  \{ X: \Omega \rightarrow \mathbb{R}_+ = [0,\infty) | X ~
     \mbox{is $G$-measurable} 
  \}.
\]
\par
The tuple of sub-$\sigma$-algebras $F_1, ~ F_2 \subseteq F$
is called {\em independent} if
${\mathbb E}[X_1 X_2 ] = {\mathbb E}[X_1] {\mathbb E}[X_2]$
for all $X_1 \in L_+(F_1)$ and all $X_2 \in L_+(F_2)$.
The definition can be extended to any finite set of
independent sub-$\sigma$-algebras, and to random variables taking values in arbitrary measurable spaces $({\cal X},B({\mathbb X}))$.

\subsection{Geometric Approach of  Gaussian Random Variables and Canonical Variable Form}
\label{ap:grvs_new}
The purpose of this section is to introduce concepts   and 
 results on the canonical variable form for a tuple of  finite-dimensional Gaussian random variables, with emphasis on the {\em geometric approach} 
of Gaussian random variables.
in the context of multi-user communication.
 Thus the spaces generated by these random variables
are the main objects of study
while the actual random variables are only representations
with respect to a basis. 
 The concept of a canonical variable decomposition
was introduced by H. Hotelling, \cite{hotelling:1936}.
%
%


 
The geometric objects are the
$\sigma$-algebras of the Gaussian random variables.
With respect to a basis, 
the probability distribution of a Gaussian random variable 
is characterized by its mean and its covariance matrix.
Because the central theme of this paper is Gaussian random variables,
the exposition is a combination of the geometric approach
and the basis representations.
\par
A {\em $\rn$-valued Gaussian random variable}
with as parameters 
the {\em mean value} $m_X \in \rn$ and 
the {\em variance}
$Q_X \in \rnn$, $Q_X = Q_X^T \geq 0$,
is a function $X: \Omega \rightarrow \rn$ 
which is a random variable and such that
the measure of this random variable equals a Gaussian 
measure described by its characteristic function,
\[
  {\bf E}[ \exp (iu^T X) ]
  = \exp( i u^T m_X - \frac{1}{2} u^T Q_X u), ~
  \forall ~ u \in \rn.
\]
Note that this definition includes the case
when the random variable is almost surely
equal to a constant in which case $Q_X = 0$.
A Gaussian random variable with these parameters
is denoted $X \in G(m_X, Q_X)$.\\
The {\it effective dimension} of the random variable is denoted by 
$\dim (X) = \rank (Q_X)$.
\par
Any tuple of random variables
$X_1, \ldots, X_k$ is called 
{\em jointly Gaussian} if
the vector $(X_1,  X_2,  \ldots,  X_k)^T$
is a Gaussian random variable.
A tuple of Gaussian random variables 
$(Y_1, Y_2)$ will be denoted this way to save space,
rather than by
\begin{align*}
  &   & \left(
        \begin{array}{l}
          Y_1 \\ Y_2
        \end{array}
        \right).
\end{align*}
Then the variance matrix of this tuple is denoted by,
\begin{align*}
   & (Y_1,Y_2) \in G(0,Q_{(Y_1,Y_2)}), ~~
        Q_{(Y_1,Y_2)}
     =  \left(
        \begin{array}{ll}
          Q_{Y_1} & Q_{Y_1,Y_2} \\
          Q_{Y_1,Y_2}^T & Q_{Y_2}
        \end{array}
        \right) \in \mathbb{R}^{(p_1+p_2) \times (p_1+p_2)}.
\end{align*}
The reader should distinguish the variance matrices
$Q_{(Y_1,Y_2)}$ and $Q_{Y_1,Y_2} \in \mathbb{R}^{p_1 \times p_2}$.
Any such tuple of Gaussian random variables
is independent if and only if $Q_{Y_1,Y_2} = 0$.
\par
The conditional expectation of 
a Gaussian random variable $X: \Omega \rightarrow \rn$
conditioned on the $\sigma$-algebra generated
by another Gaussian random variable
$Y: \Omega \rightarrow \rp$
with $(X,Y) \in G(m,Q_{(X,Y)})$
and with $Q_{Y} > 0$
is, as is well known, again a Gaussian random variable
with characteristic function,
\begin{align}
      {\bf E}[ \exp ( iu^T X) | F^Y] 
   = & \exp (i u^T {\bf E}[X|F^Y] - \frac{1}{2} u^T Q_{X|Y} u ), \ \
        \forall ~ u \in \rn, ~ 
        \mbox{where,} \label{ccha_1} \\
      Q_{(X,Y)}
  = & \left(
        \begin{array}{ll}
          Q_X & Q_{X,Y} \\
          Q_{X,Y}^T & Q_Y
        \end{array}
        \right), \nonumber \\
      {\bf E}[X|F^Y]
  = & m_X +  Q_{X,Y} Q_Y^{-1} (Y - m_Y), \\
      Q_{X|Y}
   = & Q_X - Q_{X,Y} Q_Y^{-1} Q_{X,Y}^T. \label{cchar_4}
\end{align}
\par
Next the geometric approach is defined.
A random variable as defined above
is always defined with respect to a particular basis
of the linear space.
The underlying geometric object 
of a Gaussian random variable $Y: \Omega \rightarrow \rp$ 
is the $\sigma$-algebra $F^Y$.
In this paper the authors prefer the $\sigma$-algebra
as the geometric object 
rather than the linear space generated by 
the random variable.
A {\em basis transformation} of such a random variable
is then the transformation defined by 
a nonsingular matrix $S \in \rpp$, which implies  $F^Y = F^{S Y}$
and the random variable $SY$ is Gaussian.
\par
Next consider a tuple of jointly Gaussian random variables
$(Y_1, Y_2)$.
A basis transformation of this tuple consists of
a matrix $ S = \blockdiag ( S_1 , S_2 ) $ with
$S_1 , S_2 $ square and nonsingular matrices, which then implies   that spaces satisfy
$F^{ Y_1 } = F^{S_1 Y_1}$ and $F^{Y_2} = F^{ S_2 Y_2 } $, and $S_1Y_1$ and $S_2Y_2$ are Gaussian.
This transformation introduces an equivalence relation
on the representation of the tuple of random variables $(Y_1, Y_2)$. Further, it is known that mutual information, 
 is invariant with respect to such transformations \cite{pinsker:1964}.
Thus one can speak about a canonical form for these spaces
which is introduced next.

\par
Below the following problem is considered
and the solution provided.

\begin{problem}
\label{hotelling}
Consider the tuple of jointly Gaussian random variables
$ Y_1 : \Omega \to \mathbb{R}^{ p_1 } $
and $Y_2 : \Omega \to \mathbb{R}^{ p_2 } $,
with $(Y_1 , Y_2 ) \in G(0,Q_{(Y_1,Y_2)})$.
Determine a canonical form for the spaces
$F^{ Y_1 } , \  F^{ Y_2 } $
up to linear basis transformations.
\end{problem}

The above problem has been posed and solved 
by H. Hotelling \cite{hotelling:1936} as described below.
Other references are the books
\cite{anderson:1958,gittens:1985}. The reader may identify the elements of equations (\ref{cvf_1})-(\ref{cvf_3}) with the next definition.
  Below without loss of generality the assumption is used that the variances of the random variables
  $Y_1, Y_2$ are  such that $Q_{Y_1} > 0, Q_{Y_2} > 0 $ (see \cite{charalambous:schuppen:2019:arxiv} for justification).

%
\begin{definition}\label{def:grvcommoncorrelatedprivateinfo}
The canonical variable form.\\
Consider a tuple of Gaussian random variables 
$Y_i: \Omega \rightarrow \mathbb{R}^{p_i}$, 
with $Q_{Y_i} > 0$, for $ i=1, 2$, $(Y_1 , Y_2 ) \in G(0,Q_{(Y_1,Y_2)})$.
Define the {\em canonical variable form}
of these random variables 
if a basis has been chosen and a transformation
of the random variables to this basis has been carried out
such that with respect to the new basis one has the representation,
\begin{align}
   & (Y_1,Y_2) \in  G(0,Q_{\cvf}), \ \ \mbox{where,} \nonumber \\
   &   Q_{\cvf}
   =  \left(
        \begin{array}{lll|lll}
          I_{p_{11}} & 0          & 0          & I_{p_{21}} & 0          & 0 \\
          0          & I_{p_{12}} & 0          & 0          & D          & 0 \\
          0          & 0          & I_{p_{13}} & 0          & 0          & 0 \\ \hline
          I_{p_{21}} & 0          & 0          & I_{p_{21}} & 0          & 0 \\
          0          & D          & 0          & 0          & I_{p_{22}} & 0 \\
          0          & 0          & 0          & 0          & 0          & I_{p_{23}} 
        \end{array}
        \right) \in \mathbb{R}^{p \times p}, 
        \label{eq:qcanonicalvariance} \\
    & \ \ \ \  \ \ p, ~ p_1, ~ p_2, ~
        p_{11}, ~ p_{12}, ~ p_{13}, ~
        p_{21}, ~ p_{22}, ~ p_{23} \in \mathbb{N}, \nonumber \\    
    & \ \   \ \  \ \ p = p_1 + p_2, ~
        p_1 = p_{11} + p_{12} + p_{13}, ~
        p_2 = p_{21} + p_{22} + p_{23}, ~
        p_{11} = p_{21}, ~ p_{12} = p_{22},  \nonumber \\
    &  D
  =  \diag ( d_1, \ldots, d_{p_{12}} ), ~~
          1 > d_1 \geq d_2 \geq \ldots \geq d_{p_{12}} > 0, \\
     & Y
   =  \left(
        \begin{array}{l}
          Y_1 \\ \hline
          Y_2 
        \end{array}
        \right) 
        =
        \left(
        \begin{array}{l}
          Y_{11} \\ Y_{12} \\ Y_{13} \\ \hline
          Y_{21} \\ Y_{22} \\ Y_{23}
        \end{array}
        \right), ~~
        Y_{ij}: \Omega \rightarrow \mathbb{R}^{p_{ij}}, ~
        i =1, 2, ~ j = 1, 2, 3.
\end{align}
  One then says that
  $(Y_{11} , \ldots , Y_{ 1k_1 } )$,
  $( Y_{21} , \ldots , Y_{ 2k_2 } )$
  are the  canonical variables
  and $( d_1 , \ldots , d_{ k_{12} } )$
  the  canonical correlation coefficients. 
\end{definition}

\ \


The elements of Method 2 discussed in Section~\ref{sect:mo} are identified from Definition~~\ref{def:grvcommoncorrelatedprivateinfo}, that is,
\begin{align}
&S_1Y_1 = (V_1,Y_1^\prime) = ( (Y_{11},Y_{12}),Y_{13}), \ \  S_2Y_2 = (V_2,Y_2^\prime) = ( (Y_{21},Y_{22}),Y_{23}), \\
&  Y_{11}=Y_{21}-a.s., \ \   {\bf E}[ Y_{12} Y_{22}^T ] = D.
\end{align}

Appendix~\ref{sect:cvf_e}   states Theorem~\ref{thm-exi-cvf} on the existence of the canonical variable form and properties.

The precise relation with  
Definition~\ref{def:grvcommoncorrelatedprivateinfo} to the bivarate Gaussian  random variables  (\ref{cvf_s})  of  Theorem~\ref{the:scalar-ci},  i.e.,  $p_1=p_2=1$, is discussed  in  Remark~\ref{rem-con-scalar}, by first introducing a list of important  properties. 

%

\begin{proposition}\label{prop:grvcpproperties}
Properties of components of the canonical variable form.\\
Consider a tuple $(Y_1,Y_2) \in G(0,Q_{\cvf})$ of Gaussian random variables
in the canonical variable form.
\begin{itemize}
\item[(a)]
The three components 
$Y_{11},Y_{12},Y_{13}$ of $Y_1$ are independent random variables.
Similarly,
the three components 
$Y_{21},Y_{22},Y_{23}$ of $Y_2$ are independent random variables.
\item[(b)]
The equality $Y_{11} = Y_{21}$ 
of these random variables holds almost surely.
\item[(c)]
The tuple of random variables $(Y_{12},Y_{22})$
is correlated as shown by the formula
\begin{equation}
  {\bf E}[Y_{12} Y_{22}^T ] = D = \diag (d_1, \ldots, d_{p_{12}} ).
\end{equation}
Note that the different components 
of $Y_{12}$ and of $Y_{22}$ are
independent random variables;
thus $Y_{12,i}$ and $Y_{12,j}$ are independent,
and $Y_{22,i}$ and $Y_{22,j}$ are independent, and
$Y_{12,i}$ and $Y_{22,j}$ are independent,
for all $i \neq j$;
and that
$Y_{12,j}$ and $Y_{22,j}$ for $j = 1, \ldots, p_{12}=p_{22}$
are correlated.
\item[(d)]
The random variable $Y_{13}$ is independent of $Y_2$.
Similarly,
the random variable $Y_{23}$ is independent of $Y_1$
\end{itemize}
\end{proposition}
{\bf Proof}
The results are immediately obvious 
from the fact that the random variables are all jointly
Gaussian and from the variance formula
(\ref{eq:qcanonicalvariance}) of the canonical variable form.
\hfill$\square$

\ \


\begin{example}
\label{rem-con-scalar}
Consider Definition~\ref{def:grvcommoncorrelatedprivateinfo},  for the bivarate Gaussian  random variables  (\ref{cvf_s})  of  Theorem~\ref{the:scalar-ci},  i.e., let $p_1=p_2=1$. 
Then the following results hold for three cases.\\
(i) 
${\bf E}[Y_{12} Y_{22}] = d_1 \in (0,1)$. 
Then $(Y_1, Y_2)\in G(0,Q_{\cvf})$ are described by  
$Y_1=Y_{12}, Y_2=Y_{22}$ and 
hence the components 
$Y_{11}, Y_{13}$ and  $Y_{21}, Y_{23}$ of $Y_1$ and $Y_2$, 
respectively, are absent, and 
${\bf E}[Y_{12} Y_{22}] = d_1$. 
Hence, 
\begin{align}
  Q_{\cvf}=Q_{(Y_{12}, Y_{22})} 
  = \left[ 
    \begin{array}{cc} 
      1 & d_1  \\
      d_1  & 1 
    \end{array} 
    \right]. \label{eq_rem-con-scalar}
\end{align} 
(ii) $d_1 = 0$. 
Then $(Y_1, Y_2)\in G(0,Q_{\cvf})$ are described by  
$Y_1=Y_{13}, Y_2=Y_{23}$ and 
hence the components 
$Y_{11}, Y_{12}, Y_{21}, Y_{22}$  are absent, and 
$Y_{13}, Y_{23}$ are independent Gaussian $G(0,1)$.\\
(iii) 
$d = 1$. 
Then $(Y_1, Y_2)\in G(0,Q_{\cvf})$ are described by  
$Y_1=Y_{11}, ~ Y_2=Y_{21}$, $Y_{11}=Y_{21}-$a.s., 
$Y_1, ~ Y_2 \in G(0,1)$,  and 
hence the components $Y_{12}, Y_{13}, Y_{22}, Y_{23}$  are absent.
\end{example}

\begin{remark} The previous example illustrates that, 
even  for the bivarate Gaussian  random variables  
(\ref{cvf_s})  of  Theorem~\ref{the:scalar-ci},  
the main elements are the canonical correlation coefficients 
$\{ d_i \in (0,1), ~ \forall ~ i \in \mathbb{Z}_{p_{12}} \}$, and 
not the correlation coefficients 
$\{ \rho_i \in (-1,+1), \forall ~ i \in \mathbb{Z}_{p_{12}} \}$. 
\end{remark}

Next the interpretation of the various components
of the canonical variable form is  defined, as in 
 \cite{schuppen:2015:c4cbook:ch26}.
\begin{definition}\label{def:infocommonprivate}
Interpretation of components of the canonical variable form.\\  
Consider a tuple of jointly-Gaussian random variables
$(Y_1, Y_2) \in G(0,Q_{\cvf})$
in the canonical variable form of 
Definition~\ref{def:grvcommoncorrelatedprivateinfo}.
Call the various components as defined in the next table.
\par\vspace{1\baselineskip}\par\noindent
\begin{center}
\begin{tabular}{|l|l|}
\hline 
$Y_{11} = Y_{21}-\as$ & {\em identical information}  of $Y_1$ and $Y_2$ \\
$Y_{12}$          & {\em correlated information} of $Y_1$ with respect to $Y_2$ \\
$Y_{13}$          & {\em private information}    of $Y_1$ with respect to $Y_2$ \\ \hline
$Y_{21} = Y_{11}-\as$ & {\em identical information}  of $Y_1$ and $Y_2$ \\
$Y_{22}$          & {\em correlated information} of $Y_2$ with respect to $Y_1$ \\
$Y_{23}$          & {\em private information}    of $Y_2$ with respect to $Y_1$ \\ \hline
\end{tabular}
\end{center}
\par\vspace{1\baselineskip}\par\noindent
For $Y_{11} = Y_{21}-a.s.$ the term {\em identical information} is used.
\end{definition}

\ \

The next remark recalls the formula of mutual information $I(Y_1;Y_2)$ between two scalar-valued Gaussian random variables, expressed in terms of their correlation coefficient. Then the formula of $I(Y_1;Y_2)$ for a tuple of multivariate random variables, expressed in terms of the canonical correlation coefficients is presented.

\begin{remark} 
\label{rem_smi}
Consider two jointly Gaussian scalar-valued  random variables $(Y_1, Y_2)$ with  variance matrix defined by (\ref{cvf_s}). Then mutual information, for any $\rho \in [-1,1]$, is given by 
\begin{align}
I(Y_1;Y_2)=- \frac{1}{2}\ln\Big(1-\rho^2\Big)= \left\{ \begin{array}{ccc} +\infty, & \mbox{if } & \rho\in \{-1, 1\}, \\
 - \frac{1}{2}\ln\Big(1-\rho^2\Big)\in (0,\infty), & \mbox{if} & \rho \in (-1,1), \rho \neq 0,\\
  0, & \mbox{if} & \rho=0.
\end{array} \right. \label{mi_s}
\end{align}
In particular, (\ref{mi_s}) shows that for Gaussian random variables $(Y_1,Y_2)$ then $I(Y_1;Y_2)=+\infty$ if and only if the correlation coefficient takes the values $\rho=+1, \rho=-1$. In the case, $Y_1=Y_2$ with probability one then $I(Y_1;Y_2)=+\infty$.
\end{remark}

Remark~\ref{rem_smi} is now  extended to a general tuple of finite-dimensional Gaussian random variables
$(Y_1, Y_2) \in G(0,Q_{(Y_1,Y_2)})$. It should be mentioned that a version of Theorem~\ref{them_mi} is discussed  in \cite{gelfand-yaglom:1957}.

\begin{theorem}\label{them_mi}
Consider a tuple of finite-dimensional Gaussian random variables
$(Y_1, Y_2) \in G(0,Q_{(Y_1,Y_2)})$.\\
Compute the canonical variable form of the tuple
of Gaussian random variables
according to Algorithm \ref{alg:canonicalvariabledecomposition}.
This yields the indices
$p_{11} = p_{21}$, $p_{12} = p_{22}$, $p_{13}$, $p_{23}$, and 
$n = p_{11} + p_{12} = p_{21} + p_{22}$
and the diagonal matrix $D$ with canonical correlation coefficients or singular values
$d_i \in (0,1)$ for $i = 1, \ldots, n$.\\
Then mutual information  $I(Y_1; Y_2)$ is computed according to the formula,
\begin{align}
      I(Y_1;Y_2)
   = & \left\{
        \begin{array}{lll}
            0, 
          &   \mbox{if} & 
              0 = p_{11} = p_{12} = p_{21} = p_{22}, ~
              p_{13} > 0, ~
              p_{23} > 0, \\
            -\frac{1}{2} \sum_{i=1}^n 
            \ln
            \left(1-d_i^2
            \right), 
          &   \mbox{if} & 
              0 = p_{11} = p_{21}, ~
              p_{12} = p_{22} > 0, ~
              p_{13} \geq 0, ~ p_{23} \geq 0, \\
            \infty,  
          &   \mbox{if} & 
              p_{11} = p_{21} > 0, ~
              p_{12} = p_{22} \geq 0, ~
              p_{23} \geq 0, ~ p_{23} \geq 0. 
        \end{array}
        \right. \label{eq:mi}
\end{align}
where $d_i$ are the canonical correlation coefficients, i.e.,
\begin{align}
d_i=d_i(Y_{12,i}Y_{22,i})=\frac{{\bf E}\big\{Y_{12,i} Y_{22,i}\big\}}{\sqrt{ {\bf E}\big\{Y_{12,i}\big\}^2 {\bf E}\big\{Y_{22,i}\big\}^2}}={\bf E}\big\{Y_{12,i} Y_{22,i}\big\}, \ \ i=1, \ldots, n.\label{eq:ccc_mi}
\end{align}
\end{theorem}
{\bf Proof} See Appendix~\ref{proof_app_thm-exi-cvf}.  
\hfill$\square$

\ \

Appendix~\ref{app_sect:gk-w}  describes the connection of the canonical variable form to the computation of the information definition
of Gacs and K\"{o}rner \cite{gacs-korner:1973} common information, and Wyner's information definition of common information $C(Y_1, Y_2)$.  It is remarked that the canonical variable form is directly
applicable to the lossy extension of G\'{a}cs and K\"{o}rner common information derived by Viswanatha, Akyol
and Rose \cite{viswanatha:akyol:rose:2014} in Section IV. Specifically, the information theoretic characterization of Theorem 3 in \cite{viswanatha:akyol:rose:2014}.

In Section~\ref{ap:ci}, it will be shown how to construct  a probability measure that carries an auxiliary Gaussian random variable $W$, such that conditional independence holds: ${\bf P}_{S_1Y_1,S_2Y_2|W}={\bf P}_{S_1Y_1|W}{\bf P}_{S_2Y_2|W}$, from which $C(Y_1, Y_2)=C(S_1Y_1, S_2Y_2)$ can be computed. \\

The algorithm that generates $Q_{\cvf}$ is presented below.

\begin{algorithm}\label{alg:canonicalvariabledecomposition}
  Transformation of a variance matrix to its canonical variable form.\\
  Data: $p_1 , p_2 \in Z_+ $,
  $Q \in \mathbb{R}^{ ( p_1 + p_2 ) \times ( p_1 + p_2 ) } $,
  satisfying\footnote{It is noted that $Q>0$ implies $Q_{11}>0, Q_{22}>0$.} $Q=Q^T  > 0 $, with decomposition
  \begin{eqnarray*}
        Q  
    =  \left( 
          \begin{array}{ll}
            { Q_{11} } & { Q_{12} }  \\
            { Q_{12}^T } & { Q_{22} } 
          \end{array}
          \right), ~~
          Q_{11} \in \mathbb{R}^{ p_1 \times p_1 }, ~
          Q_{22} \in \mathbb{R}^{ p_2 \times p_2 }, ~
          Q_{12} \in \mathbb{R}^{ p_1 \times p_2 }.
  \end{eqnarray*}
  \begin{enumerate}
    \item
      Perform singular-value decompositions:
      \[
        Q_{11} \  = \  U_1 D_1 U_1^T , \ \ 
        Q_{22} \  = \  U_2 D_2 U_2^T , 
      \]
      with $U_1 \in \mathbb{R}^{ p_1 \times p_1 } $
      orthogonal ($ U_1 U_1^T = I = U_1^T U_1$)
      and
      \[
        D_1 =  
        \diag ( d_{1,1} ,..., d_{ 1,p_1 } )
        \in \mathbb{R}^{ p_1 \times p_1 } , \ \ 
        d_{1,1} \geq d_{1,2} \geq \ldots \geq d_{ 1,p_1 } 
        > 0 ,
      \]
      and $U_2 , D_2 $ satisfying corresponding conditions.
    \item
      Perform a singular-value decomposition of
      \[
        D_1^{ - \frac{1}{2}  } U_1^T Q_{12}
        U_2 D_2^{ - \frac{1}{2} } 
        \  = \ 
        U_3 D_3 U_4^T ,
      \]
      with $U_3 \in \mathbb{R}^{ p_1 \times p_1 } $,
      $U_4 \in \mathbb{R}^{ p_2 \times p_2 } $ orthogonal and
      \begin{align*}
         D_3 = & 
            \left( 
            \begin{array}{lll}
              I_{p_{11}} & 0   & 0 \\
              0          & D_4 & 0  \\
              0          & 0   & 0 
            \end{array}
            \right)
            \in \mathbb{R}^{ p_1 \times p_2 } , \\
        D_4  = &  \diag ( d_{4,1} ,..., d_{ 4,p_{12} } ) \in 
          \mathbb{R}^{ p_{12} \times p_{12} }  , \ \ 
          1 > d_{4,1} \geq d_{4,2} \geq \ldots \geq d_{4,p_{12}} > 0 .
      \end{align*}
    \item
      Compute the new variance matrix according to,
      \begin{align*}
            Q_{\cvf}
         =  \left(
              \begin{array}{ll}
                I_{p_1} & D_3 \\
                D_3^T   & I_{p_2}
              \end{array}
              \right).
      \end{align*} 
    \item
      The transformation to the canoncial variable representation\\
      $( Y_1 \mapsto S_1 Y_1 , \  Y_2 \mapsto S_2 Y_2 )$
      is then 
      \[
        S_1 \  = \ 
        U_3^T D_1^{ - \frac{1}{2} } U_1^T , \ \ 
        S_2 \  = \ 
        U_4^T D_2^{ - \frac{1}{2} } U_2^T. 
      \]
  \end{enumerate}
\end{algorithm}

\begin{remark}
\label{rem:cvf_1}
 The material discussed in Section~\ref{sect:mo}, related to Method 2 are  applications of the concepts of this section. The main point to be made is that in lossy source coding problems, the source distribution is fixed, while the optimal reproduction distribution needs to be found and its realization.
  In source coding problems one application of Method 2 is the pre-encoder, which is constructed by  invoking Algorithm~\ref{alg:canonicalvariabledecomposition}. \\
\end{remark}

\subsection{Weak Realization of Conditional Independent Gaussian Random Variables}\label{ap:ci}
%
The purpose of this section is to introduce concepts   and 
 results as stated in Section~\ref{sec:problem}, under 2),  on conditional independence of  a tuple of  finite-dimensional Gaussian random variables $(Y_1,Y_2)$, conditioned on another Gaussian random variable $W$, and the  weak realization of these random variables.

The concept of conditional independence and its connection  to 
the weak  realization problem is  stated below. 
It is shown in the companion paper \cite{cdc:jhv:2019}  that the rate region ${\cal  R}_{GW}(\Delta_1,\Delta_2)$ is parametrized by a random variable $W$ that makes $Y_1$ and $Y_2$ conditionally independent. 
The fact that, for the calculation of  $C_{GW}(Y_1,Y_2;\Delta_1, \Delta_2)$ and $C_W(Y_1,Y_2)$,  it is sufficient to consider only random variables  $W$ that make $Y_1$ and $Y_2$ conditionally independent is discussed in Section~\ref{sect:wciinformation}.

However, the current paper is only concerned with the use of the material of this section (and last section) to compute the information quantity $C(Y_1,Y_2)$, which can be shown to be Wyner's  lossy common information, for certain distortion regions.

%
%
\begin{definition} Conditional independence.\\
Consider a probability space $(\Omega,F,{\mathbb P})$ and 
three sub-$\sigma$-algebras $F_1, F_2, G \subseteq F$.
Call the sub-$\sigma$-algebras $F_1$ and $F_2$
{\em conditionally independent} given, or conditioned on,
the sub-$\sigma$-algebra $G$ if the following
factorization property holds,
\begin{eqnarray}
       {\bf E}[Y_1 Y_2 | G] 
   =  {\bf E}[Y_1|G] E[Y_2|G], \ \ \forall \ \
         Y_1 \in L_+(F_1), \ \ Y_2 \in L_+(F_2).
\end{eqnarray}
Denote this property by
$(F_1,F_2|G) \in CI$.
\end{definition}
Examples of triples of sub-$\sigma$-algebras
which are conditional independent are
(1) $(F_1,F_2|F_1) \in \ci$;  \\
(2) $(F_1,F_2|F_2) \in \ci$; 
(3) $(F_1,F_2|\{\emptyset,\Omega\}) \in \ci$
if $F_1$ and $F_2$ are independent.
\ \

For Gaussian random variables the definition of minimality of a Gaussian random variable $X$ that makes two  Gaussian random variables $(Y_1,Y_2)$ conditionally independent is needed. The definition is introduced below.

\begin{definition} Minimality of conditional indepedence of Gaussian random variables.\\
\label{def_minci}
Consider three random variables,
$Y_i: \Omega \rightarrow \mathbb{R}^{p_i}$
for $i = 1,~ 2$ 
and $X: \Omega \rightarrow \mathbb{R}^n$.\\
Call the random variables $Y_1$ and $Y_2$
{\em Gaussian conditionally independent}
conditioned on or given $F^X$ if \\
(1) $(F^{Y_1},F^{Y_2}|F^X) \in \ci$ and \\
(2) $(Y_1,Y_2,X)$ are jointly Gaussian random variables.\\
The notation $(Y_1,Y_2|X) \in \cig$ is used to denote this property.\\
Call the random variables $(Y_1, Y_2 | X)$
{\em minimally Gaussian conditionally independent} if \\
(1) they are Gaussian conditionally independent and \\
(2) there does not exist another tuple
$(Y_1,Y_2|X_1)$ with $X_1: \Omega \rightarrow \mathbb{R}^{n_1}$
such that $(Y_1,Y_2|X_1) \in \cig$ and $n_1 < n$.\\
This property is denoted by
$(Y_1,Y_2|X_1) \in \cig_{min}$.
\end{definition}
There exists a simple equivalent condition
for conditional independence of tuple of Gaussian random variables by a third Gaussian random variable. This condition is expressed in terms of parametrizing the variance matrix of the tuple as presented in the next proposition. 
\begin{proposition}\label{prop:cigrvs}
\cite[Prop. 3.4]{putten:schuppen:1983} Equivalent condition
for conditional independence of tuple of Gaussian random variables.\\
Consider a triple of jointly Gaussian random variables
denoted as 
$(Y_1,Y_2,X) \in G(0,Q)$ with $Q_X > 0$.
This triple is Gaussian conditionally independent
if and only if
\begin{equation}
  Q_{Y_1,Y_2} = Q_{Y_1,X} Q_X^{-1} Q_{X,Y_2}.
\end{equation}
It is minimally Gaussian conditionally-independent
if and only if,
in addition,
$n = \dim (X) = \rank (Q_{Y_1,Y_2})$.
\end{proposition}

As mentioned earlier, the  calculation of $C_{GW}(Y_1,Y_2;\Delta_1, \Delta_2)$ and $C_W(Y_1,Y_2), C(Y_1,Y_2)$ is directly related to  the solution of the weak stochastic realization problem of random variables $(W, Y_1,Y_2, \hat{Y}_1, \hat{Y}_2)$, that achieve   the  rate distortion functions $R_{Y_i}(\Delta_i)$, $R_{Y_i|W}(\Delta_i), i=1,2$, and $R_{Y_1,Y_2}(\Delta_1, \Delta_2)$, that is, the construction of  the joint distribution ${\bf P}_{W, Y_1,Y_2,\widehat{Y}_1, \widehat{Y}_2}$, with marginal the source distribution ${\bf P}_{Y_1,Y_2}$,  which achieves these rate distortion functions. These rate distortion functions are much easier to calculate, if  
a tuple of Gaussian random variables 
$Y_i: \Omega \rightarrow \mathbb{R}^{p_i}$, 
with $Q_{Y_i} > 0$, for $ i=1, 2$, $(Y_1 , Y_2 ) \in G(0,Q_{(Y_1,Y_2)})$, is transformed to the  canonical variable form of Definition~ \ref{def:grvcommoncorrelatedprivateinfo}, i.e.$(Y_1,Y_2) \in  G(0,Q_{\cvf})$, and then Gray and Wyner's lossy rate region ${\cal R}_{GW}(\Delta_1,\Delta_2)$ is parametrized by the random variable $W$ that makes $Y_1$ and $Y_2$ conditionally independent. 


%
\begin{problem} Weak stochastic realization.\\
\label{prob_real}
{\em Weak stochastic realization problem of a Gaussian random variable.}
Consider a Gaussian measure ${\bf P}_0=G(0,Q_0)$ on the space
$(\mathbb{R}^{p_1+p_2}, B(\mathbb{R}^{p_1+p_2}))$.
Determine the integer $n \in \mathbb{N}$ and
construct all Gaussian measures on the space
$(\mathbb{R}^{p_1+p_2+n},(B(\mathbb{R}^{p_1+p_2+n}))$
such that, 
if ${\bf P}_1=G(0,Q_1)$ is such a measure with
$(Y_1,Y_2,X) \in G(0,Q_1)$, 
then \\
(1) $G(0,Q_1)|_{\mathbb{R}^{p_1+p_2}} = G(0,Q_0)$ and\\
(2) $(Y_1,Y_2|X) \in \cig_{min}$.\\
Here the indicated random variables $(Y_1,Y_2,X)$
are constructed having the measure $G(0,Q_1)$
with the dimensions $p_1,  p_2,  n \in \zpos$ respectively.
\end{problem}
Note that for the weak stochastic realization Problem~\ref{prob_real}
one is asked to construct a measure that carries the random variables $(Y_1, Y_2, X)$.

\begin{remark}
\label{rem_WSR}
 For the calculation of rates that lie on the  Gray-Wyner lossy rate region, the concept of weak stochastic realization is required.
 In such applications, Proposition~\ref{prop:cigrvs} is a characterization of conditional independence, while one is asked to solve the weak stochastic realization Problem~\ref{prob_real}, subject to the average distortion constraints.  
\end{remark}

Further to Remark~\ref{rem_WSR},    the next 
definition and proposition are 
about the weak Gaussian stochastic realization of a tuple of
jointly Gaussian multivariate random variables and its weak stochastic realization.
\begin{definition}\label{def:wgsrgrvs}
Minimality of weak stochastic realization of Gaussian random variables.\\
	Consider a Gaussian measure ${\bf P}_0=G_0(0,Q_{(y_1,Y_2)})$ with zero mean values
	for a tuple $(Y_1, Y_2)$ of random variables on the product space
$(\mathbb{R}^{p_1} \times \mathbb{R}^{p_1}, 
  B(\mathbb{R}^{p_1}) \otimes B(\mathbb{R}^{p_2})$ 
for $p_1, ~ p_2 \in \mathbb{Z}_+$ with
	\begin{eqnarray*}
		Q_{(Y_1,Y_2)}
		& = & \left(
	              \begin{array}{ll}
			      Q_{Y_1}       & Q_{Y_1,Y_2} \\
			      Q_{Y_1,Y_2}^T & Q_{Y_2} 
		      \end{array}
		      \right), ~~
		      Q_{Y_1} > 0, ~ Q_{Y_2} > 0.
	\end{eqnarray*}
\begin{itemize}
\item[(a)]
A {\em weak Gaussian stochastic realization} of the Gaussian measure $G_0(0,Q_{(y_1,Y_2)})$
is defined
to be a Gaussian measure ${\bf P}_1=G_1$ 
if there exists an integer $n \in \mathbb{Z}_+$
such that the Gaussian measure $G_1$ is defined on the space
$(\mathbb{R}^{p_1} \times \mathbb{R}^{p_1} \times \mathbb{R}^n, 
   B(\mathbb{R}^{p_1}) \otimes B(\mathbb{R}^{p_2}) \otimes B(\mathbb{R}^n))$,
associated with random variables in the three spaces denoted respectively 
by $Y_1$, $Y_2$, and $X$,
and such that:\\
(1) $G_1|_{\mathbb{R}^{p_1} \times \mathbb{R}^{p_2}} =G_0(0,Q_{(Y_1,Y_2)}) $; \\
(2) $Q_X > 0$; and\\
(3) conditional independence holds:
${\bf P}_{Y_1,Y_2|X} = {\bf P}_{Y_1|X} {\bf P}_{Y_2|X},$ where these are Gaussian  measures, 
 with means which are linear functions of the random variable $X$
and deterministic variance matrices, i.e., similar to (\ref{ccha_1})-(\ref{cchar_4}) with appropriate changes.
\item[(b)]
The weak Gaussian stochastic realization is called {\em minimal}
if the dimension $n$ of the random variable $X$ 
is the smallest possible over all weak Gaussian stochastic realizations as defined in (a).
\item[(c)]
A {\em Gaussian random variable representation} 
of a weak Gaussian stochastic realization  $G_1$
is defined as a triple of random variables
satisfying the following relations,
\begin{align}
   & (Y_1, Y_2, X, V_1, V_2), ~
	p_{V_1}, ~ p_{V_2} \in \mathbb{Z}_+, ~
	p_{V_1} \geq p_1, ~ p_{V_2} \geq p_2, \label{wgsr_1} \\
  & Y_1: \Omega \rightarrow \mathbb{R}^{p_1}, ~
        Y_2: \Omega \rightarrow \mathbb{R}^{p_2}, ~
        V_1: \Omega \rightarrow \mathbb{R}^{p_{v_1}}, ~
        V_2: \Omega \rightarrow \mathbb{R}^{p_{v_2}}, ~
        X: \Omega \rightarrow \mathbb{R}^n,  \nonumber \\
   & (V_1, V_2, X) \in G, ~ 
	\mbox{and these are zero mean independent random variables},  
          \nonumber \\
   & Q_{V_1} > 0, ~ 
        Q_{V_2} > 0, ~
        Q_X > 0;  \nonumber \\
   & C_1 \in \mathbb{R}^{p_1 \times n}, ~
        C_2 \in \mathbb{R}^{p_2 \times n}, ~
        N_1 \in \mathbb{R}^{p_1 \times p_{V_1}}, ~
        N_2 \in \mathbb{R}^{p_2 \times p_{V_2}}, \label{wgsr_2} \\
     & Y_1
   =  C_1 ~ X + N_1 ~ V_1, ~ \label{eq:yoneinxvone} \\
    &  Y_2
   =  C_2 ~ X + N_2 ~ V_2, ~ \label{eq:ytwoinxvtwo} \\
    &  Q_{Y_1} 
   =  C_1 Q_X C_1^T + N_1Q_{V_1}N_1^T, \\
   &   Q_{Y_2} 
   =  C_2 Q_X C_2^T + N_2Q_{V_2}N_2^T, \\
   &   Q_{Y_1,Y_2} 
   =  C_1 Q_X C_2^T, \label{eq:q12factorization} \\ 
   &   G_0(0,Q_{(Y_1,Y_2)})
   =  G_1|_{\mathbb{R}^{p_1} \times \mathbb{R}^{p_2}}. \nonumber 
\end{align}
From the assumptions then follows that $(Y_1,Y_2)$ are Gaussian random variables, 
hence the last equality makes sense.
\item[(d)]
A {\em minimal Gaussian random variable representation}
of a weak Gaussian stochastic realization 
is defined as a triple of random variables
as in (c) except that in addition it is required that,
\begin{eqnarray}
	&   & \rank(C_1) = n = \rank (C_2). \label{eq:observabilityconditions}
\end{eqnarray}
\end{itemize}
\end{definition}

\ \

%

The next proposition shows equivalence of  weak Gaussian stochastic realizations of  Definition~\ref{def:wgsrgrvs}.(a), (b) to Definition~\ref{def:wgsrgrvs}.(c), (d), respectively.\\

\begin{proposition}\label{prop:equivalentgmeasure2grepresentation}
Consider the setting of Definition~\ref{def:wgsrgrvs}
with $(Y_1, Y_2 ) \in G(0,Q_{(Y_1,Y_2)})$  with the representation
of  (\ref{eq:yoneinxvone}), (\ref{eq:ytwoinxvtwo}).
\begin{itemize}
\item[(a)]
A weak Gaussian stochastic realization 
in terms of a measure ${\bf P}_1=G_1$ as defined 
in Definition~\ref{def:wgsrgrvs}.(a)
is equivalent with a Gaussian random variable representation 
of Definition~\ref{def:wgsrgrvs}.(c).
\item[(b)]
The minimal weak Gaussian stochastic realization 
of Definition~\ref{def:wgsrgrvs}.(b)
is equivalent to 
a minimal weak Gaussian random variable representation 
of Definition~\ref{def:wgsrgrvs}.(d).
\end{itemize}
\end{proposition}
{\bf Proof}  See Appendix~\ref{proof_app_prop:equivalentgmeasure2grepresentation}. 
\hfill$\square$.
\section{Wyner's Information Common Information}
\label{sect:wciinformation}

\subsection{Solution Procedure}
\label{sect:sol-pro}

From the Definition~\ref{wynercommoninforvs} 
follows directly the procedure to compute  Wyner's common information (information definition) $C(Y_1, Y_2)$.

\begin{procedure} Computation of Wyner's common information $C(Y_1,Y_2)$.
\label{procedure_1}
\begin{enumerate}
\item 
Determine a parametrization of all random variables $W$
which make the two components of the tuple $(Y_1, Y_2)$
conditionally independent, 
thus so that $(F^{Y_1}, F^{Y_2} | F^W) \in \cig$, according to the weak stochastic realization of Problem~\ref{prob_real}.(a).

\item Solve Wyner's common information  problem (information definition)
\begin{align}
  C(Y_1, Y_2) = \inf_{(Y_1, Y_2, W) \in G, ~ (Y_1,Y_2|W) \in {\rm CIG}} I(Y_1, Y_2; W)
  \end{align}
over the set of all measures determined in Step 1.
\end{enumerate}
\end{procedure}

\par In Procedure~\ref{procedure_1}, use is made of the concepts of Sections~\ref{ap:grvs_new}, \ref{ap:ci}, and of a  result that is found in a paper co-authored by the second-named author  
\cite{putten:schuppen:1983}.


\label{ap:grvs}

\subsection{Preliminary Characterization of Conditional Independence}
\label{sect:ci}
Consider the pre-encoder  
of Fig. \ref{fig:gw-real}. The two signals $Y_1,Y_2$ are to be reproduced at the two decoders by $\hat{Y}_1, \hat{Y}_2$
subject to the square-error distortion functions. 
According to
Gray and Wyner, the characterization of lossy rate region is described by a single coding scheme that uses
the auxiliary random variable $W$, which is common to both $Y_1,Y_2$. Below this engineering interpretation
is further detailed in terms of the mathematical framework of weak stochastic realization such that $(F^{Y_1}, F^{Y_2} | F^W) \in \cig$.

\par 
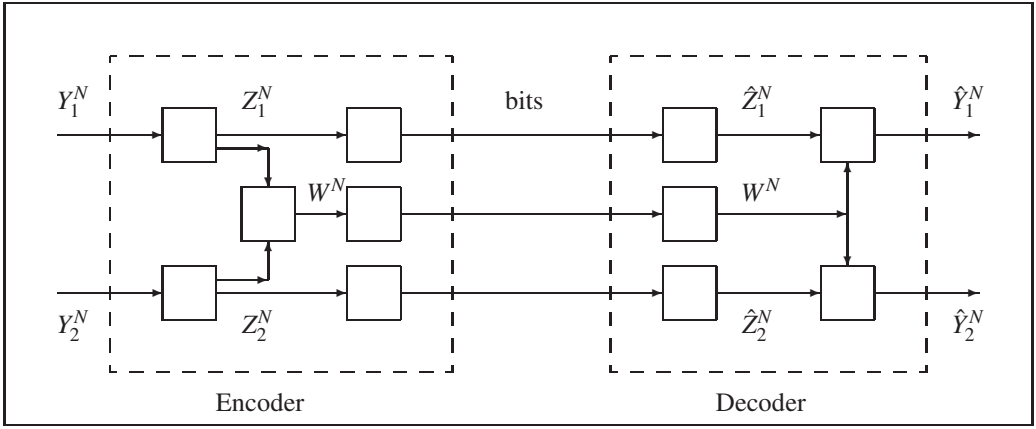
\begin{figure}
\setlength{\unitlength}{0.35cm}
\begin{center}
\begin{picture}(39,16)(0,0)
\put(0,0){\framebox(39,16)}
\put(6,4){\framebox(2,2)}
\put(6,10){\framebox(2,2)}
\put(9,7){\framebox(2,2)}
\put(13,4){\framebox(2,2)}
\put(13,7){\framebox(2,2)}
\put(13,10){\framebox(2,2)}
\put(25,4){\framebox(2,2)}
\put(25,7){\framebox(2,2)}
\put(25,10){\framebox(2,2)}
\put(31,4){\framebox(2,2)}
\put(31,10){\framebox(2,2)}
\put(4,2){\dashbox{.5}(13,12)}
\put(23,2){\dashbox{.5}(12,12)}
\put(2,5){\vector(1,0){4}}
\put(2,11){\vector(1,0){4}}
\put(8,11){\vector(1,0){5}}
\put(8,5){\vector(1,0){5}}
\put(8,10.5){\vector(1,0){2}}
\put(8,5.5){\vector(1,0){2}}
\put(10,10.5){\vector(0,-1){1.5}}
\put(10,5.5){\vector(0,1){1.5}}
\put(11,8){\vector(1,0){2}}
\put(15,5){\vector(1,0){10}}
\put(15,8){\vector(1,0){10}}
\put(15,11){\vector(1,0){10}}
\put(27,5){\vector(1,0){4}}
\put(27,8){\vector(1,0){5}}
\put(27,11){\vector(1,0){4}}
\put(32,8){\vector(0,-1){2}}
\put(32,8){\vector(0,1){2}}
\put(33,5){\vector(1,0){4}}
\put(33,11){\vector(1,0){4}}
\put(2,3.5){$Y_2^N$}
\put(2,12){$Y_1^N$}
\put(9,3.5){$Z_2^N$}
\put(9,12){$Z_1^N$}
\put(11.5,8.5){$W^N$}
\put(36,3.5){$\hat{Y}_2^N$}
\put(36,12){$\hat{Y}_1^N$}
\put(28,3.5){$\hat{Z}_2^N$}
\put(28,12){$\hat{Z}_1^N$}
\put(28,8.5){$W^N$}
\put(19,12){bits}
\put(8,0.5){Encoder}
\put(27,0.5){Decoder}
\end{picture}
\end{center}
\caption{Weak stochastic realization of $(Y_{1,i},Y_{2,i}) \sim {\bf P}_{Y_1,Y_2}, i = 1, \ldots, N$ and $(\hat{Y}_{1,i},\hat{Y}_{2,i}),  i = 1, \ldots, N$ 
at the encoder and decoder with respect to the common and private random variables $(W^N, Z_1^N, Z_2^N), (W^N, \hat{Z}_1^N, \hat{Z}_2^N)$.}\label{fig:gw-real}
\end{figure}

\begin{definition}\label{def:cituple}
The model for a {\em triple of Gaussian random variables}.\\
Consider a tuple of Gaussian random variables specified by 
$Y = (Y_1, Y_2) \in G(0,Q_{(Y_1,Y_2)})$ with
$Y_i: \Omega \rightarrow \mathbb{R}^{p_i}$ for $i = 1, 2$. Take a jointly Gaussian measure $G(0,Q_{(Y_1,Y_2,W)})$ for the triple $(Y_1, Y_2, W)$, $W: \Omega \rightarrow \mathbb{R}^n$,
$W \in G(0,Q_W)$,  such that the marginal measure on $(Y_1, Y_2)$
equals the considered measure, and      the  conditional independence holds,
$(F^{Y_1}, F^{Y_2} | F^W) \in \cig$.
Denote the joint measure by $(Y_1, Y_2, W) \in G(0,Q_{(Y_1,Y_2,W)})$ with
\[
  Q_{(Y_1,Y_2,W)} =
  \left(
  \begin{array}{lll}
    Q_{Y_1}   & Q_{Y_1,Y_2}   & Q_{Y_1,W} \\
    Q_{Y_1,Y_2}^T & Q_{Y_2}   & Q_{Y_2,W} \\
    Q_{Y_1,W}^T   & Q_{Y_2,W}^T   & Q_W
  \end{array}
  \right).
\]
In the two following sections it will be shown how to construct
such a random variable $W$ in a number of cases, starting with the simplest case.
\end{definition}

The algorithm that generates the joint measure by $(Y_1, Y_2, W) \in G(0,Q_{(Y_1,Y_2,W)})$   via weak stochastic realization
is given below.

\begin{algorithm}\label{alg:grvtuplecommunication}
Consider the model of a tuple of Gaussian random variables
of Def. \ref{def:cituple}.
Assume that 
$Q_W > 0$. 
\begin{enumerate}
\item
At the encoder, compute first the variables,
\begin{align}
      Z_1
   = & Y_1 - {\bf E}[Y_1 | F^W] = Y_1 - Q_{Y_1,W} Q_W^{-1} W, \\
      Z_2
  = & Y_2 - {\bf E}[Y_2 | F^W ] = Y_2 - Q_{Y_2,W} Q_W^{-1} W; \label{eq:y2z}
\end{align}
then the triple $(Z_1, Z_2, W)$ of jointly Gaussian random variables are independent. 
\item The tuple of random variables $(Y_1, Y_2)$ are represented according to,
\begin{align}
      Y_1
   = Q_{Y_1,W} Q_W^{-1} W +Z_1,  \ \
        Y_2
        = Q_{Y_2,W} Q_W^{-1} W +Z_2. \label{eq:hatyequalsy}
\end{align}
\end{enumerate}
\end{algorithm}

\ \

The validity of the statements of the algorithm follow from  the next proposition.

\begin{proposition}\label{prop:grvtuplecomm}
Consider the model of a tuple of Gaussian random variables
of Def. \ref{def:cituple}.
\begin{itemize}
\item[(a)]
At the encoder,
the conditional expectations are correct and
the definitions of $Z_1$ and of $Z_2$ are well defined.
\item[(b)] 
The three random variables
$(Z_1, Z_2, W)$ are independent. Consequently, 
the three sequences  \\
$(W^N, Z_1^N, Z_2^N)$, and messages generated by the Gray-Wyner encoder, \\
$f^{(E)}(Y_1^N, Y_2^N)= \overline{f}^{(E)}(W^N, Z_1^N, Z_2^N)=(S_0,S_1,S_2)$  are independent.
\end{itemize}
\end{proposition}
{\bf Proof}  This is a specific application of Proposition~\ref{prop:equivalentgmeasure2grepresentation}.
\hfill$\square$
%


\ \

For the definition of $C(Y_1, Y_2)$, use is made of the construction of the
actual family of measures     such that $(Y_1, Y_2| W) \in \cig$ holds, and the weak strochastic realization. These  are  presented
in Theorem~\ref{them_putten:schuppen:1983} and Corollary~\ref{cor:commoninfogrvcorrelated}.

\subsection{The Expression of Wyner's Information Common Information of Multivariate Gaussian Random Variables}
\label{sec:commoninfogrv}
%

First, the calculation of the formulae for $C(Y_1,Y_2)$ is stated in the form of an algorithm. The algorithm
makes use of the canonical variable form of Section~\ref{ap:grvs_new}.

\begin{algorithm}\label{alg:wcommoninfogrvs}
Consider a tuple of finite-dimensional Gaussian random variables
$(Y_1, Y_2) \in G(0,Q_{(Y_1,Y_2)})$
as defined in Section~\ref{ap:grvs}.
\begin{enumerate}
\item
Compute the canonical variable form of the tuple
of Gaussian random variables
according to Algorithm \ref{alg:canonicalvariabledecomposition}.
This yields the indices
$p_{11} = p_{21}$, $p_{12} = p_{22}$, $p_{13}$, $p_{23}$, and 
$n = p_{11} + p_{12} = p_{21} + p_{22}$
and the diagonal matrix $D$ with canonical singular values
$d_i \in (0,1)$ for $i = 1, \ldots, n$.
\item
Compute  the information quantity $C(Y_1, Y_2)$ according to the formula,
\begin{align}
      C(Y_1,Y_2)
   = & \left\{
        \begin{array}{lll}
            0, 
          &   \mbox{if} & 
              0 = p_{11} = p_{12} = p_{21} = p_{22}, ~
              p_{13} > 0, ~
              p_{23} > 0, \\
            \frac{1}{2} \sum_{i=1}^n 
            \ln
            \left(
              \frac{1+d_i}{1-d_i}
            \right), 
          &   \mbox{if} & 
              0 = p_{11} = p_{21}, ~
              p_{12} = p_{22} > 0, ~
              p_{13} \geq 0, ~ p_{23} \geq 0, \\
            \infty,  
          &   \mbox{if} & 
              p_{11} = p_{21} > 0, ~
              p_{12} = p_{22} \geq 0, ~
              p_{23} \geq 0, ~ p_{23} \geq 0. 
        \end{array}
        \right. \label{eq:wcigeneral}
\end{align}
\end{enumerate}
\end{algorithm}
\begin{theorem}\label{th:commoninfogrvs}
Consider a tuple of Gaussian random variables.
Algorithm \ref{alg:wcommoninfogrvs} is correct and 
produces the quantity $C(Y_1,Y_2)$ defined by (\ref{w_ic}) of   the tuple $(Y_1, Y_2)$.
\end{theorem}
{\bf Proof} 
The proof of the theorem is provided as follows: (i) in Theorem~\ref{th:commoninfogrvcorrelated_new}, for the special case of a tuple $(Y_1,Y_2)$ of Gaussian random variables
in the canonical variable form of  Def. \ref{def:grvcommoncorrelatedprivateinfo}, with the restriction to the  correlated parts of these random variables, and (ii) in 
Section \ref{subsec:wciarbitrarygrvs},  Theorem~\ref{thm_cia} (where the weak realization is also given), without the restriction to the correlated parts of these random variables, and 
after several special cases of the result have been derived, i.e., when $p_{11}=p_{21}=0$, Section~\ref{sect:main}.
\hfill$\square$

\ \

\par
The computation of the information quantity $C(Y_1,Y_2)$ 
is structured by the concepts of
 identical, correlated, and private components 
of the two vectors considered; see Section~\ref{ap:grvs} for the definitions of these concepts. The quantity $C(Y_1,Y_2)$\\
(i) in the first case of equation 
(\ref{eq:wcigeneral})
covers the case in which the random variables  $(Y_1,  Y_2)$
are independent random variables
and there are neither identical nor correlated components, \\
(ii) in the second case of equation (\ref{eq:wcigeneral})
covers the case in which there is no identical component,
but there are nontrivial correlated components,
and there may be independent components, and\\
(iii) in  the last case of equation (\ref{eq:wcigeneral})
covers the case when there is a nontrivial identical component and, possibly,
correlated and independent components.

\par
Successively the reader will be shown the
computation of  information quantity $C(Y_1;Y_2)$ for:\\
(1) the correlated components;\\
(2) the private components; and\\
(3) the identical information components.\\
The general case is then a combination of the above three special cases.
%
%
%
%

\subsection{Proof of Wyner's Information Common Information of Correlated Gaussian Vectors}
\label{sect:main}
This section is devoted to the application of Method 1 (a) and Method 2 to calculate 
$C(Y_1, Y_2)$, and to present the weak stochastic realization of $(Y_1,Y_2,W^*)$ that achieve this.
The identification of the random variable $W^*$, such that $(Y_1,Y_2, W^*)$ achieves $C(Y_1,Y_2)$, is given in the next theorem. The   theorem utilizes  the parametrization of  the family of Gaussian probability distributions
\begin{align}
{\cal P}^{CIG} =& \Big\{ {\bf P}_{Y_1, Y_2, W}(y_1,y_2,w)\Big| \ \ {\bf P}_{Y_1, Y_2|W}(y_1,y_2|w)={\bf P}_{Y_1|W}(y_1|w) {\bf P}_{Y_2|W}(y_2|w), \nonumber \\
&     {\bf P}_{Y_1, Y_2, W}(y_1,y_2,\infty)={\bf P}_{Y_1, Y_2}(y_1, y_2), \ \ (Y_1, Y_2,W) \ \ \mbox{jointly Gaussian}   \Big\}.
\end{align}
A subset of the set $P^{CIG}$ is the set of distributions $P_{min}^{CIG}$, with the additional constraint  that the dimension of the random variable $W$
is minimal while all other conditions hold, defined by 
\begin{align}
{\cal P}_{min}^{CIG} = \Big\{ {\bf P}_{Y_1, Y_2, W}(y_1,y_2,w) \in  {\cal P}^{CIG} \Big| \ \  (Y_1,Y_2|W) \in \cig_{min}  \Big\}.
\end{align}
Use is made of the canonical variable form
as defined in Def.~\ref{def:grvcommoncorrelatedprivateinfo}, since $C(Y_1,Y_2)$ is invariant with respect to the transformation of initial random variables $(Y_1, Y_2)$ to their canonical variable form. Further, use is made of  minimal Gaussian conditional independence  as defined in Def.~\ref{def_minci}, and of related results of Section~\ref{ap:ci}.

\par  Def.~\ref{def_minci} is needed, because,  
for the computation of $C(Y_1, Y_2)$, the 
attention can be  restricted to those state variables $W$
which are of miminal dimension.  

\par The parametrization of  the family of Gaussian probability distributions
${\cal P}^{CIG}$ and ${\cal P}_{min}^{CIG}$ require the solution of the weak stochastic realization problem of Gaussian random variables defined by Problem~\ref{prob_real}. This problem is solved in  \cite[Th. 4.2]{putten:schuppen:1983}. For the readers convenience it is stated below. 

\begin{theorem}\cite[Theorem 4.2]{putten:schuppen:1983}
\label{them_putten:schuppen:1983}
Consider a tuple $(Y_1,Y_2)$ of Gaussian random variables
in the canonical variable form of  Def. \ref{def:grvcommoncorrelatedprivateinfo}.
Restrict attention to the correlated parts of these random variables.
Thus, the random variables $Y_1, ~ Y_2$ have the same dimension 
$n = p_1 = p_2$,
and their covariance matrix $D \in \mathbb{R}^{n \times n}$
is a nonsingular diagonal matrix with on the diagonal
ordered real-numbers in the interval $(0,1)$.
Hence,
\begin{align}
    & (Y_1,Y_2) \in G(0,Q_{(Y_1,Y_2)}) = {\bf P}_0, \ \
         Y_1,Y_2: \Omega \rightarrow \rn, \ \ n \in \zpos,   \label{ci_par_1}   \\
     & Q_{(y_1,y_2)}
   = \left(
        \begin{array}{ll}
          I & D \\
          D & I
        \end{array}
        \right),   \label{ci_par_2}  \\
      &D
 = \diag (d_1, d_2, \ldots, d_n) \in \rnn, ~~
        1 > d_1 \geq d_2 \geq \ldots \geq d_n > 0. 
        \label{ci_par_3}
\end{align}
That is, $p_{11}=p_{21}=0, p_{13}=p_{23}=0.$
\begin{itemize}
\item[(a)] There exists a probability measure ${\bf P}_1$, and  a triple of Gaussian random variables
$Y_1,Y_2,~~ W: \Omega \rightarrow \rn$ defined on it, such that (i) 
${\bf P}_1|_{(Y_1,Y_2)} = {\bf P}_0$ and (ii)
$(F^{Y_1},F^{Y_2}|F^W) \in \cigmin$.

\item[(b)] There exist a family of Gaussian measures denoted by ${\bf P_{ci}}\subseteq {\cal P}_{min}^{CIG}$, that satisfy (i) and (ii) of (a), and moreover this family is  parametrized by the matrices and sets, as follows.
\begin{align}
   & G(0,Q_s(Q_W)), ~
       Q_W \in {\bf Q_W},  \label{eq:gtriple} \\
    & Q_s
   =  Q_s(Q_W) =
        \left(
        \begin{array}{lll}
          I       & D           & D^{1/2} \\
          D       & I           & D^{1/2} Q_W \\
          D^{1/2} & Q_W D^{1/2} & Q_W 
        \end{array}
        \right),\label{eq:gtriple_1} \\
     & {\bf Q_W}
   =  \Big\{ Q_W \in \rnn \Big|\; Q_W = Q_W^T, ~
                          0 < D \leq Q_W \leq D^{-1} 
        \big\}, \label{eq:pci_1} \\
     &   {\bf P_{ci}}
  =  \Big\{ G(0,Q_s(Q_W)) ~ \mbox{on} ~  
           (\mathbb{R}^{3n}, B(\mathbb{R}^{3n})) \Big|\;  Q_W \in {\bf Q_W} 
        \Big\} \subseteq {\cal P}_{min}^{CIG}. \label{eq:pci}
\end{align}
Further, for any measure $P_1 \in {\cal P}_{min}^{CIG}$
there exists a triple of state transformation of the form
$(Y_1, Y_2, W) \mapsto (S_1 Y_1, S_2 Y_2, S_W W)$
for nonsingular square matrices $S_1, ~ S_2, S_W$
such that the corresponding measure of the three transformed
variables belongs to ${\bf P_{ci}}$.
\end{itemize}
\end{theorem}
%

The application of Theorem~\ref{them_putten:schuppen:1983}  is discussed in the next remark, in the context of parametrizing any rate-triple on the Gray-Wyner lossy rate region $(R_0,R_1,R_2) \in {\cal R}_{GW}(\Delta_1,\Delta_2)$, that lies on the Pangloss plane.
in the context of parametrizing the rate region of the Gaussian MACs of Section~\ref{sect:lite}.

\begin{remark} Applications of Theorem~\ref{them_putten:schuppen:1983}.\\
(a) Theorem~\ref{them_putten:schuppen:1983} is a parametrization of  the familiy of Gaussian measures ${\bf P_{ci}}\subseteq {\cal P}_{min}^{CIG}$ by the entries of the covariance matrix $Q_W$. Hence, it is at most,  an $n(n+1)/2-$dimensional parametrization. \\
(b) It is expected that  the achievable rate region ${\cal R}_{GW}(\Delta_1, \Delta_2)={\cal R}_{GW}^{*}(\Delta_1, \Delta_2)$ is generated from distributions  ${\bf P_{ci}}\subseteq {\cal P}_{min}^{CIG}\subseteq {\cal P}$. 
\end{remark}

The next corollary  is preliminary to a subsequent theorem (i.e., Theorem~\ref{th:commoninfogrvcorrelated_new}), 
that shows $C(Y_1,Y_2)$ is achieved by the distribution ${\bf P}_{Y_1, Y_2, W^*}\in {\bf P_{ci}}\subseteq {\cal P}_{min}^{CIG}$, corresponding to  $W^* \in G(0,I)$, i.e.,  with covariance the identity matrix. The next corollary gives the realization of $(Y_1, Y_2)$, expressed in terms of an arbitrary Gaussian random variable  $W\in G(0, Q_W)$.

\begin{corollary}\label{cor:commoninfogrvcorrelated}
Consider a tuple $(Y_1,Y_2)$ of Gaussian random variables
in the canonical variable form of  Def. \ref{def:grvcommoncorrelatedprivateinfo}.
Restrict attention to the correlated parts of these random variables, as defined in Theorem~\ref{them_putten:schuppen:1983},  by (\ref{ci_par_1})-(\ref{ci_par_3}).\\
Then a realization of the random variables $(Y_1, Y_2)$  which induce the family of  measures  ${\bf P_{ci}}\subseteq {\cal P}_{min}^{CIG}$, defined by (\ref{eq:gtriple})-(\ref{eq:pci}), is 
\begin{eqnarray}
      Y_1 
  & = & Q_{Y_1, W}Q_W^{-1}W+Z_1   \label{rep_g_s1}\\
      Q_{Y_1, W}
  & = & D^{1/2}, \ \   Z_1 \in G(0,(I-D^{1/2}Q_W^{-1}D^{1/2}  )), 
           \\
      Y_2 
  & = & Q_{Y_2,W} Q_W^{-1} W + Z_2 \label{rep_g_s2}\\
      Q_{Y_2, W}
  & = & D^{1/2}Q_W, \ \   Z_2 \in G(0,(I-D^{1/2}Q_WD^{1/2}  )),   \\
  &   & (Z_1,Z_2,W), ~ \mbox{are independent.} \label{rep_g_s3} 
\end{eqnarray}  
Further, the  mutual information $I(Y_1,Y_2; W)$ is given by
\begin{align}
 I(Y_1,Y_2;W)
   = & H(Y_1,Y_2) -  H(Y_1|W) - H(Y_2|W)  \\
   = & \frac{1}{2} \sum_{i=1}^n \ln (1-d_i^2) 
        - \frac{1}{2} \ln (\det ( [I - D^{1/2} Q_W^{-1} D^{1/2} ]
                                  [ I - D^{1/2} Q_W D^{1/2} ] 
                                )
                          )
\end{align}
and it is parametrized by $Q_W \in {\bf Q_W}$, where ${\bf Q_W}$ is  defined by the set of equation  (\ref{eq:pci_1}).
\end{corollary}
{\bf Proof} The correctness of the  realization is due to  Proposition~\ref{prop:cigrvs} and Theorem~\ref{them_putten:schuppen:1983}. 
 The calculation of mutual information follows from the realization.  
 \hfill$\square$

\begin{remark} 
\label{rem_lit_2}
Further to Remark~\ref{rem_lit_1} it should be apparent that the proof  of the formula $C(Y_1,Y_2;W)$ in 
 \cite{satpathy:cuff:2012} and \cite{veld-gastpar:2016} did not account for the optimization over the set of distributions ${\bf P}_{Y_1, Y_2, W}$ achieving conditional independence ${\bf P}_{Y_1, Y_2|W}={\bf P}_{Y_1|W} {\bf P}_{ Y_2|W}$, i.e., the optimization over the    parametrized family of Gaussian measures of   Theorem~\ref{them_putten:schuppen:1983} is missing. That is, both \cite{satpathy:cuff:2012} and \cite{satpathy:cuff:2012,veld-gastpar:2016} did not prove that the minimum of $I(Y_1,Y_2;W)$ over     ${\bf P}_{Y_1, Y_2|W} ={\bf P}_{Y_1|W} {\bf P}_{ Y_2|W}$ is achieved by  $W=W^*\in G(0,I))$. Rather, the authors assumed  $W=W^*\in G(0,I))$ and calculated  $I(Y_1,Y_2;W^*)$ (see \cite{satpathy:cuff:2012}, Sections~IV.A, B). To justify, the above, note that, for a triple of jointly Gaussian random variables each having zero mean
value
one can apply a scaling such that the variance of each variable is the
unit matrix.
There are then three off-diagonal blocks in the variance matrix of the
triple
which three blocks are in general fully filled matrices.
One can then apply two transformations by orthogonal matrices to the
first two components
to make the covariance of the first two vectors a diagonal matrix as is
done in the paper.
But it is then impossible to apply any further changes by orthogonal
matrices
so that the two remaining off-diagonal blocks also become diagonal matrices
while the unit matrices of the variances of the three components is
maintained.
Therefore one has to determine a parametrization of the variance matrix
of a triple of jointly Gaussian random variables as is done in our paper.
\end{remark}

In the next theorem the family of  measures  ${\bf P_{ci}}\subseteq {\cal P}_{min}^{CIG}$, defined by (\ref{eq:gtriple})-(\ref{eq:pci}), which leads to realization of  $(Y_1, Y_2)$, given in Corollary~\ref{cor:commoninfogrvcorrelated}, is 
 ordered for the determination of a single joint distribution ${\bf P}_{Y_1, Y_2, W^*} \in {\bf P_{ci}}\subseteq {\cal P}_{min}^{CIG}$, which achieves $C(Y_1, Y_2)$. This leads to the realization of  $(Y_1, Y_2)$ expressed in terms of $W^*$ and vectors of independent Gaussian random variables $(Z_1, Z_2)$, one for each realization, each having  independent components.   

\begin{theorem}\label{th:commoninfogrvcorrelated_new}
Consider a tuple $(Y_1,Y_2)$ of Gaussian random variables
in the canonical variable form of  Def. \ref{def:grvcommoncorrelatedprivateinfo}.
Restrict attention to the correlated parts of these random variables, as defined in Theorem~\ref{them_putten:schuppen:1983},  defined by (\ref{ci_par_1})-(\ref{ci_par_3}). \\ 
The following hold.
\begin{itemize}
\item[(a)]
The  information quantity $C(Y_1, Y_2)$ is given by 
\begin{align}
     C(Y_1,Y_2)
   =  \frac{1}{2} \sum_{i=1}^n 
        \ln
        \left(
        \frac{1+d_i}{1-d_i}
        \right) 
        =
        \frac{1}{2} \sum_{i=1}^n 
        \ln
        \left(
          1 + \frac{2d_i}{1-d_i}
        \right) \in (0,\infty).   \label{lci_1}
\end{align}
\item[(b)]
The realizations of the random variables $(Y_1, Y_2, W^*)$ that achieve $C(Y_1,Y_2)$ are represented by 
\begin{align}
  & V: \Omega \rightarrow \mathbb{R}^{n}, ~
        V \in G(0,I), ~~ \mbox{the vector $V$ has independent components,} \nonumber \\
 & F^V, ~ F^{Y_1} \vee F^{Y_2}, \ \
         \mbox{are independent $\sigma$-algebras,} \nonumber \\
   &   L_1
   =  L_2 = D^{1/2} (I+D)^{-1} \in \rnn, \label{eq:l1l2} \\
    &  L_3
   =  (I-D)^{1/2} (I+D)^{-1/2} \in \rnn, \ \
        L_1, ~ L_2, ~ L_3, ~ 
        \mbox{are diagonal matrices,} 
        \label{eq:l3} \\
     & W^*
   =  L_1 Y_1 + L_2 Y_2 + L_3 V, \ \
        W^*: \Omega \rightarrow \mathbb{R}^n, \\
     & Z_1 
 =  Y_1 - D^{1/2} W^*, \ \
        Z_1: \Omega \rightarrow \rn, \\
     & Z_2 
   =  Y_2 - D^{1/2} W^*, \ \
        Z_2: \Omega \rightarrow \rn.
\end{align}
Then
\begin{align}
   & Z_1 \in G(0,(I-D)), \ \
        Z_2 \in G(0,(I-D)), \ \
        W^* \in G(0,I); \label{rep_s_1}    \\
    & (Z_1,Z_2,W^*), ~ \mbox{are independent and} \label{rep_s_2} \\
    & Y_1
  =  D^{1/2} W^* + Z_1, \ \
        Y_2 =D^{1/2} W^* + Z_2 \label{rep_s_3}
\end{align}
hence the variables $(Y_1, Y_2, W^*)$ induce a distribution ${\bf P}_{Y_1, Y_2, W^*}\in {\bf P_{ci}}\subseteq {\cal P}_{min}^{CIG}$.    
Note that, in addition,
each of the random variables $Z_1$, $Z_2$, and $W^*$
has independent components.
\item[(c)]  The variables $(Y_1, Y_2, W^*)$ defined in (b)  induce a distribution ${\bf P}_{Y_1, Y_2, W^*}\in {\bf P_{ci}}\subseteq {\cal P}_{min}^{CIG}$   which achieves $C(Y_1,Y_2)$,  
\begin{eqnarray}
     C(Y_1,Y_2)
   =  I(Y_1, Y_2;W^*).
\end{eqnarray}
\end{itemize}
\end{theorem}
{\bf Proof} Since mutual information $I(Y_1,Y_2; W)$ is invariant with respect to nonsingular transformations,  then by Theorem~\ref{thm-exi-cvf}, (b),  it suffices to consider the canonical variable form of  Def. \ref{def:grvcommoncorrelatedprivateinfo}, and to construct a measure that carries a triple of jointly Gaussian random variables $Y_1,Y_2,~~ W: \Omega \rightarrow \rn$ such that 
$(F^{Y_1},F^{Y_2}|F^W) \in \cig$.\\
(a) 
(1) Take a probability measure ${\bf P}_1$ such that
there exists a triple of Gaussian random variables
$Y_1,Y_2,~~ W: \Omega \rightarrow \rn$ with
${\bf P}_1|_{(y_1,y_2)} = {\bf P}_0$ and
$(F^{Y_1},F^{Y_2}|F^W) \in \cig$.
It will first be proven that attention can be restricted
to those state random variables $W$
of which the dimension equals $n = p_{12} = p_{22}$.
\par
Suppose that there exists a state random variable
$W: \Omega \rightarrow \mathbb{R}^{n_1}$ such that
$(F^{Y_1},F^{Y_2}|F^W) \in \cig$ and $n_1 > n$.
Hence $W$ does not make $(Y_1,Y_2)$ minimally conditionally independent.
Construct a minimal vector which makes the tuple 
minimally conditionally independent according to the procedure
of \cite[Proposition 3.5]{putten:schuppen:1983}.
Thus,
\begin{align*}
      W_1
  = & {\bf E}[ Y_1 | F^W] = L_{11} W,  \ \ 
        L_{11} \in \mathbb{R}^{n \times n_1}, \\
      W_2
  = & {\bf E} [Y_2 | F^{W_1}] = L_{12} W_1, \ \ 
        L_{12} \in \mathbb{R}^{n \times n}.
\end{align*}
Then $(F^{Y_1}, F^{Y_2} | F^{W_2}) \in \cigmin$
and the dimension of $W_2$ is $n = p_{12} = p_{22}$.
Determine a linear transformation of $W_2$ by a matrix
$L_{15} \in \mathbb{R}^{n \times n}$ such that,
\begin{align*}
      W_3
   = L_{15} W_2 = L_{15} L_{12} L_{11} W = L_{13} W,  \ \
        L_{13} = L_{15} L_{12} L_{11}, \ \
        W_3 \in G(0,Q_3), \ \
        Q_3 = I_n = L_{13} Q_W L_{13}^T.
\end{align*}
It is then possible to construct a matrix 
$L_{14} \in \mathbb{R}^{(n_1-n) \times n_1}$ such that,
\begin{align*}
     & W_4
   =  L_{14} W,  \ \ W_4 \in G(0,Q_4), \ \  Q_4 = I, \ \
        L_{14} Q_W L_{13}^T = 0; \\
  &   \left(
        \begin{array}{l}
          W_3 \\ W_4
        \end{array}
        \right) \in G(0,I_{n_1}), ~
        \rank
          \left(
            \left(
            \begin{array}{ll}
              L_{13} \\ L_{14}
            \end{array}
            \right) 
          \right) = n_1,
\end{align*}
and, due to $L_{14} Q_W L_{13}^T =0$,
$W_3$, $W_4$ are independent random variables.
See \cite[Th. 4.9]{noble:1969} for a theorem with which 
the existence of $L_4$ can be proven.
Note further that $F^{W} = F^{W_3, W_4}$.
\par
Hence the random variables $W_3,  W_4$ are independent,
$(F^{Y_1},F^{Y_2}|F^{W_3}) \in \cigmin$, and \\
$I(Y_1,Y_2;W) = I(Y_1,Y_2;W_3, W_4)$.
\par
By properties of mutual information now follows that,
\begin{align*}
     & I(Y_1,Y_2; W_3, W_4) - I(Y_1,Y_2;W_3) \\
   &=  H(Y_1,Y_2) 
        + H(W_3,W_4)
        - H(Y_1,Y_2,W_3,W_4) 
        - H(Y_1,Y_2) 
        - H(W_3)
        + H(Y_1, Y_2, W_3) \\
   &= H(Y_1,Y_2,W_3)
        + H(W_4)
        - H(Y_1, Y_2, W_3, W_4), ~
        \mbox{by independence of $W_3$ and $W_4$;} \\
   &=  I(Y_1, Y_2, W_3;W_4) \geq 0.
\end{align*}
Thus,
for the computation of $C(Y_1,Y_2)$,
attention can be restricted to those state variables $W$
which are of miminal dimension.\\
(2) Take a probability measure ${\bf P}_1$ such that
there exists a triple of Gaussian random variables
$Y_1,Y_2,~~ W: \Omega \rightarrow \rn$ with
${\bf P}_1|_{(Y_1,Y_2)} = {\bf P}_0$ and
$(F^{Y_1},F^{Y_2}|F^W) \in \cigmin$.
\par
According to \cite[Th. 4.2]{putten:schuppen:1983}
there exist in general many such measures
which are parametrized by the matrices and the sets, as stated in Theorem~\ref{them_putten:schuppen:1983}, (b), and  defined by  (\ref{eq:gtriple})-(\ref{eq:pci}).\\
(3) Next the mutual information of 
the triple of Gaussian random variables 
is calculated, using Theorem~\ref{them_putten:schuppen:1983}.(b),  for any choice
of $Q_W \in {\bf Q_W}$, where ${\bf Q_W}$ is given by  (\ref{eq:pci_1}). Then 
\begin{eqnarray*}
      I(Y_1,Y_2;W)
   =  H(Y_1,Y_2) -  H(Y_1|W) - H(Y_2|W).
\end{eqnarray*}
\par
The following calculations are then obvious,
\begin{align*}
     \det(Q_{(Y_1,Y_2)})
   = & \det 
        \left(
        \begin{array}{ll}
          I & D \\
          D & I
        \end{array}
        \right)
        =  \det (I - D^2) = \prod_{i=1}^n ( 1 - d_i^2); \\
      H(Y_1,Y_2)
   = & \frac{1}{2} \ln ( \det (Q_{(y_1,y_2)})) + \frac{1}{2} (2n) \ln (2\pi e) 
        = \frac{1}{2} \sum_{i=1}^n \ln (1-d_i^2) + n \ln (2\pi e);\\
     &  \ \ \ \   {\bf P}_{Y_1|W}(y_1|w) \in G({\bf E}[Y_1|F^W], Q_{Y_1|W}), \\
      {\bf E}[Y_1|F^W]
  = & Q_{Y_1,W} Q_W^{-1} W = D^{1/2} Q_W^{-1} W;  \ \ \mbox{by (\ref{eq:gtriple_1})}      \\
      Q_{Y_1|W}
  = & I - Q_{Y_1,W} Q_W^{-1} Q_W Q_W^{-1} Q_{Y_1,W}^T 
        = I - D^{1/2} Q_W^{-1} D^{1/2}; \ \ \mbox{by (\ref{eq:gtriple_1})}        \\
      H(Y_1|W)
   = & \frac{1}{2} \ln (\det ( I - D^{1/2} Q_W^{-1} D^{1/2} ) ) 
        + \frac{1}{2} n \ln (2\pi e);\\
      {\bf E}[Y_2|F^W]
   = & Q_{Y_2,W} Q_W^{-1} W= D^{1/2} Q_W Q_W^{-1} W = D^{1/2} W; \\
      Q_{Y_2|W}
   = & I - Q_{Y_2,W} Q_W^{-1} Q_W Q_W^{-1} Q_{Y_2,W}^T 
        = I - D^{1/2} Q_W D^{1/2}; \\
      H(Y_2|W)
   = & \frac{1}{2} \ln (\det ( I - D^{1/2} Q_W D^{1/2} ) ) 
        + \frac{1}{2} n \ln (2\pi e).
\end{align*}
From the above calculations it then follows,        
 \begin{align}       
      I(Y_1,Y_2;W)
   = & H(Y_1,Y_2) -  H(Y_1|W) - H(Y_2|W)  \\
   = & \frac{1}{2} \sum_{i=1}^n \ln (1-d_i^2) + n \ln (2\pi e) \nonumber \\
     & - \frac{1}{2} \ln (\det ( I - D^{1/2} Q_W^{-1} D^{1/2} ) ) 
        - \frac{1}{2} n \ln (2\pi e) \nonumber \\
    & - \frac{1}{2} \ln (\det ( I - D^{1/2} Q_W D^{1/2} ) ) 
        - \frac{1}{2} n \ln (2\pi e)  \\
   = & \frac{1}{2} \sum_{i=1}^n \ln (1-d_i^2) 
        - \frac{1}{2} \ln (\det ( [I - D^{1/2} Q_W^{-1} D^{1/2} ]
                                  [ I - D^{1/2} Q_W D^{1/2} ] 
                                )
                          ).   \label{com_inf_g}
\end{align}
The above calculations verify the statements of Corollary~\ref{cor:commoninfogrvcorrelated}. \\
(4) The computation of $C(Y_1,Y_2)$ requires
the solution of an optimization problem.
\begin{align}
      C(Y_1,Y_2)
   = & \inf_{P_1 \in {\bf P_{ci}}} I(Y_1,Y_2;W) \nonumber \\
  = & \inf_{Q_W \in {\bf Q_W}} 
        \left\{
          \frac{1}{2} \sum_{i=1}^n \ln (1-d_i^2) 
           - \frac{1}{2} \ln (
                \det (
                      [I - D^{1/2} Q_W^{-1} D^{1/2}] [I - D^{1/2} Q_W D^{1/2}]
                     ) 
                ) 
        \right\}. \label{eq:wcicriterion}
\end{align}
Since 
the first term in (\ref{eq:wcicriterion}), 
$\frac{1}{2} \sum_{i=1}^n \ln (1-d_i^2)$,
does not depend on $Q_W$
and  the natural logarithm 
is a strictly increasing function, then          
  \begin{align}
C(Y_1,Y_2) \ \ \mbox{is equivalent to  the  problem:} \ \
    \sup_{Q_W \in {\bf Q_W}} 
          \det
          \left[
            ( I - D^{1/2} Q_W^{-1} D^{1/2}) (I - D^{1/2} Q_W D^{1/2})
          \right]. 
\end{align}

Define,
\begin{align}
      L_1(Q_W)
  = & ( I - D^{1/2} Q_W^{-1} D^{1/2}) (I - D^{1/2} Q_W D^{1/2}), 
        \label{eq:l1qx} \\
      f_1(Q_W)
  = & \det (L_1(Q_W)).
\end{align}
Note that the expression $L_1(Q_W) \in \rnn$
is a non-symmetric square matrix in general.

It will be proven that,
\begin{align}
      f_1(Q_W) 
   = & \det (L_1(Q_W)) 
        \leq \det ([I-D]^2), \ \
        \forall ~ Q_W \in {\bf Q_W}, \label{eq:inequalityvalue} \\
      \det (L_1(Q_W))
   = & \det ([I-D]^2) \ \ 
        \mbox{if and only if} \ \ Q_W = I. \label{eq:equalityvalue}
\end{align}
From these two relations follows that 
$Q_W^* = I \in \rnn$ 
is the unique solution of the supremization problem.
\par
The inequality in (\ref{eq:inequalityvalue})
follows from Proposition \ref{proposition:matrixineqdqx}.
The equality of (\ref{eq:equalityvalue})
is proven in two steps.
If $Q_W = I$ then equality of  (\ref{eq:equalityvalue}) holds
as follows from direct substitution in (\ref{eq:l1qx}).
The converse is proven by contradiction.
Supppose that $Q_W \neq I$.
Then it follows again from 
Proposition \ref{proposition:matrixineqdqx}
that strict inequality holds in  (\ref{eq:inequalityvalue}).
Hence the equality is proven.\\
(5) Finally the value of $C(Y_1,Y_2)$ is computed
for $Q_W^* = I$.
\begin{align*}
      C(Y_1,Y_2)
  = & \frac{1}{2} \sum_{i=1}^n \ln (1- d_1^2) 
        - \frac{1}{2} \ln ( \det ( I - D^{1/2} (Q_W^*)^{-1} D^{1/2} ) )
        - \frac{1}{2} \ln ( \det ( I - D^{1/2} (Q_W^*) D^{1/2} ) ) \\
  = & \frac{1}{2} \sum_{i=1}^n \ln (1- d_i^2) 
        - \frac{1}{2} 2 \ln ( \det (I - D) ) \\
  = & \frac{1}{2} \sum_{i=1}^n \ln (1- d_i^2) 
        - \frac{1}{2} \sum_{i=1}^n  \ln ( (1 - d_i)^2 ) 
        = 
        \frac{1}{2} \sum_{i=1}^n \ln 
           \left(
           \frac{1 - d_i^2}{(1-d_i)^2}
           \right) \\
  = & \frac{1}{2} \sum_{i=1}^n \ln ( \frac{1+d_i}{1-d_i})
        =
        \frac{1}{2} \sum_{i=1}^n \ln 
           \left(
             1 + \frac{2 d_i}{1-d_i}
           \right). 
\end{align*}
(b) 
It follows from part (a) of the theorem that $C(Y_1,Y_2)$ is attained
as the mutual information $I(Y_1,Y_2;W)$
for a random variable $W$ with $Q_W = Q = I$.
Consider now a triple of random variables
$(Y_1,Y_2, W) \in G(0,Q_s(I))$
as defined in  (\ref{eq:gtriple})-(\ref{eq:pci}),
hence $Q_W = I$.
Denote the random variable $W$ from now on by
$W^*$ to indicate that it achieves 
the infimum of the definition of $C(Y_1,Y_2)$.
Thus $Q_{W^*} = I$ and,
\begin{align}
   & (Y_{12},Y_{22},W^*) \in G(0,Q_s(I)), \nonumber \\
    &  Q_s(I)
   =  \left(
        \begin{array}{lll}
          I & D & D^{1/2} \\
          D & I & D^{1/2} \\
          D^{1/2} & D^{1/2} & I
        \end{array}
        \right) > 0. \label{eq:qstar}
\end{align}
Let $V: \Omega \rightarrow \mathbb{R}^{n_{12}}$
be a Gaussian random variable with $V \in G(0,I)$
which is independent of $(Y_1,Y_2,W)$.
\par
Define the new state variable
$\overline{W} = L_1 Y_1 + L_2 Y_2 + L_3 V$.
Then $(Y_1,Y_2,V,W^*)$ are jointly Gaussian
and it has to be shown that then
$Q_{\overline{W}} = I$, 
$Q_{Y_1, \overline{W}} = D^{1/2}$, and 
$Q_{Y_2, \overline{W}} = D^{1/2}$.
These equalities follow from simple calculations using
the expressions of $L_1$, $L_2$, and $L_3$ which calculations
are omitted.
It then follows from those calculations and the definition
of the Gaussian measure $G(0,Q_s(I))$
that $\overline{W} = W^*$ almost surely.
\par
The signals are then represented by,
\begin{align}
      Z_1
   = & Y_1 - {\bf E}[Y_1|F^{W^*}] 
        = Y_1 - Q_{Y_1,W^*} (Q_{W^*})^{-1} W^*
        = Y_1 - D^{1/2} W^*, \label{eq:zone} \\
      Z_2
   = & Y_2 - {\bf E}[Y_2|F^{W^*}] 
        = Y_2 - Q_{Y_2,W^*} (Q_{W^*})^{-1} W^*
        = Y_2 - D^{1/2} W^*.
\end{align}
It is proven that the triple of random variables
$(Z_1,  Z_2, W^*)$ are independent. 
\begin{align*}
     {\bf  E}[Z_1 (W^*)^T ]
   = & {\bf E}[Y_1 (W^*)^T ] - D^{1/2} {\bf E}[W^* (W^*)^T ] = D^{1/2} - D^{1/2} = 0, \ \
        {\bf E}[Z_2 (W^*)^T ] = 0, \\
      {\bf E}[ Z_1 Z_2^T ]
   = & {\bf E}[ (Y_1 - D^{1/2} W^*) ( Y_2 - D^{1/2} W^*)^T ] = 0.
\end{align*}
Hence,  the original signals are represented
as shown by the formulas,
\begin{align*}
      Y_1
  = & Z_1 + D^{1/2} W^*, \ \
        \mbox{by (\ref{eq:qstar}), 
                $Q_{y_1,W^*} Q_{W^*}^{-1} = D^{1/2}$, and by def. of $Z_1$,
              } \\
      Y_2
   = & Z_2 + D^{1/2} W^*, \ \
        \mbox{similarly.}
\end{align*}
\hfill$\square$

%

\ \

Example~\ref{ex-b-ci} (below)  is introduced to illustrate some subtle issues related to Theorem~\ref{the:scalar-ci} that computes the lossy common information for the bivarate Gaussian  random variables  (\ref{cvf_s}), i.e., $p_1=p_2=1$, given in Theorem~\ref{the:scalar-ci}, and found in  many references, such as, \cite{gray-wyner:1974,viswanatha:akyol:rose:2014,xu:liu:chen:2016:ieeetit}.

%
 
\ \

\begin{example} 
\label{ex-b-ci}
Consider an application of Theorem~\ref{th:commoninfogrvcorrelated_new} to  the bivarate Gaussian  random variables  (\ref{cvf_s})  of  Theorem~\ref{the:scalar-ci}, i.e., $p_1=p_2=1$. 
Assume that the random variables have been transformed 
to the canonical variable form with in this case a single
canonical correlation coefficient $d_1 \in [0,1]$, 
see Example~\ref{rem-con-scalar}.(i).
Note that if the correlation coefficient of $Y_1$ and $Y_2$
is negative, $\rho_{Y_1,Y_2} < 0$
then that representation is not in the canonical variable
form so the theorem above does not apply.
The theorem above requires the canonical variable form
with a positive correlation coefficient.
\par
Distinguish the cases:
 \\
(i) $d_1 \in (0,1)$. 
Hence, the Gaussian measure in canonical variable form is 
\begin{eqnarray}
  &   & (Y_{12},Y_{22}) \in G(0,Q_{(Y_{12},Y_{22})}), \ \
      Q_{(Y_{12},Y_{22})}
      = \left(
        \begin{array}{ll}
          1 & d_1 \\
          d_1 & 1
        \end{array}
        \right), ~\\
  &   & Y_{12}: \Omega \rightarrow \mathbb{R}^{p_{12}}, ~
        Y_{22}: \Omega \rightarrow \mathbb{R}^{p_{22}}, \ \ p_{12}=p_{22}=1.
          \nonumber
\end{eqnarray}
\begin{itemize}
\item[(a)] The minimal $\sigma$-algebra which makes $Y_{12}$ and
$Y_{22}$ Gaussian conditional-independent is 
$F^{W}$, where $W:  \Omega \rightarrow \mathbb{R}, W \in G(0, Q_W)$, i.e., $W$ is the state variable. 
\item[(b)] Then  the random variable which achieves $C(Y_1, Y_2)$ is $W=W^* \in G(0, 1)$, and 
\begin{align}
C(Y_1,Y_2)=C(Y_{12}, Y_{22})=I(Y_{12}, Y_{22};W^*)
   =  \frac{1}{2} 
        \ln
        \left(
        \frac{1+d_1}{1-d_1}
        \right). \label{ex_brv_1}
\end{align}
\item[(c)] The weak stochastic realization that achieves (\ref{ex_brv_1}) is 
\begin{align}
&Y_{12}= \sqrt{d_1}W^* + \sqrt{1-d_1} Z_1,\\ 
&Y_{22}= \sqrt{d_1}W^* + \sqrt{1-d_1} Z_2, \\
&Z_1 \in G(0,1), \ \   Z_2 \in G(0,1), \ \ W^* \in G(0,1),\\
&W^*, Z_1, Z_2 \ \ \mbox{independent}.
\end{align}
\end{itemize}
(ii) $d_1 = 0$. This follows from the special cases discussed in Proposition~\ref{prop:wcicaseprivateparts}. \\
(iii) $d_1 = 1$.  This follows from the special cases discussed in Proposition~\ref{prop:wcicasecommonpart}. 
\end{example}


\subsection{Special Cases of Wyner's Information Common Information}\label{subsec:wcigrvsprivateparts}
The special case of canonical variable form with only private parts is presented below.

\ \

\begin{proposition}\label{prop:wcicaseprivateparts}
Consider the case of a tuple of Gaussian vectors with only private parts.
Hence the Gaussian measure is
\begin{eqnarray}
  &   & (Y_{13},Y_{23}) \in G(0,Q_{(Y_{13},Y_{23})}), \ \
      Q_{(Y_{13},Y_{23})}
      = \left(
        \begin{array}{ll}
          I & 0 \\
          0 & I
        \end{array}
        \right), ~
        Y_{13}: \Omega \rightarrow \mathbb{R}^{p_{13}}, ~
        Y_{23}: \Omega \rightarrow \mathbb{R}^{p_{23}}.
\end{eqnarray}
\begin{itemize}
\item[(a)]
The minimal $\sigma$-algebra $F^W$ which makes 
$Y_{13},  Y_{23}$ conditionally independent
is the trivial $\sigma$-algebra
denoted by $F_0 = \{\emptyset, \Omega\}$.
Thus, 
$(F^{Y_{13}},F^{Y_{23}}|F_0) \in \ci$.
The random variable $W$ in this case is the constant
$W_3 = 0 \in \mathbb{R}$, 
hence $F^{W_3} = F_0$.
\item[(b)]
Then $W^*=W_3$ and 
\begin{eqnarray}
  C(Y_1,Y_2)=    C(Y_{13}, Y_{23})
  =  I(Y_{13},Y_{23};W_3) = 0.
\end{eqnarray}
\item[(c)]
The weak stochastic realization that achieves $C(Y_{13}, Y_{23})=0$ is 
\begin{eqnarray}
      Z_1
  =  Y_{13}, \ \
        Z_2 = Y_{23}, \ \
        W_3 = 0.
\end{eqnarray}
\end{itemize}
\end{proposition}

%
The special case of canonical variable form with only identical  parts is presented below.

\ \

\begin{proposition}\label{prop:wcicasecommonpart}
Consider the case of a tuple of Gaussian vectors with only the identical part.
Hence the Gaussian measure is,
\begin{align}
    & Y_{11}: \Omega \rightarrow \mathbb{R}^{p_{11}}, \ \
        Y_{21}: \Omega \rightarrow \mathbb{R}^{p_{21}}, \ \
        p_{11} = p_{21}, \nonumber \\
     & (Y_{11},Y_{21}) \in G(0,Q_{(Y_{11},Y_{21})}), \ \
      Q_{(y_{11},y_{21})}
      = \left(
        \begin{array}{ll}
          I & I \\
          I & I
        \end{array}
        \right), \ \
        Y_{11} = Y_{21} ~ \as
\end{align}
\begin{itemize}
\item[(a)]
The only minimal $\sigma$-algebra which makes $Y_{11}$ and
$Y_{21}$ Gaussian conditional-independent is 
$F^{Y_{11}} = F^{Y_{21}}$.
The state variable is thus,
$W_1 = Y_{11} = Y_{21}$ and $F^W = F^{Y_{11}} = F^{Y_{21}}$.
\item[(b)] Then  
$C(Y_1, Y_2)=C(Y_{11},Y_{21}) = + \infty$.
See the comment below.
\item[(c)] The weak stochastic realization is again simple,
the variable $W$ equals the identical 
component  and there is no need to use
the signals $Z_1$ and $Z_2$.
Thus the representations are,
\begin{align}
      Z_1
   =  0 \in \mathbb{R}, \ \
        Z_2 = 0 \in \mathbb{R},  \ \
        W = Y_{11} = Y_{21}.
\end{align}
\end{itemize}
\end{proposition}
\ \

The amount of Wyner's common information of $C(Y_1,Y_2)$ for this case
is $+\infty$ and this requires comments.
Formally, to compute $I(Y_1,Y_2;W)$,
one needs to evaluate the expression (\ref{eq:mutualinfogrv}).
But in this case the determinant $\det(Q(Y_1,Y_2,W))$ is zero.
Thus,
\begin{eqnarray*}
      C(Y_1,Y_2) 
   =  I(Y_1, Y_2;W) 
        = - \ln ( \frac{\det(Q_{(Y_1,Y_2,W)})}{\det(Q_{Y_1,Y_2}) \det(Q_W)}
        = - \ln(0).
\end{eqnarray*}
One may extend the definition of the natural logarithm 
from the domain $(0,+\infty)$ to the value at zero by the limit
$\ln(0) = \lim_{u \downarrow 0} \ln(u) = - \infty$.
The authors therefore propose to define that $C(Y_1,Y_2) $  of identical Gaussian random variables
takes the value $+ \infty$,
hence $C(Y_{11},Y_{21}) = + \infty$.
This definition also makes sense
considered from an information theoretic interpretation, in which mutual information is defined via relative entropy which admits the value of $+ \infty$.

\subsection{Proof of Wyner's Information Common Information  of Arbitrary Gaussian Vectors}\label{subsec:wciarbitrarygrvs}
Theorem~\ref{thm_cia} is now proved, that corresponds to a tuple of arbitrary
Gaussian random variables, {\it not necessarily  restricted to the correlated parts of these random variables of  Theorem~\ref{them_putten:schuppen:1983},  by (\ref{ci_par_1})-(\ref{ci_par_3})}.
It is  shown that  $C(Y_1, Y_2)$  is computed 
by a decomposition and by the use
of the formulas obtained earlier in  Section~\ref{sect:main}.\\

{\bf Proof of Theorem~\ref{thm_cia}}
(a)
\begin{align}
        C(Y_1,Y_2)
  = & \inf_{(Y_1,Y_2,W) \in {\rm CIG}} I(Y_1,Y_2;W) \nonumber \\
   = & \inf ~ \Big\{ I(Y_{11},Y_{21};W_1) + I(Y_{12},Y_{22};W_2) + I(Y_{13},Y_{23};0)\Big\}, \ \
        \mbox{by Proposition \ref{proposition:mutualinfoindependent},} \nonumber \\
   \geq & \inf_{(Y_1,Y_2,W) \in {\rm CIG}} I(Y_{11},Y_{21};W_1)
           + \inf_{(Y_1,Y_2,W) \in {\rm CIG}} I(Y_{12},Y_{22};W_2)
           + \inf_{(Y_1,Y_2,W) \in {\rm CIG}} I(Y_{13},Y_{23};0) \nonumber \\
   = & \inf_{(Y_{11},Y_{21},W_1) \in {\rm CIG}} I(Y_{11},Y_{21};W_1)
           + \inf_{(Y_{12},Y_{22},W_2) \in {\rm CIG}} I(Y_{12},Y_{22};W_2)
           + I(Y_{13},Y_{23};0) \nonumber \\
  = & C(Y_{11},Y_{21};W_1) 
        + C(Y_{12}, Y_{22}; W_2)
        + C(Y_{13}, Y_{23};0) \nonumber \\
   = & \left\{
        \begin{array}{lll}
          0, & \mbox{if} &  p_{13} > 0, ~ p_{23} > 0, ~
                p_{11} = p_{12} = p_{21} = p_{22} = 0, \\
          \frac{1}{2} \sum_{i=1}^n \ln 
            \left(
              \frac{1+d_i}{1-d_i}
            \right), & \mbox{if} & p_{12} = p_{22} >0, ~ p_{11} = p_{21} =0, ~
                        p_{13} \geq 0, ~ p_{23} \geq 0, \\
          +\infty, & \mbox{else.} & 
        \end{array}
        \right. \label{eq:wcigrvexpression}
\end{align}
The latter equality follows from, respectively,
Proposition \ref{prop:wcicasecommonpart},
Theorem \ref{th:commoninfogrvcorrelated_new}, and
Proposition \ref{prop:wcicaseprivateparts}.\\
(a \& b) 
It will be shown that $C(Y_1,Y_2)$ is less or equal to the right-hand side
of equation (\ref{eq:wcigrvexpression}).
From the latter inequality and the above inequality
then follows the expression according to equation (\ref{eq:wcigrvexpression}).
\par
To be specific,
it will be proven that
$C(Y_1,Y_2)$ is less than the expression
$I(Y_1,Y_2;W^*)$
where $W^*$ is defined in statement (b) of the proposition.
It then follows from the proof of Theorem \ref{th:commoninfogrvcorrelated_new}
that $(F^{Y_{12}}, F^{Y_{22}} | F^{W_2^*}) \in \cigmin$.
\par
Then,
\begin{align*}
         C(Y_1,Y_2)
   = & \inf_{(Y_1,Y_2|W) \in \cig} I(Y_1,Y_2;W) 
        \leq I(Y_1,Y_2;W^*) \\
   = &  I(Y_{11}, Y_{21}; W_1^*) 
           + I(Y_{12}, Y_{22}; W_2^*)
           + I(Y_{13}, Y_{23}; \emptyset ) \\
   = & \left\{
        \begin{array}{lll}
          0, & \mbox{if} & p_{13} > 0, ~ p_{23} > 0, ~
                p_{11} = p_{12} = p_{21} = p_{22} = 0, \\
          \frac{1}{2} \sum_{i=1}^n \ln 
            \left(
              \frac{1+d_i}{1-d_i}
            \right), & \mbox{if} & p_{12} = p_{22} >0, ~ p_{11} = p_{21} =0, ~
                        p_{13} \geq 0, ~ p_{23} \geq 0, \\
          +\infty, & \mbox{else.}  &
        \end{array}
        \right.
\end{align*}
The latter equality is proven as follows.
In the first case, when
$p_{13} > 0$, $p_{23} > 0$, and $p_{11} = p_{12} = p_{21} = p_{22} = 0$,
then $Y_1 = Y_{13}$ and $Y_2 = Y_{23}$ are independent random variables.
It then follows from 
Proposition \ref{prop:wcicaseprivateparts}
that $I(Y_1,Y_2;0) = I(Y_{13},Y_{23};0) = 0$.
In the second case,
when $p_{12} = p_{22} > 0$, $p_{13} \geq 0$, $p_{23} \geq 0$, and
$p_{11} = p_{21} = 0$,
it follows from Proposition \ref{proposition:mutualinfoindependent}
and from Theorem \ref{th:commoninfogrvcorrelated_new} that
\begin{align*}
      I(Y_1,Y_2;W^*) 
  =  I(Y_{12},Y_{22};W_2^*) + I(Y_{13},Y_{23};0) 
        = \frac{1}{2} \sum_{i=1}^n \ln 
            \left(
              \frac{1+d_i}{1-d_i}
            \right).
\end{align*}
In the third case, 
when $p_{11} = p_{21} > 0$ and the other $p_{ij}$ indices
are arbitrary,
then $I(Y_1,Y_2;W^*) = + \infty$.
Hence the inequality $C(Y_1,Y_2) \leq \mbox{right-hand side}$
is proven and hence equality holds.\\
(c) This follows directly from Proposition \ref{prop:grvtuplecomm}.\\
\hfill$\square$
\par\vspace{1\baselineskip}\par\noindent
\subsection{Numerical Calculations of Wyner's Information Common Information}
The reader is advised to carry out an approximation
of the covariance matrix based on the computations
of Algorithm \ref{alg:canonicalvariabledecomposition}
as explained next.
Consider the case in which the dimensions of the two vectors
are ordered as $p_1 \geq p_2$.
The case of $p_1 < p_2$ can be obtained by interchanging the random variables.
The outcome of the computation of the algorithm is then 
of the form,
\begin{align*}
     Q_{\cvf}
  = & \left(
        \begin{array}{ll}
          I_{p_1} & Q_{12} \\
          Q_{12}^T & I_{p_2}
        \end{array}
        \right), ~
        Q_{12} =
        \left(
        \begin{array}{l}
          D \\ 0
        \end{array}
        \right) \in \mathbb{R}^{p_1 \times p_2}, \\
      D
   = & \diag (d_1, \ldots, d_{p_2}) \in \mathbb{R}^{p_2 \times p_2}, ~
        1 \geq d_1 \geq d_2 \geq \ldots \geq d_{p_2} \geq 0.
\end{align*}
Next the values on the diagonal of the matrix $D$ are 
partitioned in three groups depending
on two threshold values $h_1, ~ h_2 \in (0,1)$ with $h_1 > h_2$,
according to,
\begin{align*}
      \overline{d}_i
  =  \left\{
        \begin{array}{lll}
          1.0000, & \mbox{if} & d_i \in (h_1, 1], \\
          0.0000, & \mbox{if} & d_i \in [0,h_2), \\
          d_i,    & \mbox{if} & d_i \in [h_2,h_1],
        \end{array}
        \right. 
\end{align*}
for all $i \in \mathbb{Z}_{p_2}$.
For example, 
if $h_1 = 0.9990$ and $d_1 = 0.9993$
then set $\overline{d}_1 = 1.0000$;
and if $h_2 = 0.0001$ and $d_{p_2} = 0.00006$
then set $\overline{d}_{p_2} = 0.0000$.
Denote then the number of elements of the vector $\overline{d}$ 
which are equal to $1.0000$ by $p_{11}$ and
the number of elements of that vector 
which are equal to $0.00000$ by $p_{23}$.
Define further,
$p_{21} = p_{11}$, 
$p_{22} = p_2 - p_{21} - p_{23}$,
$p_{12} = p_{22}$, and
$p_{13} = p_1 - p_{11} - p_{12}$,
\begin{eqnarray*}
      \overline{D}
  & = & \diag (\overline{d}_{p_{11}+1}, \ldots, \overline{d}_{p_{11}+p_{12}}) 
           \in \mathbb{R}^{p_{12} \times p_{12}}, \\
      \overline{Q}_{12}
  & = & \left(
        \begin{array}{lll}
          I_{p_{11}} & 0            & 0 \\
          0          & \overline{D} & 0 \\
          0          &              & 0_{p_{23} \times p_{23}} \\
          0          &              & 0_{(p_{13}-p_{23}) \times p_{23}}
        \end{array}
        \right), ~~
      \overline{Q}_{\cvf}
      = \left(
        \begin{array}{ll}
          I_{p_1}             & \overline{Q}_{12} \\
          \overline{Q}_{12}^T & I_{p_2}
        \end{array}
        \right).
\end{eqnarray*}
The indices 
$(p_{11}, p_{12}, p_{13})$ and $(p_{21}, p_{22}, p_{23})$
now satisfy the relations,
$p_1 = p_{11} + p_{12} + p_{13}$,
$p_2 = p_{21} + p_{22} + p_{23}$,
$p_{11} = p_{21}$, and
$p_{12} = p_{22}$.
The reader can use either $Q_{\cvf}$ 
or its approximation $\overline{Q}_{\cvf}$.
\par\noindent
\begin{example}\label{ex:wcigrvex1}
Consider the tuple of Gaussian random variables,
\begin{align*}
   (Y_1, Y_2) \in & G(0,Q_{(Y_1,Y_2)}, \ \
         p_1 = 3, ~ p_2 = 3, \\
      Q_{(Y_1,Y_2)}
  = & \left(
        \begin{array}{ll}
          I_{p_1}     & Q_{Y_1,Y_2} \\
          Q_{Y_1,Y_2}^T & I_{p_2}
        \end{array}
        \right), ~~
        Q_{Y_1,Y_2} =
        \left(
        \begin{array}{lll}
          0.8 & 0   & 0 \\
          0   & 0.5 & 0 \\
          0   & 0   & 0.1 
        \end{array}
        \right) \in \mathbb{R}^{p_1 \times p_2}.
\end{align*}
A computation then yields,
\begin{align*}
    & (p_{11}, p_{12}, p_{13}) = (0,3,0), \ \ 
        (p_{21}, p_{22}, p_{23}) = (0,3,0), \\ 
     & D
   =  \left(
        \begin{array}{lll}
          0.8 & 0   & 0 \\
          0   & 0.5 & 0 \\
          0   & 0   & 0.1 
        \end{array}
        \right) \in \mathbb{R}^{p_{12} \times p_{22}}, \\
    &  C(Y_1,Y_2)
   =  5.0444 ~ \mbox{bits.}
\end{align*}
\end{example}
\begin{example}\label{ex:wcigrvex2}
Consider the tuple of Gaussian random variables,
\begin{align*}
  (Y_1, Y_2) \in & G(0,Q_{(Y_1,Y_2)}, \ \
         p_1 = 6, ~ p_2 = 5, \\
      Q_{(Y_1,Y_2)}
   = & \left(
        \begin{array}{ll}
          I_{p_1}     & Q_{Y_1,Y_2} \\
          Q_{Y_1,Y_2}^T & I_{p_2}
        \end{array}
        \right), \\
      Q_{Y_1,Y_2}
   = & \left(
        \begin{array}{lllll}
          0.999998 & 0        & 0   & 0   & 0 \\
          0        & 0.999992 & 0   & 0   & 0 \\
          0        & 0        & 0.8 & 0   & 0 \\
          0        & 0        & 0   & 0.3 & 0 \\
          0        & 0        & 0   & 0   & 0.000004\\
          0        & 0        & 0   & 0   & 0
        \end{array}
        \right) \in \mathbb{R}^{p_1 \times p_2}.
\end{align*}
A computation then yields,
\begin{align*}
    & (p_{11}, p_{12}, p_{13}) = (2,2,2), \ \
        (p_{21}, p_{22}, p_{23}) = (2,2,1), ~~
        D =
        \left(
        \begin{array}{ll}
          0.8 & 0 \\
          0   & 0.3 
        \end{array}
        \right) \in \mathbb{R}^{p_{12} \times p_{22}}, \\
     & C(Y_1,Y_2)
  =  + \infty, ~ \mbox{bits}, \\
     & C(Y_{12}, Y_{22})
   =  4.0630 ~ \mbox{bits.}
\end{align*}
\end{example}
\begin{example}\label{ex:wcigrvex3a}
Consider a tuple of Gaussian random variables
for which the variance matrix is generated by a random number generator.
Generate the matrix $L \in \mathbb{R}^{p \times p}$
such that every element has a normal distribution with parameters $G(0,1)$ and 
such that all elements of the matrix are independent.
Then define $Q = L L^T$ to guarantee that the matrix $Q$
is semi-positive-definite. Then,
\begin{align*}
    & (Y_1, Y_2) \in G(0,Q_{(y_1,y_2)}), ~
         p_1 = 5, ~ p_2 = 4, ~ p = p_1 + p_2 = 9; \\
    & Q_{(Y_1,Y_2)}, ~
        \mbox{randomly generated as described above, values not displayed.}
\end{align*}
The outcome of a computation is then that,
\begin{align*}
    & (p_{11}, p_{12}, p_{13}) = (0, 4, 1);
        (p_{21}, p_{22}, p_{23}) = (0, 4, 0); \\
     & C(Y_{12}, Y_{22})
   =  13.1597 ~ \mbox{bits.}
\end{align*}
\end{example}
\begin{example}\label{ex:wcigrvex3b}
Consider a tuple of Gaussian random variables
of which the variance is generated as in the previous example.
\begin{align*}
    & (y_1, y_2) \in G(0,Q_{(y_1,y_2)}, \ \
         p_1 = 5, ~ p_2 = 4, ~ p = p_1 + p_2 = 9; \\
    & Q_{(y_1,y_2)}, ~
        \mbox{randomly generated as described above.}
\end{align*}
Then
\begin{align*}
    & (p_{11}, p_{12}, p_{13}) = (0, 4, 1); \ \
        (p_{21}, p_{22}, p_{23}) = (0, 4, 0); \\
    &  C(Y_{12}, Y_{22})
  =  13.9962 ~ \mbox{bits.}
\end{align*}
\end{example}
\section{Parametrization of Gray and Wyner Rate Region and Wyner's Lossy Common Information}
\label{sect:wcilossy}
\subsection{Definition of Wyner's Lossy Common Information}
\label{subsect:wcilossyconcepts}
For the calculatation of $C_{GW}(Y_1,Y_2;\Delta_1,\Delta_2)$ 
via Theorem~\ref{theorem_4}, and $ C_W(Y_1, Y_2)$ via Theorem~\ref{them_5_xlc},  it is sufficient
to impose the conditional independence $(F^{Y_1}, F^{Y_2} | F^W) \in \ci$, that is,  ${\bf P}_{Y_1, Y_2|W}={\bf P}_{Y_1|W} {\bf P}_{Y_2|W}$.    This is
due to the following. \\
(1) The well-known inequality 
\begin{align}
I(Y_1,Y_2;W) = H(Y_1,Y_2)-H(Y_1,Y_2|W) \geq  H(Y_1,Y_2)-H(Y_1|W)-H(Y_2|W)
\end{align}
which is achieved if ${\bf P}_{Y_1, Y_2|W}={\bf P}_{Y_1|W} {\bf P}_{Y_2|W}$, and \\
(2)  a   necessary condition for the equality constraint (\ref{equality_1}) to hold is  (see Appendix B in \cite{xu:liu:chen:2016:ieeetit}) is 
\begin{align}
R_{Y_1, Y_2|W}(\Delta_1, \Delta_2)=R_{Y_1|W}(\Delta_1)+ R_{Y_2|W}(\Delta_2). \label{nec_equality_1}
\end{align}
Further, a sufficient condition for (\ref{nec_equality_1}) to hold is the conditional independence condition \cite{xu:liu:chen:2016:ieeetit}: ${\bf P}_{Y_1,Y_2|W}={\bf P}_{Y_1|W} {\bf P}_{Y_2|W}$.

Hence, a sufficient condition for  any rate $(R_0,R_1,R_2) \in  {\cal R}_{GW}(\Delta_1,\Delta_2)$ to lie on the Pangloss plane is the  conditional independence. 

Further, \\
(3) for jointly Gaussian random variables $(Y_1,Y_2)$ with square-error distortion, then by the maximum
entropy principle the optimal joint distribution ${\bf P}_{Y_1, Y_2,\hat{Y}_1,\hat{Y}_2,W}$ of the optimization problem $C_{GW}(Y_1,Y_2;\Delta_1,\Delta_2)$  is 
confined to a jointly Gaussian distribution.

%
%
%
%
%

\par Thus one arrives at  the definition of Wyner's lossy common information given below.

\begin{definition}\label{wynercommoninforvs}
{\em Wyner's lossy common information} of a tuple of Gaussian multivariate random variables.
Consider a tuple of jointly Gaussian random variables 
$Y_1: \Omega \rightarrow \mathbb{R}^{p_1} \equiv {\mathbb Y}_1$, 
$Y_2: \Omega \rightarrow \mathbb{R}^{p_2}\equiv {\mathbb Y}_2$, in terms of the notation $(Y_1, Y_2) \in G(0,Q_{(y_1,Y_2)})$, and square error distortion functions between $(y_{1}, y_{2})$,  and its reproduction $(\hat{y}_{1}, \hat{y}_{2})$, given by 
\begin{align}
D_{{Y}_1}(y_1, \hat{y}_1) =  ||{y}_{1}- \hat{{y}}_{1}||_{{\mathbb R}^{p_1}}^2, \ \   \ \   D_{{Y}_2}(y_2, \hat{y}_2) =  ||{y}_{2}- \hat{{y}}_{2}||_{{\mathbb R}^{p_2}}^2 \label{pr_2}
\end{align}
where $||\cdot||_{{\mathbb R}^{p_i}}^2$ denotes Euclidean distances on $\mathbb{R}^{p_i}, i=1,2$.  \\
(a) Wyner's lossless common information (information definition)
of the tuple of Gaussian random  variables $(Y_1, Y_2)$ is defined  by the expression,
\begin{eqnarray}
      C(Y_1,Y_2)
  =  \inf_{W: \Omega \rightarrow \mathbb{R}^n, ~ 
              (F^{Y_1}, F^{Y_2}| F^W) \in {\rm CIG}
             } ~
        I(Y_1,Y_2; W) \in [0,\infty]. \label{w_ic}
\end{eqnarray}
Call any random variable $W$ as defined above such that
$(Y_1, Y_2, W) \in G$ and $(F^{Y_1}, F^{Y_2} | F^W) \in \cig$
a {\em state} of the tuple $(Y_1, Y_2)$.\\
If there exists a random variable 
$W^*: \Omega \rightarrow \mathbb{R}^{n^*}$
with $n^* \in \mathbb{Z}_+ = \{1,2,\ldots,\}$
which attains the infimum,
thus if
$C(Y_1;Y_2) = I(Y_1, Y_2; W^*)$,
then call that random variable a {\em minimal information state} 
of the tuple $(Y_1, Y_2)$.\\
(b) Wyner's lossy common information (operational definition) is defined for a strictly positive numbers $\gamma = (\gamma_1, \gamma_2)\in {\mathbb R}_{++}\times {\mathbb R}_{++}=(0,\infty)\times (0,\infty)$ such that, for all $0\leq (\Delta_1, \Delta_2) \leq \gamma$,  
\begin{align}
C_{GW}(Y_1, Y_2; \Delta_1, \Delta_2)=&C_{W}(Y_1, Y_2) = C(Y_1, Y_2), \nonumber \\ 
&\mbox{for} \ \ (\Delta_1,\Delta_2) \in  {\cal D}_W = \Big\{(\Delta_1,\Delta_2) \in [0,\infty]\times [0,\infty]\Big|  0\leq (\Delta_1, \Delta_2) \leq \gamma\Big\}   \label{eq_WCI}
\end{align} 
provided identity (\ref{equality_1}) holds, i.e., $R_{Y_1|W}(\Delta_1)+R_{Y_2|W}(\Delta_2)+ I(Y_1, Y_2; W)=R_{Y_1, Y_2}(\Delta_1, \Delta_2)$.
\end{definition}

\ \

By the above definition,   the problem of calculating Wyner's lossy common information via (\ref{eq_202}) is decomposed into the characterization of $C(Y_1, Y_2)$ such that identity (\ref{equality_1}) is satisfied. This follows from the fact that the only difference between $C_W(Y_1,Y_2)$ and $C(Y_1,Y_2)$ is the specification of the region ${\cal D}_W$ such that $C_{GW}(Y_1, Y_2; \Delta_1, \Delta_2)=C_W(Y_1,Y_2) = C(Y_1, Y_2)$, i.e., it is constant for $(\Delta_1, \Delta_2) \in {\cal D}_W$. 

The current paper is mainly devoted to the optimization problem $C(Y_1,Y_2)$   defined by (\ref{w_ic}),  in terms of the joint distribution ${\bf P}_{Y_1, Y_2, W^*}$ which achieves the infimum in (\ref{w_ic}). It is then shown that   the weak realization of the tuple $(Y_1, Y_2)$, expressed in terms of the random variable $W^*$, ensures the validity of identity (\ref{equality_1})  of Theorem~\ref{theorem_4}, with $W$ replaced by $W^*$.  Hence, by Theorem~\ref{them_5_xlc}, there exists a ${\cal D}_W$, such that Wyner's lossy common information (operational definition) is computed from  (\ref{eq_WCI}). \\
Alternatively, the reader can verify  that the weak realization of the tuple $(Y_1, Y_2)$, expressed in terms of the random variable $W^*$, which is obtained from  the characterization of Wyner's common information $C(Y_1,Y_2)$    defined by (\ref{w_ic}), ensures that all conditions of Theorem 1 in \cite{viswanatha:akyol:rose:2014} are satisfied.


\par In general there are many random variables $W$
which make the tuple of random variables $(Y_1, Y_2)$
conditionally independent, i.e., as stated in Theorem~\ref{them_putten:schuppen:1983}.
Therefore one wants to infimize the mutual information
overall random variables $W$ which make $Y_1$ and $Y_2$
conditionally independent.

\par
In the next sections
the Wyner lossy common information of 
finite-dimensional Gaussian random variables will be computed. 

%
\subsection{Proof of Wyner's Lossy Common Information of Correlated Gaussian Vectors}
\label{sect:com_r0}
Consider a tuple $(Y_1,Y_2)$ of Gaussian random variables
in the canonical variable form of  Def. \ref{def:grvcommoncorrelatedprivateinfo}.
Restrict attention to the correlated parts of these random variables, as defined in Theorem~\ref{them_putten:schuppen:1983},  defined by (\ref{ci_par_1})-(\ref{ci_par_3}). \\

Consider the  tuple of jointly Gaussian random variables 
$Y_1: \Omega \rightarrow \mathbb{R}^{p_1}$, $Y_2: \Omega \rightarrow \mathbb{R}^{p_2}$ with square error distortion functions of Definition~\ref{wynercommoninforvs}. Then the following statements hold.
%
%
%
\begin{itemize}
\item[(a)] The minimization of  $I(Y_1, Y_2; W)$ over joint distributions ${\bf P}_{Y_1, Y_2, W}(y_1,y_2,w)$ having a marginal distribution  ${\bf P}_{Y_1, Y_2, W}(y_1,y_2,\infty)={\bf P}_{Y_1, Y_2}(y_1,y_2) $  can be confined to  Gaussian random variables $W$ such that $(Y_1,Y_2, W)$ are jointly Gaussian distributed, and $(Y_1,Y_2|W) \in {\rm CIG}$. 

\item[(b)] The calculation of the rate distortion functions $R_{Y_i}(\Delta_i), R_{Y_i|W}(\Delta_i), R_{Y_i}(\Delta_i), i=1,2$ and $R_{Y_1, Y_2}(\Delta_1, \Delta_2)$ can be confined to jointly Gaussian random variables $(Y_1, Y_2, \hat{Y}_1, \hat{Y}_2,W)$.

\item[(c)] The characterization of the Gray-Wyner lossy rate region ${\cal R}_{GW}(\Delta_1, \Delta_2)$ of Theorem~\ref{theorem_8}, in terms of the joint distributions ${\bf P}_{Y_1, Y_2, W} \in {\cal P}$, can be confined to  Gaussian random variables $W$ such that $(Y_1,Y_2, W)$ are jointly Gaussian distributed.  

\item[(d)] The characterization of Wyner's lossy common information $C_W(Y_1, Y_2)$   is invariant with respect to nonsingular basis transformation   
\begin{align}      
      S = \blockdiag (S_{1,} S_2 ) \in \mathbb{R}^{(p_1+p_2) \times (p_1+p_2)}
      \end{align}
such that with respect to the new basis
      $(S_1 Y_1 , S_2 Y_2 )$ is jointly Gaussian.
\item[(e)] Any rate triple $(R_0,R_1, R_2)$ that belongs to the characterization of the Gray-Wyner lossy rate region ${\cal R}_{GW}(\Delta_1, \Delta_2)$ of Theorem~\ref{theorem_8}, is equivalently computed by transforming the tuple $(Y_1,Y_2)$ of Gaussian random variables
into  their canonical variable form  of  Def. \ref{def:grvcommoncorrelatedprivateinfo}.       
\end{itemize}

The proof of Theorem~\ref{thm:r0} makes use  of the characterization of the joint RDF  $R_{Y_1, Y_2}(\Delta_1, \Delta_2)$ given in the next theorem.

\begin{theorem}
\label{th:jrdf_g}
Consider a tuple of Gaussian random variables 
$Y_i: \Omega \rightarrow \mathbb{R}^{p_i}$, 
with $Q_{Y_i} > 0$, for $ i=1, 2$, $(Y_1 , Y_2 ) \in G(0,Q_{(Y_1,Y_2)})$, not necessarily in canonical variable form. Consider the  joint RDF $R_{Y_1,Y_2}(\Delta_1, \Delta_2)$ with square error distrortion functions $D_{Y_1} (y_1, \hat{y}_1)= ||y_{1}-\hat{y}_{1}||_{{\mathbb R}^{p_1}}^2,  \  D_{Y_2} (y_2, \hat{y}_2)= ||y_{2}-\hat{y}_{2}||_{{\mathbb R}^{p_2}}^2$. Then the following hold.\\
\indent (a) The mutual information $I(Y_1, Y_2; \hat{Y}_1, \hat{Y}_2)$ satisfies
\begin{align}
I(Y_1, Y_2; \hat{Y}_1, \hat{Y}_2)\geq I(Y_1, Y_2; {\bf E}\big\{Y_1\Big|F^{\hat{Y}_1, \hat{Y}_2}\big\}, {\bf E}\big\{Y_2\Big|F^{\hat{Y}_1, \hat{Y}_2}\big\}) \label{jrdf_G1}
\end{align}
and the mean square error satisfies 
\begin{align}
{\bf E}\Big\{||Y_{1}-\hat{Y}_{1}||_{{\mathbb R}^{p_1}}^2\Big\} \geq {\bf E}\Big\{||Y_{1}-{\bf E}\big\{Y_1\Big|F^{\hat{Y}_{1}}\big\}||_{{\mathbb R}^{p_1}}^2\Big\},  \ \  {\bf E}\Big\{||Y_{2}-\hat{Y}_{2}||_{{\mathbb R}^{p_2}}^2\Big\} \geq {\bf E}\Big\{||Y_{2}-{\bf E}\big\{Y_2\Big|F^{\hat{Y}_{2}}\big\}||_{{\mathbb R}^{p_2}}^2\Big\}. \label{jrdf_G2}
\end{align}
Moreover, inequalities in (\ref{jrdf_G1}), (\ref{jrdf_G2}) hold with equality, if there exists a jointly Gaussian realization of $(\hat{Y}_1, \hat{Y}_2)$ or  a Gaussian test channel   distribution ${\bf P}_{\hat{Y}_1, \hat{Y}_2|Y_1, Y_2}$,  such that  the joint distribution  ${\bf P}_{\hat{Y}_1, \hat{Y}_2,Y_1, Y_2}$ is jointly Gaussian, and such that the identity holds.
\begin{align}
 {\bf E}\big\{Y_1\Big|F^{\hat{Y}_{1}, \hat{Y}_2}\big\}={\bf E}\big\{Y_1\Big|F^{\hat{Y}_{1}}\big\}=\hat{Y}_1, \ \ \ \ {\bf E}\big\{Y_2\Big|F^{\hat{Y}_1, \hat{Y}_{2}}\big\}={\bf E}\big\{Y_2\Big|F^{\hat{Y}_{2}}\big\}=\hat{Y}_2. \label{jrdf_G3}
\end{align}
\indent (b) A realization that achieves the lower bounds of part (a), i.e., satisfies  (\ref{jrdf_G3}), is the Gaussian realization of $(Y_1, Y_2, \hat{Y}_1, \hat{Y}_2)$ given by 
\begin{align}
&\left( \begin{array}{ll} \hat{Y}_1 \\\hat{Y}_{2}  \end{array} \right) = H \left( \begin{array}{ll} Y_1 \\ Y_2\end{array} \right) + \left( \begin{array}{ll}  V_1\\V_2   \end{array} \right)  \label{jrdf_G4} \\
&(V_1,V_2) \in G(0, Q_{(V_1, V_2)}), \ \ (V_1, V_2) \ \ \mbox{independent of} \ \ (Y_1, Y_2), \\
&H= I_{p_1+p_2} - Q_{(E_1, E_2)}Q_{(Y_1, Y_2)}^{-1},\label{jrdf_G5}\\
&Q_{(V_1,V_2)}=Q_{(E_1,E_2)}H^T= H Q_{(E_1,E_2)}= Q_{(E_1,E_2)}-Q_{(E_1,E_2)}Q_{(Y_1,Y_2)}^{-1} Q_{(E_1,E_2)}\geq 0, \label{jrdf_G6}
\end{align}
where $(E_1, E_2)$ is the  error tuple, 
\begin{align}
E_1 = Y_1 - {\bf E}\big\{Y_1\Big|F^{\hat{Y}_{1}}\big\}=Y_1-\hat{Y}_1, \ \ E_2= {\bf E}\big\{Y_2\Big|F^{\hat{Y}_{2}}\big\}=Y_2-\hat{Y}_2,\label{jrdf_G7}
\end{align}
and the variance matrix of this tuple is,
\begin{align*}
   & (E_1,E_2) \in G(0,Q_{(E_1,E_2)}), ~~
        Q_{(E_1,E_2)}
     =  \left(
        \begin{array}{ll}
          Q_{E_1} & Q_{E_1,E_2} \\
          Q_{E_1,E_2}^T & Q_{E_2}
        \end{array}
        \right) \in \mathbb{R}^{(p_1+p_2) \times (p_1+p_2)}.
\end{align*}
\indent (c) The  joint RDF $R_{Y_1,Y_2}(\Delta_1, \Delta_2)$ is characterized by 
\begin{align}
&R_{Y_1,Y_2}(\Delta_1, \Delta_2)=\inf_{{\bf E}\big\{||E_1||_{{\mathbb R}^{p_1}}^2\big\} \leq \Delta_1, \:   {\bf E}\big\{||E_2||_{{\mathbb R}^{p_2}}^2\big\} \leq \Delta_2} \frac{1}{2} \ln ( \frac{\det(Q_{(Y_1,Y_2)})}{\det(Q_{(E_1,E_2)})}),\label{jrdf_G8} \\
&\mbox{such that} \ \ Q_{(\hat{Y}_1,\hat{Y}_2)}=Q_{(Y_1,Y_2)}- Q_{(E_1,E_2)}\geq 0,\label{jrdf_G9}
\end{align}
where the test channel distribution ${\bf P}_{\hat{Y}_1, \hat{Y}_2|Y_1, Y_2}$ or the joint distribution  ${\bf P}_{\hat{Y}_1, \hat{Y}_2,Y_1, Y_2}$ is induced by the realization of part (b). \\
\indent (d) Compute the canonical variable form of the tuple
of Gaussian random variables $(Y_1 , Y_2 ) \in G(0,Q_{(Y_1,Y_2)})$,
according to Algorithm \ref{alg:canonicalvariabledecomposition}.
This yields the indices
$p_{11} = p_{21}$, $p_{12} = p_{22}$, $p_{13}$, $p_{23}$, and 
$n = p_{11} + p_{12} = p_{21} + p_{22}$,
 the diagonal matrix $D$ with canonical singular values
$d_i \in (0,1)$ for $i = 1, \ldots, n$, and decompositions  (see Algorithm \ref{alg:canonicalvariabledecomposition}, 1-5)
  \begin{align}
        Q_{Y_1} \  = \  U_1 D_1 U_1^T , \ \ 
        Q_{Y_2} \  = \  U_2 D_2 U_2^T , \label{jrdf_G10}
      \end{align}
       with $U_i \in \mathbb{R}^{ p_i \times p_i } $
      orthogonal ($ U_i U_i^T = I_{p_i} = U_i^T U_i$), $i=1,2$, 
     singular-value decomposition of
      \begin{align}
        D_1^{ - \frac{1}{2}  } U_1^T Q_{Y_1Y_2}
        U_2 D_2^{ - \frac{1}{2} } 
        \  = \ 
        U_3 D_3 U_4^T ,
     \end{align}
      with $U_3 \in \mathbb{R}^{ p_1 \times p_1 } $,
      $U_4 \in \mathbb{R}^{ p_2 \times p_2 } $ orthogonal, 
            \begin{align}
         D_3 = & 
            \left( 
            \begin{array}{lll}
              I_{p_{11}} & 0   & 0 \\
              0          & D_4 & 0  \\
              0          & 0   & 0 
            \end{array}
            \right)
            \in \mathbb{R}^{ p_1 \times p_2 } , \\
        D_4  = &  \diag ( d_{4,1} ,..., d_{ 4,p_{12} } ) \in 
          \mathbb{R}^{ p_{12} \times p_{12} }  , \ \ 
          1 > d_{4,1} \geq d_{4,2} \geq \ldots \geq d_{4,p_{12}} > 0 .
      \end{align}
  Define  the  new variance matrix of $Q_{(Y_1,Y_2)}$ according to,
      \begin{align}
            Q_{\cvf}
         =  \left(
              \begin{array}{ll}
                I_{p_1} & D_3 \\
                D_3^T   & I_{p_2}
              \end{array}
              \right).
      \end{align}
Compute the canonical variable form of the tuple
of Gaussian error random variables $(E_1 , E_2 ) \in G(0,Q_{(E_1,E_2)})$ of part (b),
according to Algorithm \ref{alg:canonicalvariabledecomposition}.
This yields the indices
$\overline{p}_{11} = \overline{p}_{21}$, $\overline{p}_{12} = \overline{p}_{22}$, $\overline{p}_{13}$, $\overline{p}_{23}$, and 
$\overline{n} = \overline{p}_{11} + \overline{p}_{12} = \overline{p}_{21} + \overline{p}_{22}$
and the diagonal matrix $\overline{D}$ with canonical singular values
$\overline{d}_i \in (0,1)$ for $i = 1, \ldots, n$, and decompositions (see Algorithm \ref{alg:canonicalvariabledecomposition}, 1-5), 
 \begin{align}
      &  Q_{E_1} \  = \  \overline{U}_1 \overline{D}_1 \overline{U}_1^T , \ \ 
        Q_{E_2} \  = \  \overline{U}_2 \overline{D}_2 \overline{U}_2^T , \\
  & \overline{D}_i  =   \diag ( \overline{d}_{i,1} ,..., \overline{d}_{i,p_{i} } ) \in 
          \mathbb{R}^{ p_{i} \times p_{i} }  , \ \ 
          \overline{d}_{i,1} \geq \overline{d}_{i,2} \geq \ldots \geq \overline{d}_{i,p_{i}} \geq 0, \ \ i=1,2,      
      \end{align}
      with $\overline{U}_i \in \mathbb{R}^{ p_i \times p_i } $
      orthogonal ($ \overline{U}_i \overline{U}_i^T = I_{p_i} = \overline{U}_i^T \overline{U}_i$), $i=1,2$, 
     singular-value decomposition of
      \begin{align}
        \overline{D}_1^{ - \frac{1}{2}  } \overline{U}_1^T Q_{E_1E_2}
        \overline{U}_2 \overline{D}_2^{ - \frac{1}{2} } 
        \  = \ 
        \overline{U}_3 \overline{D}_3 \overline{U}_4^T ,
     \end{align}
      with $\overline{U}_3 \in \mathbb{R}^{ p_1 \times p_1 } $,
      $\overline{U}_4 \in \mathbb{R}^{ p_2 \times p_2 } $ orthogonal, 
            \begin{align}
         \overline{D}_3 = & 
            \left( 
            \begin{array}{lll}
              I_{\overline{p}_{11}} & 0   & 0 \\
              0          & \overline{D}_4 & 0  \\
              0          & 0   & 0 
            \end{array}
            \right)
            \in \mathbb{R}^{ p_1 \times p_2 } , \\
        \overline{D}_4  = &  \diag ( \overline{d}_{4,1} ,..., \overline{d}_{ 4,\overline{p}_{12} } ) \in 
          \mathbb{R}^{ \overline{p}_{12} \times \overline{p}_{12} }  , \ \ 
          1 > \overline{d}_{4,1} \geq \overline{d}_{4,2} \geq \ldots \geq \overline{d}_{4,\overline{p}_{12}} > 0 .
      \end{align}
  Define  the  new variance matrix of $Q_{(E_1,E_2)}$ according to,
      \begin{align}
            \overline{Q}_{\cvf}
         =  \left(
              \begin{array}{ll}
                I_{p_1} & \overline{D}_3 \\
                \overline{D}_3^T   & I_{p_2}
              \end{array}
              \right).\label{jrdf_G19}
      \end{align}
Then with $U_1=\overline{U}_1, U_2=\overline{U}_2$, the joint RDF $R_{Y_1,Y_2}(\Delta_1, \Delta_2)$ of part (c)  is equivalently characterized by 
\begin{align}
&R_{Y_1,Y_2}(\Delta_1, \Delta_2)=\inf_{ \sum_{i=1}^{p_1} \overline{d}_{1,i}   \leq \Delta_1, \:  \sum_{i=1}^{p_2} \overline{d}_{2,i}   \leq \Delta_2  } \frac{1}{2} \ln ( \frac{\det(D_1) \det(D_2)\det(Q_{\cvf})}{\det(\overline{D}_1) \det(\overline{D}_2) \det( \overline{Q}_{\cvf}) }),\label{jrdf_G20} \\
&\mbox{such that} \ \ Q_{(\hat{Y}_1,\hat{Y}_2)}=Q_{(Y_1,Y_2)}- Q_{(E_1,E_2)}\geq 0,\label{jrdf_G21}
\end{align}
where 
\begin{align}
         \det(Q_{\cvf})
   = & \det(I_{p_1}-D_3 D_3^T) \label{jrdf_G22}\\
   = & \left\{
        \begin{array}{lll}
          1, & \mbox{if} & p_{13} > 0, ~ p_{23} > 0, ~
                p_{11} = p_{12} = p_{21} = p_{22} = 0, \\
          \prod_{i=1}^n \left(
              1-d_{4,i}^2            \right), & \mbox{if} & p_{11} = p_{12} =0, ~ p_{12} = p_{22} =n, ~
                        p_{13} \geq 0, ~ p_{23} \geq 0, \\
          0, & \mbox{if}  & p_{11}=p_{21}>0, ~p_{12}=p_{22}\geq 0, ~ p_{12}\geq 0, ~p_{23} \geq 0,
        \end{array}
        \right. \label{jrdf_G23}
\end{align}
\begin{align}
         \det(\overline{Q}_{\cvf})
   = & \det(I_{p_1}-\overline{D}_3 \overline{D}_3^T) \label{jrdf_G24}\\
   = & \left\{
        \begin{array}{lll}
          1, & \mbox{if} & \overline{p}_{13} > 0, ~ \overline{p}_{23} > 0, ~
                \overline{p}_{11} = \overline{p}_{12} = \overline{p}_{21} = \overline{p}_{22} = 0, \\
          \prod_{i=1}^{\overline{n}} \left(
              1-\overline{d}_{4,i}^2            \right), & \mbox{if} & \overline{p}_{11} = \overline{p}_{12} =0, ~ \overline{p}_{12} = \overline{p}_{22} =\overline{n}, ~
                        \overline{p}_{13} \geq 0, ~ \overline{p}_{23} \geq 0, \\
          0, & \mbox{if}  & \overline{p}_{11}=\overline{p}_{21}>0, ~\overline{p}_{12}=\overline{p}_{22}\geq 0, ~ \overline{p}_{12}\geq 0, ~\overline{p}_{23} \geq 0,
        \end{array}
        \right.\label{jrdf_G25}
\end{align}
\indent  (e)  The lower bound holds:
\begin{align}
R_{Y_1,Y_2}(\Delta_1, \Delta_2)\geq &R_{Y_2|Y_1}( \Delta_2) + R_{Y_1}(\Delta_1) \label{g_in_a} \\
=& \inf_{ {\bf E}\big\{||E_2||_{{\mathbb R}^{p_2}}^2\big\} \leq \Delta_2} \frac{1}{2} \ln ( \frac{\det(Q_{Y_2|Y_1})}{\det(Q_{E_2})}) +  \inf_{{\bf E}\big\{||E_1||_{{\mathbb R}^{p_1}}^2\big\}  \leq \Delta_1} \frac{1}{2} \ln ( \frac{\det(Q_{Y_1})}{\det(Q_{E_1})}) \label{g_in}\\
=& \inf_{ \sum_{i=1}^{p_2} \overline{d}_{2,i}   \leq \Delta_2  } \frac{1}{2} \ln ( \frac{ \det(D_2)\det(Q_{\cvf})}{ \det(\overline{D}_2)})+  \inf_{ \sum_{i=1}^{p_1} \overline{d}_{1,i}   \leq \Delta_1  } \frac{1}{2} \ln ( \frac{\det(D_1)}{\det(\overline{D}_1) }) \label{g_in_b}
\end{align}
where $D_1, D_2, \overline{D}_1, \overline{D}_2, Q_{\cvf}$ are defined in part (d).\\
Moreover,  if $\overline{p}_{11}=\overline{p}_{21}=\overline{p}_{12}=\overline{p}_{22}=0$, there exists a strictly positive surface ${\cal D}_C(Y_1, Y_2)\subseteq [0,\infty)\times [0,\infty)$ such that the inequalities  (\ref{g_in}), (\ref{g_in_b}) hold with equalities, and ${\cal D}_C(Y_1, Y_2)=\Big\{(\Delta_1, \Delta_2)\in [0,\infty] \times [0,\infty]\Big| Q_{(Y_1, Y_2)}- Q_{(E_1, E_2)}>0\Big\}$. 
\end{theorem}

{\bf Proof}  (a) It is known that the value of $I(Y_1, Y_2; \hat{Y}_1, \hat{Y}_2)$ evaluated at any arbitrary joint distribution induced by the random variables $(Y_1,Y_2, \hat{Y}_1, \hat{Y}_2)$ with a given variance, is larger or equal to the value of $I(Y_1, Y_2; \hat{Y}_1, \hat{Y}_2)$ evaluated at a jointly Gaussian distribution, ${\bf P}_{Y_1,Y_2, \hat{Y}_1, \hat{Y}_2}$ induced by  jointly Gaussian random variables $(Y_1,Y_2, \hat{Y}_1, \hat{Y}_2)$, with the same variance (as the arbitrary distributed random variables).  Hence, restrict attention to jointly Gaussian distributed random variables, $(Y_1,Y_2, \hat{Y}_1, \hat{Y}_2)$. 
By property of mutual information, for any measurable functions $g_i(\hat{y}_1,\hat{y}_2), i=1,2$, then 
\begin{align}
I(Y_1, Y_2; \hat{Y}_1, \hat{Y}_2)=&I(Y_1, Y_2; \hat{Y}_1, \hat{Y}_2, g_1(\hat{Y}_1,\hat{Y}_2), g_2(\hat{Y}_1,\hat{Y}_2)) \\
 =&       I(Y_1, Y_2; \hat{Y}_1, \hat{Y}_2| g_1(\hat{Y}_1,\hat{Y}_2), g_2(\hat{Y}_1,\hat{Y}_2)) + I(Y_1, Y_2; g_1(\hat{Y}_1,\hat{Y}_2), g_2(\hat{Y}_1,\hat{Y}_2))\\
   \geq & I(Y_1, Y_2; g_1(\hat{Y}_1,\hat{Y}_2),  g_2(\hat{Y}_1,\hat{Y}_2))    \ \ \mbox{by} \ \   I(Y_1, Y_2; \hat{Y}_1, \hat{Y}_2| g_1(\hat{Y}_1,\hat{Y}_2), g_2(\hat{Y}_1,\hat{Y}_2))\geq 0   \label{jrdf_G1_p_a}  \\
   \geq & I(Y_1, Y_2; {\bf E}\big\{Y_1\Big|F^{\hat{Y}_1, \hat{Y}_2}\big\}, {\bf E}\big\{Y_2\Big|F^{\hat{Y}_1, \hat{Y}_2}\big\}), \ \ \forall  g_i(\hat{y}_1,\hat{y}_2), i=1,2. \label{jrdf_G1_p}
\end{align}
The last inequality establishes (\ref{jrdf_G1}).  Equality in (\ref{jrdf_G1_p_a}) holds if and only if $I(Y_1, Y_2; \hat{Y}_1, \hat{Y}_2| g_1(\hat{Y}_1,\hat{Y}_2), g_2(\hat{Y}_1,\hat{Y}_2))=0$. Further,  if there exists a realization of $(\hat{Y}_1, \hat{Y}_2)$ such that,  the  joint distribution  ${\bf P}_{\hat{Y}_1, \hat{Y}_2,Y_1, Y_2}$ is jointly Gaussian, and  equalities  (\ref{jrdf_G3}) hold, then (\ref{jrdf_G1_p_a}) and  (\ref{jrdf_G1_p}) hold with equality, because for $g_i(\hat{Y}_1, \hat{Y}_2)={\bf E}\big\{Y_i\Big|\hat{Y}_{1}, \hat{Y}_2\big\}=\hat{Y}_i, i=1,2$ then     $\frac{{\bf P}_{\hat{Y}_1, \hat{Y}_2| g_1(\hat{Y}_1,\hat{Y}_2), g_2(\hat{Y}_1,\hat{Y}_2,Y_1, Y_1}}{{\bf P}_{\hat{Y}_1, \hat{Y}_2| g_1(\hat{Y}_1,\hat{Y}_2), g_2(\hat{Y}_1,\hat{Y}_2)}}=1, $almost surely. Finally,  (\ref{jrdf_G2}) is a property of mean square estimation theory. (b) From part (a), restricting attention to jointly distributed Gaussian random variables $(Y_1,Y_2, \hat{Y}_1, \hat{Y}_2)$,    that satisfy (\ref{jrdf_G3}),  it follows that the test channel distribution  ${\bf   P}_{\hat{Y}_1,\hat{Y}_1|Y_1, Y_2}$ is induced  by (\ref{jrdf_G4}), where $H, Q_{(V_1, V_2)}$ satisfy the stated conditions.   (c) The expressions (\ref{jrdf_G8}), (\ref{jrdf_G9}) follow from the realization of part (b). (d) First, apply Algorithm~\ref{alg:canonicalvariabledecomposition}  to the tuple
of Gaussian random variables $(Y_1 , Y_2 ) \in G(0,Q_{(Y_1,Y_2)})$, and the Gaussian random variables $(E_1 , E_2 ) \in G(0,Q_{(E_1,E_2)})$ of part (b). This gives (\ref{jrdf_G10})-(\ref{jrdf_G19}). To obtain (\ref{jrdf_G20})-(\ref{jrdf_G25}), notice that by the error definitions of $(E_1, E_2)$, then the unitary matrices can be taken, without loss of generality,  to  satisfy $U_1=\overline{U}_1, U_2=\overline{U}_2$, i.e., $Q_{Y_i}$ and $Q_{E_i}$ have the same eigenvectors for $i=1,2$. Then (\ref{jrdf_G20}) follows from (\ref{jrdf_G8}) using (\ref{jrdf_G10})-(\ref{jrdf_G19}), and standard properties of a determinant of a matrix.  The remaining equations are obtained from (\ref{jrdf_G10})-(\ref{jrdf_G19}). (e) The lower bound (\ref{g_in_a}) is due to Gray \cite{gray:1973}. The equality is due to  (\ref{g_in_a}), and $(Y_1,Y_2, \hat{Y}_1, \hat{Y}_2)$ are  jointly Gaussian random variables. The equality (\ref{g_in_b}) follows from part (d), by simple calculations. The existence of the strictly positive surface  follows from Gray \cite{gray:1973}, while the specific definition is due to (\ref{jrdf_G20}). \hfill$\square$

\ \

It should be mentioned that prior literature that analyzes the joint RDF $R_{Y_1,Y_2}(\Delta_1,\Delta_2)$, such as,  \cite{veld-gastpar:2016}, does not include statements (a)-(c), (d), (e) of Theorem~\ref{th:jrdf_g}.


\begin{theorem}\label{th:jrdf}
Consider a tuple $(Y_1,Y_2)$ of Gaussian random variables
in the canonical variable form of  Def. \ref{def:grvcommoncorrelatedprivateinfo}.
Restrict attention to the correlated parts of these random variables, as defined in Theorem~\ref{them_putten:schuppen:1983},   by (\ref{ci_par_1})-(\ref{ci_par_3}). Further, consider  a realization of the random variables $(Y_1, Y_2)$  which induce the family of  measures  ${\bf P_{ci}}\subseteq {\cal P}_{min}^{CIG}$,  as defined in Corollary~\ref{cor:commoninfogrvcorrelated}, by (\ref{rep_g_s1})-(\ref{rep_g_s3}).\\ 
Then the following hold.
\begin{itemize}
\item[(a)] The  joint rate distortion function $R_{Y_1, Y_2}(\Delta_1, \Delta_2)$ of $(Y_1, Y_2)$  with square error distortion satisfies 
\begin{align}
R_{{Y}_1, Y_2}(\Delta_1, \Delta_2)=&  \inf_{ \sum_{j=1}^n \Delta_{1,j}\leq \Delta_1, \sum_{j=1}^n \Delta_{2,j}\leq \Delta_2}  \frac{1}{2} \sum_{j=1}^n  \ln \Big( \frac{(1-d_{j}^2)}{\Delta_{1,j} \Delta_{2,j} }\Big), \ \  \ \ (\Delta_1,\Delta_2) \in  {\cal D}_C(Y_1,Y_2),  \label{jrdf_1}\\
{\bf E}||Y_1- \widehat{Y}_1||_{{\mathbb R}^{n}}^2=&\sum_{j=1}^n \Delta_{1,j}, \ \
{\bf E}||Y_2- \widehat{Y}_2||_{{\mathbb R}^{n}}^2=\sum_{j=1}^n \Delta_{2,j}, \label{jrdf_1aa} 
\end{align}
where  ${\cal D}_C(Y_1,Y_2)$ is a strictly positive surface, defined by ${\cal D}_C(Y_1, Y_2)=\Big\{(\Delta_1, \Delta_2)\in [0,\infty] \times [0,\infty]\Big| Q_{(Y_1, Y_2)}- Q_{(E_1, E_2)}>0\Big\}$, where $Q_{(E_1, E_2)}$ is the variance of the errors $E_i=Y_i-\widehat{Y}_i, i=1,2$, with parameters  $\overline{p}_{11}=\overline{p}_{21}=\overline{p}_{12}=\overline{p}_{22}=0$, and $\overline{p}_{13}=\overline{p}_{23}=n$.\\
The  conditional rate distortion functions $R_{Y_i|W }(\Delta_i)$ of $Y_i$  conditioned on $W$ with square error distortion satisfy
\begin{align}
R_{{Y}_i|W}(\Delta_i)=& \inf_{{\bf E}||Y_i- \widehat{Y}_i||_{{\mathbb R}^{n}}^2 \leq \Delta_i}  \frac{1}{2}   \ln \Big( \frac{\det(Q_{Y_i|W})}{\det(Q_{E_i})  }\Big), \ \ i=1, 2,  \\
=&  \inf_{ \sum_{j=1}^n \Delta_{i,j}\leq \Delta_i}  \frac{1}{2} \sum_{j=1}^n  \ln \Big( \frac{(\Lambda_{Y_i|W,j}}{\Delta_{i,j}  }\Big),    \label{crdf_1}\\
{\bf E}||Y_i- \widehat{Y}_i||_{{\mathbb R}^{n}}^2=&\sum_{j=1}^n \Delta_{i,j}, \ \ i=1,2, \label{crdf_2} 
\end{align}
%
\item[(b)] The representations of reproductions $(\widehat{Y}_1, \widehat{Y}_2)$ of $(Y_1, Y_2)$ at the output of decoder 1 and decoder 2, which achieve the joint rate distortion function $R_{{Y}_1, Y_2}(\Delta_1, \Delta_2)$ of part (a),  is    
 \begin{align}
 & Y_1  
  =  D^{1/2}Q_W^{-1} W + Z_1, \label{jrdf_2} \\
&        Y_2 =D^{1/2} W + Z_2, \label{rdf_3a}  \\ 
& \widehat{Y}_1= Y_1 -Q_{E_1}(I-D^{1/2}Q_W^{-1}D^{1/2})^{-1}Z_1 +V_1,      \label{jrdf_2_a} \\
& \ \ \ \ =  D^{1/2}Q_W^{-1} W  + A_1 Z_1 + V_1,       \\
&\widehat{Y}_2 =  Y_2 -Q_{E_2}(I-D^{1/2}Q_W D^{1/2})^{-1}Z_2 +V_2,  \label{jrdf_3}     \\
& \ \ \ \ =  D^{1/2} W  + A_2 Z_2 + V_2, \\
& Z_1 \in G(0,(I-D^{1/2}Q_W^{-1}D^{1/2})), \ \     Z_2 \in G(0,(I-D^{1/2}Q_W D^{1/2})), \ \ W \in G(0,Q_W), \label{jrdf_4}  \\
&Q_{E_1}= {\bf E}\{(Y_1 -\widehat{Y}_1)(Y_1 -\widehat{Y}_1)^T\big\}, \ \ Q_{E_2}= {\bf E}\{(Y_2 -\widehat{Y}_2)(Y_2 -\widehat{Y}_2)^T\big\}, \\
&V_1 \in G(0, Q_{E_1} A_1^T), \ \   V_2 \in G(0, Q_{E_2} A_2^T), \label{jrdf_5} \\  
 &A_1= I- Q_{E_1} (I-D^{1/2}Q_W^{-1}D^{1/2})^{-1}, \ \ A_2= I- Q_{E_2} (I-D^{1/2}Q_WD^{1/2})^{-1}, \\   
 &Q_{E_i}=U_i \Lambda_i U_i^T, \ \    \Lambda_i
  =  \diag (\Delta_{i,1}, \Delta_{i,2}, \ldots, \Delta_{i,n}) \in \rnn, \ \  U_iU_i^T =U_i^TU_i=I, \ \      i=1,2,  \label{rdf_3} \\ 
  &Q_{Y_1|W}=I-D^{1/2}Q_W^{-1}D^{1/2}=U_1 \Lambda_{Y_1|W} U_1^T, \ \    \Lambda_{Y_1|W}=
    \diag (\Lambda_{Y_1|W,1}, \ldots, \Lambda_{Y_1|W,n}) \in \rnn,   \label{rdf_3_n} \\ 
    &Q_{Y_2|W}=I-D^{1/2}Q_WD^{1/2}=U_2 \Lambda_{Y_2|W} U_2^T, \ \    \Lambda_{Y_2|W}=
    \diag (\Lambda_{Y_2|W,1}, \ldots, \Lambda_{Y_2|W,n}) \in \rnn,   \label{rdf_3_nn} \\ 
   &(V_1,V_2,Z_1, Z_2, W), ~~~ \mbox{are independent} \label{jrdf_5}
\end{align}
and it is parametrized by $Q_W \in {\bf Q_W}$, where ${\bf Q_W}$ is  defined by the set of equation  (\ref{eq:pci_1}). \\
Moreover, the joint distribution ${\bf P}_{Y_1, Y_2, \widehat{Y}_1, \widehat{Y}_2, W}$ which achieves $R_{{Y}_1, Y_2}(\Delta_1, \Delta_2)$ satisfies
\begin{align}
&{\bf P}_{Y_1, Y_2, \widehat{Y}_1, \widehat{Y}_2, W}={\bf P}_{\widehat{Y}_1|Y_1, W} {\bf P}_{\widehat{Y}_2|Y_2, W}
 {\bf P}_{Y_1|W}{\bf P}_{Y_2|W}{\bf P}_{W} \label{jrdf_6}\\
& {\bf P}_{Y_1, Y_2, \widehat{Y}_1, \widehat{Y}_2, W}={\bf P}_{Y_1|\widehat{Y}_1} {\bf P}_{Y_2|\widehat{Y}_2}
 {\bf P}_{\widehat{Y}_1|W}{\bf P}_{\widehat{Y}_2|W}{\bf P}_{W}. \label{jrdf_6a}
\end{align}
\item[(c)] Suppose $Q_W=Q_{W^*} \in {\bf Q}_W$ is diagonal, i.e., $Q_{W^*}=\diag (Q_{W_1^*},\ldots, Q_{W_n^*}), \: d_i \leq Q_{W_i^*} \leq d_i^{-1}, \forall i$. Then the conditional RDFs $R_{Y_i|W^*}(\Delta_i)$ are given by 
\begin{align}
R_{Y_1|W^*}(\Delta_1)=&  \inf_{ \sum_{j=1}^n \Delta_{1,j}=\Delta_1}  \frac{1}{2} \sum_{j=1}^n  \log \Big( \frac{(1-d_{j}/Q_{W_j^*})}{\Delta_{1,j} }\Big),   \label{crdf_1-d}\\
R_{Y_2|W^*}(\Delta_2)=&  \inf_{ \sum_{j=1}^n \Delta_{2,j}=\Delta_2}  \frac{1}{2} \sum_{j=1}^n  \log \Big( \frac{(1-d_{j}Q_{W_j^*})}{\Delta_{2,j} }\Big),   \label{crdf_2-d}
\end{align}
and  the water-filling equations hold:
\begin{align}
&
\Delta_{1,j} = \left\{ \begin{array}{cc} \lambda, & \lambda < 1-d_{j}/Q_{W_j^*}  \  \\
1-d_{j}, & \lambda \geq 1-d_{j}/Q_{W_j^*}, \end{array} \right. \: \Delta_1\in [0,\infty), \\
&
\Delta_{2,j} = \left\{ \begin{array}{cc} \lambda, & \lambda < 1-d_{j}Q_{W_j^*}  \  \\
1-d_{j}, & \lambda \geq 1-d_{j}Q_{W_j^*}, \end{array} \right. \: \Delta_2\in [0,\infty).
\end{align}
and the representations of part (b) hold, with $Q_{E_i}, Q_{Y_i|W^*}$ diagonal matrices.  
\end{itemize}
\end{theorem}
{\bf Proof} (a) Since the  attention is restricted to the correlated parts of these random variables, as defined in Theorem~\ref{them_putten:schuppen:1983},   by (\ref{ci_par_1})-(\ref{ci_par_3}), then  the statements of joint RDF $R_{{Y}_1, Y_2}(\Delta_1, \Delta_2)$ of  part (a), are a special case of Theorem~\ref{th:jrdf_g}.(d).  However, these will also follow, from the derivation of part (b).  (b)
Recall that the joint rate distortion function is achieved by a jointly Gaussian distribution ${\bf P}_{Y_1,Y_2, \widehat{Y}_1, \widehat{Y}_2}$ such that the average square-error distortions are satisfied. Consider the realization of the random variables $(Y_1, Y_2)$  which induce the family of  measures  ${\bf P_{ci}}\subseteq {\cal P}_{min}^{CIG}$,  as defined in Corollary~\ref{cor:commoninfogrvcorrelated}, by (\ref{rep_g_s1})-(\ref{rep_g_s3}).  By properties of mutual information then 
 \begin{align}
I(Y_1, Y_2;\widehat{Y}_1, \widehat{Y}_2)=&H(Y_1, Y_2)- I(Y_1, Y_2|\widehat{Y}_1, \widehat{Y}_2)\\
=&H(Y_1, Y_2) - H(Y_1|\widehat{Y}_1, \widehat{Y}_2, Y_2)-  H(Y_2|\widehat{Y}_1, \widehat{Y}_2)\\
=&H(Y_1, Y_2) - H(Y_2|\widehat{Y}_1, \widehat{Y}_2, Y_1)-  H(Y_1|\widehat{Y}_1, \widehat{Y}_2)\\
\geq &H(Y_1, Y_2) - H(Y_1|\widehat{Y}_1)-  H(Y_2|\widehat{Y}_2), \ \ \mbox{by conditioning reduces entropy},  \label{jrdf_10}\\
=&\frac{1}{2} \sum_{i=1}^n \ln (1-d_i^2) + n \ln (2\pi e)- H(Y_1|\widehat{Y}_1)-  H(Y_2|\widehat{Y}_2)\\
\geq & \frac{1}{2} \sum_{i=1}^n \ln (1-d_i^2) + n \ln (2\pi e)-          \frac{1}{2}\sum_{i=1}^n \ln (\Delta_{1,i} ) 
        - \frac{1}{2} n \ln (2\pi e) \nonumber \\
        &-          \frac{1}{2}\sum_{i=1}^n \ln (\Delta_{2,i} ) 
        - \frac{1}{2} n \ln (2\pi e), \ \  \mbox{by Gaussian distribution maximizes entropy}  \\
=& \frac{1}{2} \sum_{i=1}^n \ln \Big(\frac{(1-d_i^2)}{\Delta_{1,i} \Delta_{2,i}}\Big)        
\end{align} 
where $\sum_{i=1}^n \Delta_{1,i} ={\bf E}[ \; ||{Y}_{1}- \widehat{Y}_{1}||_{{\mathbb R}^n}^2]\leq \Delta_1$ and $\sum_{i=1}^n \Delta_{2,i} ={\bf E}[ \; ||{Y}_{2}- \widehat{Y}_{2}||_{{\mathbb R}^n}^2]\leq \Delta_2$.
The average  distortion satisfy
\begin{align}
&\Delta_1 \geq  {\bf E}[ \; ||{Y}_{1}- \widehat{Y}_{1}||_{{\mathbb R}^n}^2] \geq  {\bf E}[\; ||{Y}_{1}- {\bf E}[Y_{1}|F^{\widehat{Y}_1}]||_{{\mathbb R}^n}^2],\label{jrdf_11}   \\
&\Delta_2 \geq  {\bf E}[\; ||{Y}_{1}- \widehat{Y}_{1}||_{{\mathbb R}^n}^2] \geq  {\bf E}[\; ||{Y}_{1}- {\bf E}[Y_{1}|F^{\widehat{Y}_1}]||_{{\mathbb R}^n}^2] . \label{jrdf_12} 
\end{align}
Further, 
\begin{align}
&\mbox{if} \ \  {\bf P}_{Y_1,Y_2|\widehat{Y}_1, \widehat{Y}_2}= {\bf P}_{Y_1|\widehat{Y}_1} {\bf P}_{Y_2|\widehat{Y}_2} \ \ \mbox{then inequality (\ref{jrdf_10}) holds with equality},\label{jrdf_13} \\
&\mbox{if} \ \  {\bf E}[Y_{1}|F^{\widehat{Y}_1}]=\widehat{Y}_1  \ \ \mbox{then inequality (\ref{jrdf_11}) holds with equality,} \label{jrdf_14}\\
&\mbox{if} \ \  {\bf E}[Y_2|F^{\widehat{Y}_2}]=\widehat{Y}_2  \ \ \mbox{then inequality (\ref{jrdf_12}) holds with equality}.\label{jrdf_15}
\end{align} 
It can be  verified  that the representations (\ref{jrdf_2})-(\ref{jrdf_5}) satisfy ${\bf P}_{Y_1,Y_2|\widehat{Y}_1, \widehat{Y}_2}= {\bf P}_{Y_1|\widehat{Y}_1} {\bf P}_{Y_2|\widehat{Y}_2}$, ${\bf E}[Y_{1}|F^{\widehat{Y}_1}]=\widehat{Y}_1$, ${\bf E}[Y_2|F^{\widehat{Y}_2}]=\widehat{Y}_2$, and that all inequalities become equalities. The decomposition of the joint distribution according to equation (\ref{jrdf_6}) follows from the representations of $(\widehat{Y}_1, \widehat{Y}_2)$, and similarly for (\ref{jrdf_6a}).  The conditional RDFs $R_{Y_i|W}(\Delta_i), i=1,2$ are shown as above. (c) This follows directly from parts (a), (b). 
\hfill$\square$

\ \

%
%

{\bf Proof of Theorem~\ref{thm:r0}} One   way to prove the statement is to compute the characterizations of the  rate distortion functions $R_{Y_i}(\Delta_i), R_{Y_i|W}(\Delta_i), i=1,2$ and $R_{Y_1, Y_2}(\Delta_1, \Delta_2)$, using  the  realization of the random variables $(Y_1, Y_2)$  which induce the family of  measures  ${\bf P_{ci}}\subseteq {\cal P}_{min}^{CIG}$,  as defined in Corollary~\ref{cor:commoninfogrvcorrelated}, by (\ref{rep_g_s1})-(\ref{rep_g_s3}). In view of Definition~\ref{wynercommoninforvs}.(b), it suffices to   verify that identity (\ref{equality_1}) holds, i.e., $R_{Y_1|W}(\Delta_1)+R_{Y_2|W}(\Delta_2)+ I(Y_1, Y_2; W)=R_{Y_1, Y_2}(\Delta_1, \Delta_2)$ for $(\Delta_1,\Delta_2) \in {\cal D}_W$, for the choice $W=W^* \in G(0, I)$ which achieves the minimum in  (\ref{w_ic}) (i.e., due to  Theorem~\ref{th:commoninfogrvcorrelated_new}.(b)). \\ 
Similar to Theorem~\ref{th:jrdf}, it can be shown that the conditional RDFs $R_{Y_i|W^*}(\Delta_i), i=1,2$ are given by 
\begin{align}
&R_{{Y}_1|W^*}(\Delta_1) =\inf_{ \sum_{j=1}^n \Delta_{1,j}\leq \Delta_1}  \frac{1}{2} \sum_{j=1}^n  \ln \Big( \frac{(1-d_{j})}{\Delta_{1,j}}\Big), \ \ W^* \in G(0, I) \label{rdf_a}  \\
&R_{{Y}_2|W^*}(\Delta_2)=\inf_{ \sum_{j=1}^n \Delta_{2,j}\leq \Delta_1}\frac{1}{2} \sum_{j=1}^n  \ln \Big( \frac{(1-d_{j})}{\Delta_{2,j}}\Big), \ \  W^* \in G(0, I), \label{rdf_aa} \\
&{\bf E}||Y_1- \widehat{Y}_1||_{{\mathbb R}^{n}}^2=\sum_{j=1}^n \Delta_{1,j}, \ \ \ \
{\bf E}||Y_2- \widehat{Y}_2||_{{\mathbb R}^{n}}^2=\sum_{j=1}^n \Delta_{2,j} . \label{rdf_b}
\end{align}
The pay-off of the joint RDF $R_{Y_1, Y_2}(\Delta_1, \Delta_2)$  in  (\ref{jrdf_1}) is related to the pay-offs of the conditional RDFs $R_{Y_1|W}(\Delta_1), R_{Y_2|W}(\Delta_2)$, and $C(Y_1,Y_2)=I(Y_1, Y_2; W^*)$ in (\ref{lci_1}), via the identity,
\begin{align}
 \frac{1}{2} \sum_{j=1}^n  \ln \Big( \frac{(1-d_{j}^2)}{\Delta_{1,j} \Delta_{2,j} }\Big) = \frac{1}{2} \sum_{j=1}^n  \ln \Big( \frac{(1-d_{j})}{\Delta_{1,j}}\Big)+ \frac{1}{2} \sum_{j=1}^n  \ln \Big( \frac{(1-d_{j})}{\Delta_{2,j}}\Big)+ \frac{1}{2} \sum_{i=1}^n 
        \ln
        \left(
        \frac{1+d_i}{1-d_i}
        \right) . \label{sum_rate}
\end{align}
For $(\Delta_1, \Delta_2) \in {\cal D}_W$ defined by (\ref{dist_reg_ci}), it then follows from  (\ref{sum_rate}), the identity  
\begin{align}
R_{{Y}_1, Y_2}(\Delta_1, \Delta_2)=&  \inf_{ \sum_{j=1}^n \Delta_{1,j}\leq \Delta_1, \sum_{j=1}^n \Delta_{2,j}\leq \Delta_2}  \frac{1}{2} \sum_{j=1}^n  \ln \Big( \frac{(1-d_{j}^2)}{\Delta_{1,j} \Delta_{2,j} }\Big) \\
=& R_{{Y}_1|W^*}(\Delta_1)+ R_{{Y}_2|W^*}(\Delta_2)+ I(Y_1,Y_2;W^*), \ \ \mbox{for} \ \ (\Delta_1,\Delta_2) \in  {\cal D}_W \ \ \mbox{defined by  (\ref{dist_reg_ci})}.
\end{align}
This completes the proof. 
\hfill$\square$

\ \

It is noted that Theorem~\ref{the:scalar-ci}, that corresponds to bivariate Gaussian random variables, i.e., $p_1=p_2=1$,  first derived by Gray and Wyner  \cite{gray-wyner:1974}, and subsequently in \cite{viswanatha:akyol:rose:2014,xu:liu:chen:2016:ieeetit}, is a  strict special case of the above theorem.

\ \

In \cite{charalambous:vanschuppe:2020},  versions of  Theorem~\ref{th:jrdf_g}  and Theorem~\ref{th:jrdf} are used to parametrize rate triples that lie on the Gray-Wyner rate regions, and on the pangloss plane.

The next remark illustrates the application of the results developed in this paper to the optimization problems analyzed in   \cite{satpathy:cuff:2012} and \cite{veld-gastpar:2016,sula:gastpar:2019:arxiv}.

\begin{remark} Applications to problems in \cite{satpathy:cuff:2012} and \cite{veld-gastpar:2016,sula:gastpar:2019:arxiv}\\
\label{rem_lit_rel}
(a) Consider the Gaussian secure source coding and Wyner's common information \cite{satpathy:cuff:2012}. The main optimization problem analyzed in  \cite{satpathy:cuff:2012} (see eqn(18), Section IV.B)  is
\begin{align}
\arg \min_{{\bf P}_{Y_1, Y_2, W}:  {\bf P}_{Y_1, Y_2|W}= {\bf P}_{Y_1|W}{\bf P}_{Y_2|W}  } \Big\{\lambda    I(Y_1;W)+ I(Y_2, Y_2; W)\Big\}, \ \ \lambda \in [0,\infty) \label{SC_2_a}
\end{align}
Suppose the tuple  $(Y_1, Y_2)$ are zero mean jointly Gaussian. Transform the  tuple $(Y_1,Y_2)$ 
in the canonical variable form of  Def. \ref{def:grvcommoncorrelatedprivateinfo}.
Restrict attention to the correlated parts of these random variables, as defined in Theorem~\ref{them_putten:schuppen:1983},  by (\ref{ci_par_1})-(\ref{ci_par_3}).  
By Corollary~\ref{cor:commoninfogrvcorrelated},
\begin{align}
\lambda    I(Y_1;W)+ I(Y_2, Y_2; W)=&\frac{\lambda}{2}  \ln \frac{\det(Q_W)}{\det(Q_W-D)} \nonumber \\
& + \frac{1}{2} \sum_{i=1}^n \ln (1-d_i^2)   - \frac{1}{2} \ln (\det ( [I - D^{1/2} Q_W^{-1} D^{1/2} ]
                                  [ I - D^{1/2} Q_W D^{1/2} ] 
                                )
                          )    \label{sat-cuff}
\end{align}
and it is parametrized by $Q_W \in {\bf Q_W}$, where ${\bf Q_W}$ is  defined by the set of equation  (\ref{eq:pci_1}). Further, similar to the derivation of Theoreem~\ref{th:commoninfogrvcorrelated_new}, i.e., an achievable lower bound on  $I(Y_1, Y_2;W)$ holds,  when $Q_W$ is diagonal,  and  an achievable lower bound on $I(Y_1;W)$ holds, when $Q_W$ is diagonal, i.e., because $Y_1$ has independent components. 
Then an achievable lower bound   on (\ref{sat-cuff})    is obtained, if $Q_W=D_W=\diag(q_1, q_2, \ldots, q_n) \in {\bf Q_W}$,  $q_1 \geq q_2 \geq \ldots \geq q_n>0$,  where ${\bf Q_W}$ is  defined by the set of equation  (\ref{eq:pci_1}). Thus, we obtain 
\begin{align}
&\arg \min_{{\bf P}_{Y_1, Y_2, W}:  {\bf P}_{Y_1, Y_2|W}= {\bf P}_{Y_1|W}{\bf P}_{Y_2|W}  } \Big\{\lambda    I(Y_1;W)+ I(Y_2, Y_2; W)\Big\} \nonumber \\
&= \ \   \arg \min_{ Q_W=D_W \in {\bf Q_W}: \mbox{ ${\bf Q_W}$ defined by (\ref{eq:pci_1})}   }  \Big\{     \frac{\lambda}{2} \ln \frac{\det(D_W)}{\det(D_W-D)}+    \frac{1}{2} \sum_{i=1}^n \ln (1-d_i^2) \nonumber \\
     & \ \ \ \  - \frac{1}{2} \ln (\det ( [I - D^{1/2} D_W^{-1} D^{1/2} ]
                                  [ I - D^{1/2} D_W D^{1/2} ] 
                                )
                          )   \Big\}
\label{SC_2_b}\\
&= \ \   \arg \min_{ Q_W=D_W \in {\bf Q_W}: \mbox{ ${\bf Q_W}$ defined by (\ref{eq:pci_1})}   }  \Big\{     \frac{\lambda}{2} \sum_{i=1}^n \ln ([1-\frac{d_i}{q_i}]^{-1})+    \frac{1}{2} \sum_{i=1}^n \ln (1-d_i^2) \nonumber \\
     & \ \  \ \  - \frac{1}{2} \ln ( [1 - \frac{d_i}{q_i} ]
                                  [ 1- q_i d_i ] 
                              )
                            \Big\}    .             
\end{align}
Since $Q_W=D_W \in {\bf Q_W}$ and  ${\bf Q_W}$ is defined by (\ref{eq:pci_1}), then the remaining optimization can be carried out.\\
Similarly, the methods of this paper can be applied to the problems in \cite{veld-gastpar:2016}.\\
(b) Consider the Gaussian relaxed Wyner's common information considered in  \cite{veld-gastpar:2016,sula:gastpar:2019:arxiv} (see eqn(8), page 8 of \cite{sula:gastpar:2019:arxiv})
\begin{align}
C_\gamma(Y_1,Y_2)=&\min_{ {\bf P}_{W|Y_1,Y_2}:\: I(Y_1,;Y_2|W) \leq \gamma}  I(Y_1, Y_2;W).
                                   \label{sula-gastpar}
\end{align}
By  Proposition~\ref{prop:equivalentgmeasure2grepresentation} or 
Corollary~\ref{cor:commoninfogrvcorrelated}, there exists a family of realizations of $(Y_1, Y_2)$ parametrized by a Gaussian random variable $W$, which induce conditional independence ${\bf P}
_{Y_1, Y_2|W}={\bf P}
_{Y_1|W}{\bf P}
_{ Y_2|W}$, and hence  the lower bound $I(Y_1,Y_2;W) \geq H(Y_1,Y_2)- H(Y_1|W)- H(Y_2|W)$ is achieved, i.e.,   the constraint in (\ref{sula-gastpar}) is always satisfied, because the minimizer is such that  $I(Y_1;Y_2|W)=0$. Hence, the general solution of  (\ref{sula-gastpar})  is the one given in Theorem~
\ref{thm_cia}.  Similarly, for the  problems in \cite{veld-gastpar:2016} (as shown in part (a)).
\end{remark}

\subsection{Parametrization of the Gray and Wyner Rate Region}
Finally, the following parametrization of the Gray-Wyner rate region is presented, which follows directly from Theorem~\ref{th:jrdf}.

\begin{theorem}
\label{thm_par}
 Consider the statement of Theorem~\ref{th:jrdf}.
The region  ${\cal R}_{GW}(\Delta_1, \Delta_2)$
is determined from 
\begin{align}
T(\alpha_1, \alpha_2) =& \inf_{ {\bf Q_W}} \Big\{I(Y_1, Y_2; W)+\alpha_1 R_{Y_1|W}(\Delta_1) + \alpha_2 R_{Y_2|W}(\Delta_2)\Big\} \label{g-w_reg}
\end{align}
$0\leq \alpha_i\leq 1, i=1,2,  \alpha_1+\alpha_2\geq 1$, 
where  $I(Y_1, Y_2; W)$ is given by (\ref{com_inf_g}),    $R_{Y_i|W}(\Delta_i), i=1,2$ are given in  Theorem~\ref{th:jrdf}.(a), (b),  and the infimum is taken over  $Q_W \in {\bf Q_W}$, defined by the set of equation  (\ref{eq:pci_1}).  
\end{theorem}
{\bf Proof}  The stated characterization (\ref{g-w_reg}) is application  of  \cite[(4) of page 1703,  eqn(42)]{gray-wyner:1974} and the  stated results of this paper. 
 \hfill$\square$

\ \

In view of Theorem~\ref{theorem_8} (i.e., Theorem 8 in \cite{gray-wyner:1974}), additional parametrizations of the Gray-Wyner rate region ${\cal R}_{GW}(\Delta_1, \Delta_2)$ follow directly from the expressions already derived,  i.e., $I(Y_1, Y_2; W), R_{Y_1|W}(\Delta_1), R_{Y_2|W}(\Delta_2)$.


\par

\section{Concluding Remarks}\label{sec:concludingremarks}
This paper formulates the  classical Gray and Wyner source coding for a simple network with  a tuple of multivariate, correlated  Gaussian random variables,  from   the geometric approach of a  Gaussian random variables, and  the weak stochastic realization of correlated Gaussian random variables. This approach leads to a very general treatment of such problems, and provides insight which is missing when treated  by other methods. A closed-form expression is derived for Wyner's lossy common information between two multivariate correlated Gaussian random variables, when the decoders apply mean-square error decoding; it corresponds to the minimum common message rate $R_0$  on the Gray and Wyner rate region with sum rate $R_0+R_1+R_2$ equal to the joint rate distortion function, which is constant over a specific rate region. Additional applications are described in \cite{charalambous:schuppen:2019:arxiv}.    However,  much remains to be done to exploit the new approach to other multi-user problems of information theory.

\subsubsection*{Acknowledgements}
The second author is grateful to 
H.S. Witsenhausen (formerly affiliated with Bell Laboratories)
for contacts about the problem of common information
in the early 1980's.
This paper is an answer to his questions about the problem
of Wyner's common information.
\par
The authors are very grateful to the University of Cyprus 
for the partial financial support which made their cooperation possible.
\par
The authors are also grateful to
Prof. Guo Lei (Chinese Academy of Sciences, Institute for Mathematics)
and to Dr. Xi Kaihua (Shandong University, Jinan, Shandong Province, China; 
formerly of Delft University of Technology)
for help with obtaining copies of the papers of Hua LooKeng.
\par
The second named author thanks the Department of Applied Mathematics
of Delft University of Technology for the hospitality arrangement at the
department
during which the paper was written.

\newpage
\begin{footnotesize}
\bibliography{bibliography_paper1_new}
\bibliographystyle{plain}
\end{footnotesize}

\newpage
\appendix

\newpage

\newpage

\newpage

\section{Appendix}

\subsection{Properties of Canonical Variable Form}
\label{sect:cvf_e}
The next theorem states that there always exists a nonsingular basis transformation such that with respect to the new basis
      $(S_1 Y_1 , S_2 Y_2 ) \in G(0,Q_{\cvf}) $
      has the canonical variable form presented
      in Definition~ \ref{def:grvcommoncorrelatedprivateinfo}, and gives some of the properties.

\begin{theorem} Existence of the canonical variable form and properties.\\
\label{thm-exi-cvf}
  Let $ Y_1 : \Omega \to \mathbb{R}^{ p_1 } $
  and $Y_2 : \Omega \to \mathbb{R}^{ p_2 } $
  be jointly Gaussian random variables with
  $(Y_1 , Y_2 ) \in G(0,Q_{(Y_1,Y_2)})$ and $Q_{Y_i} > 0$, for $ i=1, 2$.
  \begin{itemize}
    \item[(a)]
      \it
      Then there exists a nonsingular basis transformation 
\begin{align}      
      S = \blockdiag (S_{1,} S_2 ) \in \mathbb{R}^{(p_1+p_2) \times (p_1+p_2)}
      \end{align}
      such that with respect to the new basis
      $(S_1 Y_1 , S_2 Y_2 ) \in G(0,Q_{\cvf}) $
      has the canonical variable form presented
      in Definition~ \ref{def:grvcommoncorrelatedprivateinfo}.
      It could be that one or more components of the
      canonical variable form are not present for a particular
      probability distribution.
    \item[(b)]
      \it
      Assume that the pair $(Y_1 , Y_2 ) \in G(0,Q_{\cvf}) $
      is in canonical variable form.
      Then the basis transformation
      $S= \blockdiag (S_1 , S_2 ) $ leaves the
      canonical variable form invariant if and only if when,
      \begin{eqnarray*}
        D & = & \blockdiag ( D_1 ,..., D_m ) , ~
           \mbox{with,}\\
        D_i  & = & \diag ( d_i ,..., d_i ) \  = \ 
           d_i I , \ i \neq j \  \Rightarrow \
           d_i \neq d_j , ~ \mbox{then,}\\
        S_1 & = &  
        \blockdiag ( S_{1,1} ,..., S_{1,m} , S_{ 1, m+1 } ) , \\ 
        S_2 & = &  
        \blockdiag ( S_{2,1} ,..., S_{2,m} , S_{ 2, m+1 } ) ,
      \end{eqnarray*}
      are both compatible with the block decomposition of $D$,
      then for all $ i \in Z_m $
      \[
        S_{1,i}^T S_{1,i} \  = \  I , \ \ 
        S_{2,i}^T S_{2,i} \  = \  I , \ \ 
        D_i S_{ 2,i } \  = \ S_{1,i}  D_i ,
      \]
      and 
      $ S_{ 1,m+1 }^T S_{ 1,m+1 } = I , ~ S_{ 2,m+1 }^T S_{ 2,m+1 } = I $.
      In case the canonical variables are all distinct,
      $( \Leftrightarrow \  (i \neq j \Rightarrow d_i \neq d_j ))$,
      then
      \[
        S_1 \  = \ \blockdiag ( S_{1,1} , S_{ 1,m+1 } ) , \ \ 
        S_2 \  = \ \blockdiag ( S_{2,1} , S_{ 2,m+1 } ) , 
      \]
      with $S_{1,1} , S_{2,1} $ sign matrices.
      A {\em sign matrix} is a diagonal matrix with on the diagonal
      only elements of the set $\{ -1, ~ +1 \} \subset \mathbb{R}$.
      \rm
  \end{itemize}
\end{theorem}
{\bf Proof} The result is due to H. Hotelling,
\cite{hotelling:1936}.
A formulation may be found in the book \cite{anderson:1958}.
\hfill$\square$

\subsection{Proof of  Theorem~\ref{them_mi}}
\label{proof_app_thm-exi-cvf} 
By Theorem~\ref{thm-exi-cvf} and  Proposition~\ref{prop:grvcpproperties} there always exists a nonsingular basis transformation such that with respect to the new basis
      $(S_1 Y_1 , S_2 Y_2 ) \in G(0,Q_{\cvf}) $
      has the canonical variable form presented
      in Definition~ \ref{def:grvcommoncorrelatedprivateinfo}. Further, since the basis transformation is nonsingular then
\begin{align}      
       I(Y_1; Y_2)=I(S_1Y_1; S_2Y_2)= I(Y_{11}, Y_{12}, Y_{13}; Y_{21}, Y_{22}, Y_{23}).  \label{eq_bas_1}
       \end{align} 
If $p_{11}=p_{21}>0$ then the componets $Y_{11}, Y_{21}$ are present and since $Y_{11}=Y_{21}-a.s.$, then  the last right hand side entry in (\ref{eq:mi}) is obtained. If $0 = p_{11} = p_{12} = p_{21} = p_{22},
              p_{13} > 0,
              p_{23} > 0$, then by  (\ref{eq_bas_1}), $I(Y_1; Y_2)= I(Y_{13};  Y_{23})=0,$ by independence of $Y_{13}$ and $Y_{23}$. Hence, the first  right hand side entry in (\ref{eq:mi}) is obtained.  Suppose $p_{11}=p_{21}=0$, so that $Y_{11}, Y_{21}$ are not present. By the chain rule of mutual information, then 
 \begin{align}      
       I(Y_1; Y_2)=&I(S_1Y_1; S_2Y_2)= I(Y_{12}, Y_{13}; Y_{22}, Y_{23}) \\
      =& I(Y_{12}; Y_{22}, Y_{23}|Y_{13})+I(Y_{13}; Y_{22}, Y_{23})\\
 =& I(Y_{12}; Y_{22}|Y_{23},Y_{13})+ I(Y_{12}; Y_{23}|Y_{13})+   I(Y_{13}; Y_{22}, Y_{23})\\
 =&I(Y_{12}; Y_{22}), \ \   \ \  \mbox{by  Proposition~\ref{prop:grvcpproperties}, of mutual indep. of $Y_{12}, Y_{23}, Y_{13}$, and $Y_{13}, Y_{22}, Y_{23}$ }\\
 =& \sum_{i=1}^n I(Y_{12,i}; Y_{22,i}),   \ \ \ \ \mbox{by $(Y_{12,i}, Y_{22,i}),$ mutually independent,  Proposition~\ref{prop:grvcpproperties}.}
       \end{align} 
By Proposition~\ref{prop:grvcpproperties}, the variance matrix is 
\begin{align}
     & Q_{(Y_{12},Y_{22})}
   = \left(
        \begin{array}{ll}
          I & D \\
          D & I
        \end{array}
        \right),  \\
      &D
 = \diag (d_1, d_2, \ldots, d_n) \in \rnn, ~~
        1 > d_1 \geq d_2 \geq \ldots \geq d_n > 0.
\end{align}
Hence,  the second  right hand side entry in (\ref{eq:mi}) is obtained, and (\ref{eq:ccc_mi}) follows as well. 
\hfill$\square$

\subsection{Relation of Canonical Variable Form and G\'{a}cs and K\"{o}rner Common Randomness and Wyner's Common Information}
\label{app_sect:gk-w} 
A related notion to Wyner's common information 
is the G\'{a}cs and K\"{o}rner \cite{gacs-korner:1973} definition of common randomness between a tuple of jointly
independent and identically distributed random variables $\{(Y_{1,i},Y_{2,i}): i = 1, \ldots,\}$. Let $f_N^{(E)}, g_N^{(E)}$ be the encoder mappings 
which generate messages $(S_1, S_2)=(s_1, s_2)$, each  with values in a message set ${\cal M}$, defined   by 
\begin{align}
f_N^{(E)}(Y_1^N)=S_1,  \ \ g_N^{(E)}(Y_2^N)=S_2. \label{eq_GK_1}
\end{align} 
Define $\epsilon_N = Prob\{S_1 \neq S_2\}$ and $\rho_N= \frac{1}{N}H(S_1)$. Let $\{(f_N^{(E)}, g_N^{(E)}): N=1, \ldots\}$ be the  sequence of encoder mappings such that $\lim_{N \rightarrow \infty} \epsilon_N=0$, and  $\rho_\infty=\limsup_{N \rightarrow \infty} \frac{1}{N}H(S_1)$. Then it is possible to independently extract approximately $\rho_\infty$  bits per  symbol by 
 observing either one of the two sequences $Y_1^N,
Y_2^N$, as $N \rightarrow \infty$.  
G\'{a}cs and K\"{o}rner \cite{gacs-korner:1973}  common randomness is defined by $C_{GK}(Y_1,Y_2)=\sup \rho_\infty$, where the supremum is taken over all encoder sequences 
$\{(f_N^{(E)}, g_N^{(E)}): N=1, \ldots\}$ such that $\lim_{N \rightarrow \infty} \epsilon_N=0$. \\
It is known from  \cite{wyner:1975}, Remark E, that 
\begin{align}
C_{GK}(Y_1,Y_2) \leq  I(Y_1;Y_2) \leq C(Y_1,Y_2) \label{eq_GK_2}
\end{align}
Moreover, 
\begin{align}
&C_{GK}(Y_1,Y_2) =C(Y_1,Y_2) = I(Y_1;Y_2) \ \ \mbox{if and only if it is possible to represent}  \label{eq_GK_3}\\
&\mbox{ $Y_1 = (V,Y_1^\prime), Y_2 =
(V,Y_2^\prime)$, where $Y_1^\prime$ and $Y_2^\prime$
are conditionally independent given $V$}.\label{eq_GK_4} 
\end{align}
The above condition is equivalent to  the condition that the joint probability mass function of random variables $X,Y$ can be expressed, with a relabeling of
the rows and columns, in terms of a new joint probability mass function,  with specific block elements on its diagonal and zero on its off diagonal elements. Such a transformation is analogous to a pre-processing, by an invariance transformation that leaves  Wyner's common information of $(Y_1,Y_2)$, unchanged.  \\
In the special case discussed by Wyner  \cite{wyner:1975} throughout the paper,
\begin{align}
&\mbox{ $Y_1 = (V,Y_1^\prime), \ \ Y_2 =
(V,Y_2^\prime)$, where $Y_1^\prime, Y_2^\prime, V$  are independent, then }  \label{wyner-sc1}  \\
&  C_{GK}(Y_1,Y_2) =C(Y_1,Y_2) = H(V) \label{wyner-sc2}
\end{align} 
where $H(V)$ is the entropy of $V$.\\
Next, two special cases of  the canonical variable form of Definition~\ref{def:grvcommoncorrelatedprivateinfo}, are discussed. \\
Case (1). If  $Y_{11}$ is absent then $Y_{21}$ is absent, and vice-versa,  that is, $p_{11}=p_{21}=0$, and   then 
\begin{align}
&S_1Y_1 = (V_1,Y_1^\prime) = (Y_{12},Y_{13}), \ \  S_2Y_2 = (V_2,Y_2^\prime) = (Y_{22},Y_{23}), \\
&  Y_{13} \ \ \mbox{and}\ \  Y_{23} \ \ \mbox{are independent  and each of these has independent components},\label{cvf_2_rem1}  \\
&  Y_{12} \ \ \mbox{and}\ \  Y_{22} \ \ \mbox{are correlated and each of these has independent components},\label{cvf_2a_rem_2}  \\
& Y_{22}, Y_{23}, Y_{13}  \ \ \mbox{are independent and}, \ \  Y_{12}, Y_{13}, Y_{23}  \ \ \mbox{are independent},\\
&    {\bf E}[ Y_{12} Y_{22}^T ] = D.
\end{align}
Case (2). If  $Y_{12}, Y_{22}$ are absent, that is, $p_{12}=p_{22}=0$,  then 
\begin{align}
&S_1Y_1 = (V_1,Y_1^\prime) = (Y_{11},Y_{13}), \ \  S_2Y_2 = (V_2,Y_2^\prime) = (Y_{21},Y_{23}),\\
& Y_{11}=Y_{21}-a.s. \ \ \Longleftrightarrow \ \ V_1=V_2-a.s.\\
&  Y_{11}, Y_{13}, Y_{23} \ \  \mbox{are independent  and each of these has independent components},\label{cvf_2_rem3}  \\
&  Y_{21},   Y_{23}, Y_{13} \ \ \mbox{are independent  and each of these has independent components}. \label{cvf_2a_rem4}  
&
\end{align}
Hence, Case (2) is the analog of (\ref{wyner-sc1}), i.e., special case discussed by Wyner  \cite{wyner:1975}, hence it follows that $C_{GK}(Y_1,Y_2) =C(Y_1,Y_2) = H(Y_{11})=H(Y_{21})$, and $Y_{11} = Y_{21}-a.s.  \in G(0,I_{p_{11}})$.

It is remarked that the canonical variable form is directly
applicable to the lossy extension of G\'{a}cs and K\"{o}rner common information derived by Viswanatha, Akyol
and Rose \cite{viswanatha:akyol:rose:2014} in Section IV. Specifically, the information theoretic characterization of Theorem 3 in \cite{viswanatha:akyol:rose:2014}.

\subsection{Proof of 
Proposition~\ref{prop:equivalentgmeasure2grepresentation}}
\label{proof_app_prop:equivalentgmeasure2grepresentation}
(a) ($\Leftarrow$)
From the assumption that each of the following three random variables
$V_1, ~ V_2, X$ is Gaussian and that they are independent,
follows that $(Y_1, Y_2, X)$ are jointly Gaussian.
From the assumptions 
follows that $G_0 = G_1|_{\mathbb{R}^{p_1} \times \mathbb{R}^{p_2}}$, i.e., the specified restriction of $G_1$ is $G_0$.
\par
That conditional independence holds for
$(Y_1, Y_2 | X) \in \cig$ 
follows from,
\begin{align}
	{\bf E}[Y_1 | X]
	& = {\bf E}[ C_1 X + N_1 V_1 | X]
	      = C_1 X + N_1 {\bf E}[V_1] = C_1 X
	      = Q_{Y_1,X} Q_X^{-1} X, \\
	    C_1
	& = Q_{Y_1,X} Q_X^{-1}, ~ \mbox{because} ~ Q_X > 0,
\end{align}
where the well known formula is used for conditional expectation
of two jointly Gaussian random variables for
${\bf E}[ Y_1 | X ] = Q_{Y_1,X} Q_X^{-1} X$.
Similarly, $C_2 = Q_{Y_2,X} Q_X^{-1}$.
From this follows that,
\begin{eqnarray*}
	Q_{Y_1,X} Q_X^{-1} Q_{Y_2,X}^T 
	& = & C_1 Q_X C_2^T = Q_{Y_1,Y_2},
\end{eqnarray*}
where the last equality follows from equation (\ref{eq:q12factorization}).
From the obtained equality follows that
the conditional independence of $(Y_1, Y_2|X) \in \cig$ holds.\\
(a) ($\Rightarrow$)
Consider a triple of jointly Gaussian random variables
$(Y_1, Y_2, X)$ such that conditional independence holds,
$(Y_1, Y_2 | X) \in \cig$.
From the conditional independence and Proposition \ref{prop:cigrvs}
follows that,
$Q_{Y_1, Y_2} = Q_{Y_1,X} Q_X^{-1} Q_{Y_2,X}^T$.
\par
Because these random variables are jointly Gaussian
the following conditional expectations have the formulated
form and one can construct the indicated random variables,
\begin{align}
	    {\bf E}[Y_1|X]
	= & Q_{Y_1,X} Q_X^{-1} X = C_1 X, ~~
	      C_1 = Q_{Y_1,X} Q_X^{-1}, \\
	    N_1 V_1
	= & Y_1 - C_1 X, \\
	    {\bf E}[Y_2|X]
	= & Q_{Y_2,X} Q_X^{-1} X = C_2 X, ~~
	      C_2 = Q_{Y_2,X} Q_X^{-1}, \\
	   N_2 V_2
	= & Y_2 - C_2 X, \\
	    {\bf E}[N_1 V_1 X^T ]
	= & {\bf }E[ Y_1 X^T - C_1 X X^T] = Q_{Y_1,X} - C_1 Q_X = 0, \\
	    {\bf E}[N_2 V_2 X^T ]
	= & {\bf E}[ Y_2 X^T - C_2 X X^T] = Q_{Y_2,X} - C_2 Q_X = 0, \\
	    {\bf E}[N_1V_1 (N_2V_2)^T]
	= &{\bf  E}[ (Y_1 - C_1 X) ( Y_2 - C_2 X)^T ]
	      = Q_{Y_1,Y_2} - C_1 Q_{X,Y_2} - Q_{Y_1,X} C_2^T + C_1 Q_X C_2 
	      = 0
\end{align}
by the formulas for $C_1$, $C_2$, and the one for conditional independence.
Note that the above also establishes that
$(X, V_1, V_2)$ are independent random variables.
Thus one obtains the representation of Definition~ \ref{def:wgsrgrvs}.(b).\\
(b) ($\Rightarrow$)
By Proposition \ref{prop:cigrvs} (shown in \cite{putten:schuppen:1983}) 
 a weak Gaussian realization is minimal
if and only if
$\rank (Q_{Y_1,Y_2}) = n$.
From the proof of (a) $\Leftarrow$ or $\Rightarrow$
follows that,
\begin{eqnarray}
	 Q_{Y_1,Y_2} = Q_{Y_1,X} Q_X^{-1} Q_{Y_2,X}^T = C_1 Q_X C_2^T.
\end{eqnarray}
\par
From the above quoted characterization of minimality follows that,
\begin{eqnarray}
	n
	=  \rank (Q_{Y_1,Y_2}) 
	      = \rank( C_1 Q_X C_2^T),
\end{eqnarray}
hence that $\rank (C_1) = n = \rank(C_2)$.\\
(b) ($\Leftarrow$)
From the rank assumption and the obtained relation follows that,
\begin{eqnarray}
	&   & n = \rank (C_1 Q_X C_2^T) = \rank (Q_{Y_1,Y_2}),
\end{eqnarray}
thus minimality holds.
\hfill$\square$.

\subsection{Information Theory}
The reader finds in this appendix two formulas
of information theory which are used in the body of the paper. These are obtained from 
\cite{cover:thomas:1991,gallager:1968,yaglom:yaglom:1983}.

\par
\par
The first equality is proven in
\cite[p. 31, (2.4.26), (2.4.28)]{gallager:1968}
and \cite[p. 19, Th. 2.4.1]{cover:thomas:1991}.
\par
\begin{proposition}\label{proposition:mutualinfoindependent}
Consider random variables $Y_{1,1}, Y_{1,2}, Y_{2,1}, Y_{2,2}, X_1, X_2$
such that the following two triples are independent random variables,
$(Y_{1,1}, Y_{2,1}, X_1)$ and $(Y_{1,2}, Y_{2,2}, X_2)$.
Then the mutual information expression decomposes additively,
\begin{eqnarray}
      I(Y_{1,1}, Y_{1,2}, Y_{2,1}, Y_{2,2}; X_1, X_2)
   =  I(Y_{1,1}, Y_{2,1}; X_1) + I(Y_{1,2}, Y_{2,2}; X_2).
\end{eqnarray}
\end{proposition}
{\bf Proof}
The independence of the random variables
and the following calculations establish the equality,
\begin{align*}
        &I(Y_{1,1}, Y_{1,2}, Y_{2,1}, Y_{2,2}; X_1, X_2)
      \\
  & =  H(Y_{1,1}, Y_{1,2}, Y_{2,1}, Y_{2,2}) + H(X_1,X_2) 
        - H(Y_{1,1}, Y_{1,2}, Y_{2,1}, Y_{2,2}, X_1, X_2) \\
  & = H((Y_{1,1}, Y_{2,1}), (Y_{1,2}, Y_{2,2}) ) 
        + H(X_1) + H(X_2) - H((Y_{1,1}, Y_{2,1}, X_1), ~ ( Y_{1,2}, Y_{2,2}, X_2)) \\
  & =  \Big\{ H(Y_{1,1}, Y_{1,2}) + H(X_1) - H(Y_{1,1}, Y_{2,1}, X_1) \Big\} 
        + \Big\{ H(Y_{1,2}, Y_{2,2}) + H(X_2) - H(Y_{1,2}, Y_{2,2}, X_2) \Big\} \\
  & = I(Y_{1,1}, Y_{2,1};  X_1) 
        + I(Y_{1,2}, Y_{2,2}; X_2).
\end{align*}
\hfill$\square$
\par\vspace{1\baselineskip}\par\noindent
%



Consider a tuple of jointly Gaussian random variables
$(X,Y) \in G(0,Q_{(X,Y)})$ with 
$X: \Omega \rightarrow \mathbb{R}^n$, 
$Y: \Omega \rightarrow \mathbb{R}^p$, 
$Q_{(X,Y)} > 0$, and
$Q_{(X,Y)} = \left(
     \begin{array}{ll}
       Q_X & Q_{X,Y} \\
       Q_{X,Y}^T & Q_Y
     \end{array}
     \right)
$.
Then,
\begin{align}
     H(X)
   = & \frac{1}{2} \ln \Big(\det (Q_X)\Big) + \frac{1}{2} n \ln (2 \pi e), \\
      H(Y|X)
   = & H(X,Y) - H(X)
        = \frac{1}{2} \ln \Big(  \det(Q_Y - Q_{X,Y}^T Q_X^{-1} Q_{X,Y})\Big)
          + \frac{1}{2} p \ln( 2 \pi e), \\
      I(Y;X)
   = & - \frac{1}{2} \ln \Big( \frac{\det(Q_{(X,Y)})}{\det (Q_Y) \det (Q_X)}\Big).
         \label{eq:mutualinfogrv}
\end{align}

\subsection{An Inequality for Determinants}\label{ap:inequality}
An inequality for matrices is derived in this appendix
which is needed in the body of the paper.
\begin{lemma}\label{lemma:detinequalityasymmetric}
Consider the real-valued matrices
$A, ~ B \in \mathbb{R}^{n \times n}$.
Assume that,
\begin{eqnarray}
         0
  & \leq & I - A^T A, ~~
           0 < I - B^T B, ~ \mbox{and} ~
           \rank (B) = n.
\end{eqnarray}
Then,
\begin{eqnarray}
  &   & \det( [ I - A^T A ] [ I - B^T B ] )
        \leq ( \det (I - A^T B) )^2.
\end{eqnarray}
\end{lemma}
A related result is mentioned at
\cite[Th. 9.E.6]{marshall:olkin:1979}.
In that book the proof of corresponding result
is referred to the paper \cite{hua:1955a}.
That reference has been received by the authors 
but they cannot read it because the paper is in Chinese.
However, the formulas of the paper they can read.
Hua LooKeng developed these results to calculate
an orthonormal basis for a function of one complex variable.
A more recent reference for this inequality is
\cite[Th. 7.19]{zhang:2011}.
\par
The proof of Lemma \ref{lemma:detinequalityasymmetric}
below is analogous to that of Hua LooKeng in \cite{hua:1955a}.
The main differences are in the assumptions.
%

\begin{lemma}\label{lemma:matrixequality}
\cite[p. 464, 470]{hua:1955a}.
Consider the matrices $A,B \in \rnn$.
Assume that 
$I - B^T B$ is a nonsingular matrix and that
$\rank (B) = n$.
Then
\begin{eqnarray}
  &   & (I - A^T A) 
       - (I - A^T B) [ I - B^T B]^{-1} (I - A^T B)^T \nonumber \\
  & = & - ( A - B) [ I - B^T B]^{-1} ( A - B)^T.
\end{eqnarray}
\end{lemma}
{\bf Proof}
Obviously,
\begin{eqnarray}
  &   &  B^T ( I - B B^T) = ( I - B^T B) B^T, ~~
         (I - B^T B) B = B (I - B^T B). \label{eq:bbb}
\end{eqnarray}
From this equation, the assumptions that
$I - B^T B$ is nonsingular and that $\rank (B) = n$
follows that
$I - B B^T$ is nonsingular.
\par
Note the following calculations.
\begin{eqnarray}
  &   &  B^T ( I - B B^T) = ( I - B^T B) B^T, \nonumber \\
  & \Leftrightarrow & 
         B^T ( I - B B^T)^{-1} = ( I - B^T B)^{-1} B^T, 
         \label{eq:ab1} \\
  & \Leftrightarrow & 
         ( I - B B^T)^{-1} B = B ( I - B^T B)^{-1}; 
         \label{eq:ab2} \\
      I
  & = & I - B^T B + B^T B \nonumber \\
  & = & (I - B^T B) 
        + B^T ( I - B B^T)^{-1} ( I - B B^T) B \nonumber \\
  & = & (I - B^T B) 
        + B^T ( I - B B^T)^{-1} B ( I - B^T B), ~
        \mbox{by (\ref{eq:bbb}),} \nonumber \\
  & \Leftrightarrow & \nonumber \\
      (I - B^T B)^{-1}
  & = & I + B^T ( I - B B^T)^{-1} B; 
        \label{eq:ab3} \\
      I
  & = & I - B B^T + B B^T \nonumber \\
  & = & (I - B B^T) 
        + B (I - B^T B)^{-1} ( I - B^T B) B^T \nonumber \\
  & = & (I - B B^T) 
        + B (I - B^T B)^{-1} B^T ( I - B B^T), ~
        \mbox{by (\ref{eq:ab2}),}   \nonumber \\
  & \Leftrightarrow & \nonumber \\
      (I - B B^T)^{-1}
  & = & I + B (I - B^T B)^{-1} B^T; 
        \label{eq:ab4}\\
  &   & (I - A^T A) 
        - (I - A B^T) ( I - B B^T)^{-1} ( I - A B^T)^T \nonumber \\
  & = & I - [I - B B^T]^{-1} 
        + A B^T [ I - B B^T]^{-1} 
        + [ I - B B^T]^{-1} B A^T + \nonumber \\
  &   & - A [ I + B^T ( I - B B^T)^{-1} B ] A^T, ~ \nonumber \\
  & = & - B [ I - B^T B]^{-1} B^T
        + A [ I - B^T B]^{-1} B^T
        + B [ I - B^T B]^{-1} A^T + \nonumber \\
  &   & - A [ I - B^T B]^{-1} A^T \nonumber \\
  &   & \mbox{using respectively (\ref{eq:ab4}), (\ref{eq:ab1}), 
                (\ref{eq:ab2}), and (\ref{eq:ab3}),
             }  \nonumber \\
  & = & - (A - B) [ I - B^T B]^{-1} (A - B)^T. \nonumber
\end{eqnarray}
\hfill$\square$ 
\par\vspace{1\baselineskip}\par\noindent
%

\begin{proposition}\label{proposition:detinequality}
\cite[eq. (2)]{hua:1955a}.
Consider the symmetric positive-definite matrices
$Q_1, ~ Q_2, ~ Q \in \rnn$ such that
$Q_1 + Q_2 = Q$.
Then
\begin{eqnarray}
  \det ( Q_1) + \det (Q_2) \leq \det (Q).
\end{eqnarray}
\end{proposition}
{\bf Proof}
By \cite[Th. 12.7]{noble:1969}
and because $Q_1$ and $Q_2$ are positive-definite,
there exists a matrix $S \in \rnn$
and a diagonal matrix $D \in \rnn$ 
with $D = \diag (d_1, \ldots, d_n) \geq 0$
such that
$ Q_1 = S S^T$ and $Q_2 = S D S^T$.
Then,
\begin{eqnarray*}
    Q
  & = & Q_1 + Q_2 = S ( I + D ) S^T, \\
      \det (Q)
  & = & \det (S) \det (I + D) \det (S^T) 
        = \det (S) \prod_{i=1}^n (1 + d_i) \det (S^T) \\
  & \geq & \det (S) [ 1 + \prod_{i=1}^n d_i ] \det (S^T)
        = \det (S S^T) + \det (S D S^T)
        = \det (Q_1) + \det (Q_2).
\end{eqnarray*}
\hfill$\square$
\par\vspace{1\baselineskip}\par\noindent
{\bf Proof of Lemma~\ref{lemma:detinequalityasymmetric}.}
By the assumptions,
$0 \leq  (I - A^T A)$, 
$0 <  (I - B^T B)$, and
$\rank(B) =n$,
from Lemma~\ref{lemma:matrixequality}
follows that
\begin{eqnarray*}
  &   & (A-B) [ I - B^T B]^{-1} (A-B)^T + [ I - A^T A ] \\
  & = & (I - A^T B) [ I - B^T B]^{-1} (I - A^T B)^T; ~
        \label{eq:atb} \\
  &   & 0 \leq \det ( I - A^T A ), ~
        \mbox{by the assumption on $A$,} \\
  & \leq  & \det ( (A-B) [ I - B^T B]^{-1} (A-B)^T )
            + \det([I - A^T A]),  
            \mbox{by an assumption on $B$,} \\
  & \leq & \det ( (I-A^T B) [ I - B^T B]^{-1} (I -A^T B)^T ) ~ \\
  &   & \mbox{by Lemma~\ref{lemma:matrixequality}, 
                Proposition~\ref{proposition:detinequality}, and 
                by the assumptions,
             } \\
  & = & \left(
          \det(I - A^T B) 
        \right)^2 [ \det ([I - B^T B]) ]^{-1}, \\
  & \Rightarrow  & \det( [ I - A^T A ] [ I - B^T B ] ) 
        = \det( [ I - A^T A ]) \det( [ I - B^T B] ) 
        \leq \left(
               \det( [ I - A^T B] )
             \right)^2.
\end{eqnarray*}
\hfill$\square$
\par\vspace{1\baselineskip}\par\noindent
Another preliminary result is needed.
\begin{proposition}\label{proposition:inverseinequality}
Consider the matrix 
$Q_X \in \spd$ 
of Prop. \ref{prop:cigrvs}
and the matrix $D \in \mathbb{R}^{n \times n}$
of Def. \ref{def:grvcommoncorrelatedprivateinfo}.
Thus both $Q_X > 0$ and $D > 0$.
Then,
\begin{eqnarray}
  &   & D \leq Q_X^{-1} \leq D^{-1} ~~ 
        \Leftrightarrow ~~
        D \leq Q_X \leq D^{-1}, \\
  &   & D < Q_X^{-1} < D^{-1} ~~ 
        \Leftrightarrow ~~
        D < Q_X < D^{-1}.
\end{eqnarray}
\end{proposition}
{\bf Proof}
Consider a symmetric matrix $Q \in \mathbb{R}^{n \times n}$
satisfying $0 < Q < I$. 
Note that in general $Q \neq Q_X$.
Its square root exists and satisfies, 
$0 < Q^{1/2} = (Q^{1/2})^T \in \mathbb{R}^{n \times n}$.
Then 
\begin{eqnarray}
  &   & Q \leq I 
        \Leftrightarrow I = Q^{-1/2} Q Q^{-1/2} \leq Q^{-1} 
        \Leftrightarrow I \leq Q^{-1}. 
        \label{eq:inverseinequality}
\end{eqnarray}
Hence,
\begin{eqnarray*}
  &                 & Q_X \leq D^{-1} 
                      \Leftrightarrow ~ D^{1/2} Q_X D^{1/2} \leq I \\
  & \Leftrightarrow & I \leq D^{-1/2} Q_X^{-1} D^{-1/2},
                      \mbox{by equation (\ref{eq:inverseinequality}),}  \\
  & \Leftrightarrow & D \leq Q_X^{-1}; \\
  &                 & D \leq Q_X  ~
                      \Leftrightarrow ~ I \leq D^{-1/2} Q_X D^{-1/2} \\
  & \Leftrightarrow & D^{1/2} Q_X^{-1} D^{1/2} \leq I, ~
                      \mbox{by equation (\ref{eq:inverseinequality}),}  \\
  & \Leftrightarrow & Q_X^{-1} \leq D^{-1}.
\end{eqnarray*}
The proof for the case with strict inequalities is similar.
\hfill$\square$
\par\vspace{1\baselineskip}\par\noindent

%
\begin{proposition}\label{proposition:matrixineqdqx}
Consider the matrices defined in 
Def. \ref{def:grvcommoncorrelatedprivateinfo}.
Thus,
$D \in \mathbb{R}^{n \times n}$, 
is a diagonal matrix satisfying $0 < D$,
and the matrix $Q_X \in \mathbb{R}^{n \times n}$
satisfies $Q_X = Q_X^T$ and
$0 < D \leq Q_X \leq D^{-1}$.
Then,
\begin{eqnarray}
  &   & \det 
        \left(
           [ I - D^{1/2} Q_X^{-1} D^{1/2} ]
           [ I - D^{1/2} Q_X      D^{1/2} ]
        \right) 
        \leq 
        \det ( [ I - D ]^2 ), ~
        \forall ~ Q_X ~ \mbox{considered, while,} 
        \label{eq:detinequalithdqx} \\
  &   & \det 
        \left(
           [ I - D^{1/2} Q_X^{-1} D^{1/2} ]
           [ I - D^{1/2} Q_X      D^{1/2} ]
        \right) 
        < \det ( [ I - D ]^2 ), ~ 
        \mbox{if} ~ Q_X \neq I.
        \label{eq:detinequalitystrictdqx}
\end{eqnarray}
\end{proposition}
{\bf Proof}
(1) If $Q_X < D^{-1}$ then $\det(D^{-1} - Q_X) > 0$.
Consider the case of in which $Q_X$ satisfies
$Q_X \leq D^{-1}$ but not $Q_X < D^{-1}$.
Then $\det(D^{-1} - Q_X) = 0$.
Hence,
\begin{eqnarray*}
  &   &  \det ( I - D^{1/2} Q_X D^{1/2} ) 
         = \det ( D^{1/2} [ D^{-1} - Q_X ] D^{1/2} ) 
         = \det(D^{1/2}) \det(D^{-1} - Q_X) \det(Q^{1/2}) = 0.
\end{eqnarray*}
Then $0 < D < I$ implies that $\det([I - D]^2) \geq 0$
hence that the inequality (\ref{eq:detinequalithdqx}) holds.
\par
(2) If $D < Q_X$ then $\det(D - Q_X) < 0$.
If $D \leq Q_X$ but not $D < Q_X$
then by Proposition \ref{proposition:inverseinequality}
$Q_X^{-1} \leq D^{-1}$ but not $Q_X^{-1} < D^{-1}$.
Then $\det(D^{-1} - Q_X^{-1}) = 0$.
Hence,
\begin{eqnarray*}
  &   &  \det ( I - D^{1/2} Q_X^{-1} D^{1/2} ) 
         = \det ( D^{1/2} [ D^{-1} - Q_X^{-1} ] D^{1/2} ) 
         = \det(D^{1/2}) \det(D^{-1} - Q_X^{-1}) \det(D^{1/2})
         = 0.
\end{eqnarray*}
In this case the inequality (\ref{eq:detinequalithdqx}) also holds.
\par
(3) 
Next consider the case in which $D < Q_x < D^{-1}$. 
Lemma \ref{lemma:detinequalityasymmetric} 
will be used to prove the result.
Define therefore,
\begin{eqnarray*}
      A 
  & = & (Q_X^{- 1/2}) D^{1/2}, ~~
      B 
      = (Q_X^{1/2}) D^{1/2}.
\end{eqnarray*}
First it is proven that the assumptions of the lemma are satisfied.
Note that $0 < Q_X$ implies that $\rank (Q_X) = n$.
This and the fact that $\rank (D) = n$
imply that $\rank (B) = \rank ( Q_X^{1/2} D^{1/2}) = n$.
Note further that,
\begin{eqnarray*}
      I - A^T A
  & = & I - D^{1/2} Q_X^{-1} D^{1/2}, \\
  &   & 0 < D \leq Q_X \leq D^{-1} ~
        \mbox{by assumption,} ~
        \Rightarrow \\
      0 
  & < & D \leq Q_X^{-1} \leq D^{-1}, ~~~ 
        \mbox{by Proposition \ref{proposition:inverseinequality},} \\
      0 
  & < & D^2 
        \leq D^{1/2} Q_X^{-1} D^{1/2} \leq I ~
           \Rightarrow \\
      0
  & \leq & I - D^{1/2} Q_X^{-1} D^{1/2}
           = I - A^T A; \\
  &   & 0 < D < Q_X < D^{-1} ~
        \mbox{by the case considered,} \\
      0 
  & < & D^2 < D^{1/2} Q_X D^{1/2} < I, ~
        \Rightarrow \\
      0
  & < & I - D^{1/2} Q_X D^{1/2}
           = I - B^T B; \\
      I - A^T B
  & = & I - D^{1/2} Q_X^{-1/2} Q_X^{1/2} D^{1/2} 
        = I - D.
\end{eqnarray*}
From Lemma \ref{lemma:detinequalityasymmetric} then follows that,
\begin{eqnarray*}
  &   & \det
        \left(
          [ I - D^{1/2} Q_X^{-1} D^{1/2} ]
          [ I - D^{1/2} Q_X D^{1/2} ]
        \right) 
        \leq  \det ( [ I - D ]^2 ).
\end{eqnarray*}
(4) Next suppose that in addition $Q_X \neq I$.
Then 
\begin{eqnarray*}
      A - B
  & = & Q_X^{- 1/2} D^{1/2} - Q_X^{1/2} D^{1/2}
        = Q_x^{-1/2} [ I - Q_X ] D^{1/2} \neq 0, ~
        \mbox{using that,} ~
        Q_X \neq I, \\
      0
  & < & (A - B) [I - B^T B]^{-1} (A - B)^T, ~
        \mbox{because} ~ 0 < I - B^T B, ~
        \mbox{and} ~ A - B \neq 0, \\
  &   & \det ( I - A^T A)  \\
  & < & \det ( (A - B) [I - B^T B]^{-1} (A-B) ) 
        + \det( I - A^T A), \\
  & \leq & \det ( ( I - A^T B) [ I - B^T B]^{-1} ( I - A^T B)^T ), ~
           \mbox{by Lemma \ref{lemma:matrixequality},} \\ 
  & \Rightarrow & \det ( ( I - A^T A) (I - B^T B) )
                  = \det ( I - A^T A) \det (I - B^T B) \\
  & < & [ \det ( I - A^T B) ]^2, ~
        \mbox{by Lemma \ref{lemma:matrixequality} and its proof,} \\
  & \Rightarrow  & \det( [ I - D^{1/2} Q_X^{-1} D^{1/2} ]
              [ I - D^{1/2} Q_X D^{1/2} ]
            ) 
        < \det( [ I - D]^2), ~
              \mbox{by substitution of $A$ and $B$.}
\end{eqnarray*}
\hfill$\square$
\par\vspace{1\baselineskip}\par\noindent 
\newpage
\newpage
\tableofcontents
\end{document}